\documentclass[prb,twocolumn,amsmath,amssymb,unsortedaddress,eqsecnum,floatfix]{revtex4}
\usepackage{graphicx}
\usepackage{bm}
\usepackage{hyperref}

\begin{document}

\preprint{cond-mat/0112343}

\title{Competing orders in a magnetic field:\\
spin and charge order in the cuprate superconductors}

\author{Ying Zhang}
 \email{ying.zhang@yale.edu}
\affiliation{Department of Physics, Yale University, P.O. Box
208120, New Haven CT 06520-8120}

\author{Eugene Demler}
 \email{demler@cmts.harvard.edu}
\affiliation{Department of Physics, Harvard University, Cambridge
MA 02138}
\author{Subir Sachdev}%
 \email{subir.sachdev@yale.edu}
 \homepage{http://pantheon.yale.edu/~subir}
\affiliation{Department of Physics, Yale University, P.O. Box
208120, New Haven CT 06520-8120}

\date{April 17, 2002}

\begin{abstract}
We describe two-dimensional quantum spin fluctuations in a
superconducting Abrikosov flux lattice induced by a magnetic field
applied to a doped Mott insulator. Complete numerical solutions of
a self-consistent large $N$ theory provide detailed information on
the phase diagram and on the spatial structure of the dynamic spin
spectrum. Our results apply to phases with and without long-range
spin density wave order and to the magnetic quantum critical point
separating these phases. We discuss the relationship of our
results to a number of recent neutron scattering measurements on
the cuprate superconductors in the presence of an applied field.
We compute the pinning of static charge order by the vortex cores
in the `spin gap' phase where the spin order remains dynamically
fluctuating, and argue that these results apply to recent scanning
tunnelling microscopy (STM) measurements. We show that with a
single typical set of values for the coupling constants, our model
describes the field dependence of the elastic neutron scattering
intensities, the absence of satellite Bragg peaks associated with
the vortex lattice in existing neutron scattering observations,
and the spatial extent of charge order in STM observations. We
mention implications of our theory for NMR experiments. We also
present a theoretical discussion of more exotic states that can be
built out of the spin and charge order parameters, including spin
nematics and phases with {\em `exciton fractionalization'}.
\end{abstract}

\maketitle

\section{Introduction}
\label{sec:intro}

The determination of the ground state of the cuprate
superconductors as a function of the hole density has been one of
the central problems in condensed matter physics in the last
decade. At zero hole density, it is well established that the
ground state is a Mott insulator with long range magnetic N\'{e}el
order. At moderate hole density, it is also widely accepted that
the ground state is a $d$-wave superconductor, all of whose
important qualitative properties are identical those of the
standard BCS-BdG theory. At issue are the ground states which
interpolate between these well understood limits, and the manner
in which they influence the anomalous properties at temperatures
($T$) above $T_c$ (the critical temperature for the onset of
superconductivity).

While a plethora of interesting proposals for these intermediate
states have been made, we will focus here on (in our view) the
simplest possibility: the order parameters characterizing the
intermediate ground states are simply those of spin and charge
density waves (SDW and CDW), and superconductivity (SC) itself.
Apart from a small range at very low doping, which shall not be of
interest in this paper, we know from neutron scattering
experiments that there is SDW order collinearly polarized at the
wavevectors
\begin{equation}{\bf K}_{s x} =
\left(\frac{2\pi}{a}\right)\left(\frac{1}{2}-\vartheta,\frac{1}{2}\right)~~,~~{\bf
K}_{s y} = \left(\frac{2 \pi}{a}\right)
\left(\frac{1}{2},\frac{1}{2}-\vartheta\right) \label{Kval}
\end{equation}
where $a$ is square lattice spacing, and the wavevector shift from
two sublattice order, $0<\vartheta<1/2$, is a function of the
doping concentration. In particular, strong motivation for our
study here was provided by the remarkable experiments of Wakimoto
{\em et al.}\cite{wakimoto,wakimoto2}. They showed that the onset
of superconductivity in La$_{2-\delta}$Sr$_{\delta}$CuO$_4$ occurs
first at $\delta = 0.055$ (in a first-order
insulator-to-superconductor transition) into a state which also
has long-range spin-density-wave order at $T=0$ with a wavevector
of the form (\ref{Kval}) {\em i.e.} as $\delta$ is moved away from
the insulator at $\delta=0$, the first conducting state is a
SC+SDW state. As the ground state for large enough $\delta$ is a
SC state, it follows that there must be at least one quantum phase
transition between the SC+SDW and SC states, and we will work with
the simplest possibility that there is one direct transition at a
critical $\delta = \delta_c$. Wakimoto {\em et al.} also showed
that such a transition associated with the vanishing of the SDW
moment occurred for $\delta_c \approx 0.14$ (see Fig 1 in
Ref.~\onlinecite{wakimoto2}). We shall assume that the SC+SDW to
SC quantum phase transition is second-order: direct evidence for
critical magnetic fluctuations in
La$_{2-\delta}$Sr$_{\delta}$CuO$_4$ for $\delta \approx 0.14$ was
provided in the neutron scattering experiments of Aeppli {\em et
al.} \cite{aeppli}.

We will also discuss the appearance of local and long-range CDW
order in the above phases. It is important to note that,
throughout this paper, we use the term ``charge density wave'' (or
``charge order'') in its most general sense: such order implies
that there is a periodic spatial modulation in all observables
which are invariant under spin rotations and time reversal, such
as the electron kinetic energy, the exchange energy, or even the
electron pairing amplitude; the modulation in the site charge
density may well be unobservably small because of screening by the
long-range Coulomb interactions.

We note that the doping dependence of the magnetic order in the
cuprates can be quite complex, varies significantly between
different compounds, and is influenced  by the degree of disorder:
the magnetic order may well be spin-glass like at the lowest
energy scales at some $\delta$. The SDW order is also enhanced in
the vicinity of special commensurate values of the doping like
$\delta = 1/8$ (see {\em e.g.} Fig 1 in
Ref.~\onlinecite{wakimoto2}), along with a suppression of SC
order. In general, we do not wish to enter into most of these
complexities here, although we will mention (in
Section~\ref{sec:op}) how our theory could be extended to explain
the commensuration effects---some other relevant issues will be
discussed in Section~\ref{oldsdw}. Our primary assumption is that
the low energy collective excitations can be described using the
theory of the vicinity of a quantum critical point between the
SC+SDW and the SC phases; evidence supporting this assumption was
also reviewed in Ref.~\onlinecite{sciencereview}. This critical
point is present either as a function of $\delta$ in the material
under consideration, or in a generalized parameter space but quite
close to the physical axis.

It is also important not to confuse this magnetic quantum critical
point, with other proposals for quantum critical points near
optimal doping that have appeared in the recent literature
\cite{tallon,valla}. These latter critical points are near $\delta
\approx 0.19$, and are probably not associated with long-range
spin density wave order at a wavevector of the form (\ref{Kval}).
This paper will  discuss magnetic transitions at smaller doping.

Upon accepting the existence of a second order quantum critical
point at $T=0$ between the SC+SDW and SC phases, a powerful
theoretical tool for the analysis of experiments becomes
available\cite{book}. The structure of the critical theory, and
its associated classification of eigen-perturbations, allows a
systematic  and controlled theory of the spin excitations in the
SC and SDW phases on either side of the critical point. Such an
approach was recently exploited to study the influence of
non-magnetic Zn and Li impurities in the SC phase~\cite{sbv}. In
this paper we will use the same tools to study the influence of an
applied magnetic field, oriented perpendicular to the CuO$_2$
layers, on both the SC and the SC+SDW phases. An outline of our
results has already appeared in previous communications
\cite{prl,sns,pphmf}: here we will present the full numerical
solution of the our self-consistent equations for the dynamic spin
spectrum in an applied field, along with a number of new results.
Measurements of the spin and charge correlations in the presence
of such an applied magnetic field have appeared recently in a
number of illuminating neutron scattering, \cite{lake,boris,lake2}
NMR \cite{curro,halperin,halperin2} and STM
experiments\cite{seamus}, and we will compare their results with
our prior predictions.

\subsection{Order parameters and field theory}
\label{sec:op}

The field theory for a SC to SC+SDW transition in zero applied
magnetic field can be expressed entirely in terms of the SDW order
parameter which we will introduce in this subsection; the quantum
fluctuations of the SC order can be safely neglected, a point we
will discuss further in Section~\ref{oldsdw}. Consideration of the
applied magnetic field will appear in the following subsection.

We introduced above the wavevectors of the SDW ordering ${\bf
K}_{sx}$ and ${\bf K}_{sy}$; almost all of our analysis will apply
for general values of $\vartheta$, but the value $\vartheta=1/8$
is of particular interest above a doping of about $1/8$. To obtain
an order parameter for such a SDW, we write the spin operator
$S_{\alpha} ({\bf r}, \tau)$, $\alpha=x,y,z$, at the lattice site
${\bf r}$ as
\begin{equation}
S_{\alpha} ({\bf r}, \tau) = \mbox{Re} \left[e^{i {\bf K}_{sx}
\cdot {\bf r}} \Phi_{x \alpha} ({\bf r}, \tau)+  e^{i {\bf K}_{sy}
\cdot {\bf r}} \Phi_{y \alpha}({\bf r}, \tau)\right], \label{e1}
\end{equation}
where $\Phi_{x,y\alpha}$ are the required order parameters. Except
for the case of two sublattice order with $\vartheta =0$ (which we
exclude for now), the fields $\Phi_{x,y\alpha}$ are complex. These
fields can describe a wide variety of SDW configurations, but we
now list the two important limiting cases.
\newline
({\em i\/}) Collinearly polarized SDWs, for which
\begin{equation}
\Phi_{y \alpha} ({\bf r} , \tau) = e^{i \theta ({\bf r} , \tau)}
n_{\alpha} ({\bf r} , \tau) \label{q1}
\end{equation}
where $n_{\alpha}$ is a {\em real} vector and $\theta$ is also
real (and similarly for $\Phi_{x\alpha}$). Parameterized in this
manner, and for $n_{\alpha}^2 = \mbox{constant}$ (summation over
the repeated index $\alpha$ is implied here and henceforth), the
order parameter $\Phi_{y \alpha}$ belongs to the space $(S_2
\times S_1)/Z_2$, where $S_n$ is the $n$-dimensional surface of a
sphere in $n+1$ dimensions, and $Z_p$ is the discrete cyclic group
of $p$ elements. The $Z_2$ quotient is necessary because a shift
$\theta\rightarrow \theta+\pi$ is equivalent to a rotation which
sends $n_{\alpha} \rightarrow - n_{\alpha}$.
\newline
({\em ii\/}) Circular spiral SDWs, for which
\begin{equation}
\Phi_{y \alpha} ({\bf r} , \tau) = n_{1\alpha} ({\bf r} , \tau) +
i n_{2 \alpha}({\bf r} , \tau) \label{q2}
\end{equation} where
$n_{1,2\alpha}$ are two {\em real\/} vectors obeying
$n_{1\alpha}^2 = n_{2\alpha}^2$ and $n_{1 \alpha} n_{2 \alpha} =
0$ (and similarly for $\Phi_{x\alpha}$). Now for $n_{1 \alpha}^2 =
\mbox{constant}$, the order parameter $\Phi_{y \alpha}$ belongs to
the space SO(3)$\cong S_3/Z_2$ (see {\em e.g.\/} Section 13.3.2 in
Ref.~\onlinecite{book}).
\newline
The experimental evidence\cite{tran,ylee} supports the conclusion
the SDW ordering in the cuprates in collinear, but the present
formalism allows a common treatment of both the collinear and
spiral cases. This complex-vector formulation of the SDW order
allows treatment of the SDW quantum transition by a
straightforward generalization of the real-vector theory used for
the N\'{e}el state in the insulator; related points have been made
by Castro Neto and Hone~\cite{cnh} and Zaanen~\cite{zaanen}. The
same approach was also used by Zachar {\em et al.}\cite{zachar} to
treat the onset of SDW order at finite temperatures, as we will
indicate below.

Along with the SDW order, CDW order may also appear. We
parameterize the charge density modulation by:
\begin{equation}
\delta \rho ({\bf r}, \tau) = \mbox{Re} \left[e^{i {\bf K}_{cx}
\cdot {\bf r}} \phi_x ({\bf r}, \tau) + e^{i {\bf K}_{cy} \cdot
{\bf r}} \phi_y ({\bf r}, \tau)\right] \label{e2}
\end{equation}
where ${\bf K}_{cx,y}$ are the CDW ordering wavevectors and
$\phi_{x,y}$ the corresponding complex order parameters. The
quantum numbers of the observable $\delta \rho$ are identical to
those of $S_{\alpha}^2$, and so by squaring (\ref{e1}) we see that
associated with the SDW is a CDW with \cite{zachar} ${\bf
K}_{cx}=2 {\bf K}_{sx}$, ${\bf K}_{cy}=2 {\bf K}_{sy}$ (modulo
reciprocal lattice vectors),
\begin{equation}
\phi_x ({\bf r}, \tau) \propto \Phi_{x\alpha}^2 ({\bf r}, \tau)
,~\mbox{and}~\phi_y ({\bf r}, \tau) \propto \Phi_{y \alpha}^2
({\bf r}, \tau). \label{square}
\end{equation}
Note that this CDW is absent for the case of a circular spiral SDW
(in which case $\Phi_{x,y \alpha}^2 = 0$) but is necessarily
present for a collinear SDW. In principle, in a state with
condensates of both $\Phi_{x\alpha}$ and $\Phi_{y \alpha}$, a CDW
can also be present at wavevector ${\bf K}_{sx}+ {\bf K}_{sy}$; we
will not consider this possibility here as it does not seem to be
experimentally relevant. As was emphasized in the third paragraph
of Section~\ref{sec:intro}, we are using the term CDW here in its
broadest sense: there is a modulation at the wavevector ${\bf
K}_c$ in all observables which are invariant under spin rotations
and time reversal. The precise nature of the CDW order may be
determined from an analysis of the STM spectrum--this has been
discussed recently in Refs.~\onlinecite{dd,vojta}.

The order parameters $\Phi_{x,y\alpha}$, $\phi_{x,y}$ allow a rich
variety of phases and phase transitions in the presence of
background SC order. These will be discussed in some detail in
Section~\ref{newphases}. Central to a description of these phases
is an understanding of the symmetries respected by any effective
action for the order parameters. We describe these below and then
focus on a particular phase transition of physical interest.

An obvious symmetry is that under spin rotations; this is
described by the group SU(2), and the fields $\Phi_{x,y\alpha}$
transform as $S=1$ vectors labeled by the index $\alpha$. In
addition, there is an independent {\em sliding\/} symmetry
\begin{equation}
\Phi_{x,y\alpha} \rightarrow e^{i \theta_{x,y}} \Phi_{x,y\alpha}
\label{sliding}
\end{equation}
associated with the translational symmetry of the underlying
lattice model: translating ${\bf r}$ to ${\bf r} + (ma,0)$ ($m$
integer) in (\ref{e1}) leads to (\ref{sliding}) with $\theta_x =
m\pi (1 - 2 \vartheta)$ and $\theta_y = m\pi$ ($\vartheta$ was
defined in (\ref{Kval})). For $\vartheta$ irrational, we see that
all real values of $\theta_{x,y}$ can be generated with the
different choices for $m$, and hence the sliding symmetry is
U(1)$\times$U(1). For rational $\vartheta$, with
$1/2-\vartheta=p'/p$, and $p'$, $p$ relatively prime integers,
only integer multiples of $\theta_{x,y}=2\pi/p$ are allowed in
(\ref{sliding}); in this case the sliding symmetry is reduced to
$Z_p \times Z_p$. The difference between U(1) and $Z_p$ will not
be material to any of our results for $p>2$. In a similar manner,
we can also determine the action of other elements of the square
lattice space group on $\Phi_{x,y\alpha}$ and we mention two
important cases: under a spatial inversion we have $\Phi_{x,y
\alpha} \rightarrow \Phi_{x,y \alpha}^{\ast}$, and under the
interchange of $x$ and $y$ axes, we have $\Phi_{x \alpha}
\leftrightarrow \Phi_{y \alpha}$.

We now apply these symmetries to determine the effective action of
a physically relevant transition discussed earlier in the
introduction (and in the phase diagrams of
Section~\ref{sec:zero}): that between the SC+SDW and SC phases.
This transition is driven by the condensation of
$\Phi_{x,y\alpha}$; if the SDW order is collinear, it will drive a
concomitant CDW order, as discussed above. Supplementing the
symmetries by a renormalization group (RG) procedure which selects
terms with smaller powers of $\Phi_{x,y\alpha}$ and fewer spatial
and temporal gradients, we obtain \cite{zachar,vsvzs,sns,pphmf}
the effective action
\begin{eqnarray}
{\cal S}_{\Phi} = && \int d^2 r d \tau \Bigl[
   \left| \partial_\tau \Phi_{x \alpha} \right|^2
   + v_1^2 \left| \partial_x \Phi_{x \alpha} \right|^2
   + v_2^2 \left| \partial_y \Phi_{x \alpha} \right|^2
\nonumber \\
&& +\left| \partial_\tau \Phi_{y \alpha} \right|^2
   + v_2^2 \left| \partial_x \Phi_{y \alpha} \right|^2
   + v_1^2 \left| \partial_y \Phi_{y \alpha} \right|^2
\nonumber \\
&& + s (\left|\Phi_{x \alpha}\right|^2
   + \left|\Phi_{y \alpha}\right|^2)
   +\frac{u_1}{2} ( \left|\Phi_{x \alpha}\right|^4
   + \left|\Phi_{y \alpha}\right|^4 )
\nonumber \\
&& + \frac{u_2}{2} ( \left|\Phi_{x \alpha}^2\right|^2
   + \left|\Phi_{y \alpha}^2\right|^2 )
   + w_1 \left|\Phi_{x \alpha}\right|^2\left|\Phi_{y \alpha}\right|^2
\nonumber \\
&& + w_2 \left|\Phi_{x \alpha} \Phi_{y \alpha}\right|^2
   + w_3 \left|\Phi_{x \alpha}^* \Phi_{y \alpha} \right|^2 \Bigr].
\label{SPhi}
\end{eqnarray}
Note that first-order temporal gradient terms like
$\Phi_{x\alpha}^{\ast} \partial_{\tau} \Phi_{x \alpha}$ are
forbidden by spatial inversion symmetry\cite{vsvzs}. In principle,
first-order spatial gradient terms like $i\Phi_{x\alpha}^{\ast}
\partial_{x} \Phi_{x \alpha}$ are permitted by all symmetries;
such terms lead to a shift in the wavevector at which SDW
fluctuations are largest, and we assume that they have already
been absorbed by our choice of ${\bf K}_{sx}$. Here $v_1$ and
$v_2$ are velocities, which are expected to be of order the
spin-wave velocity, $v$, of the N\'{e}el state in the undoped
insulator. The parameter $s$ tunes the system from the SC phase
($s>s_c$) to the SC+SDW phase ($s<s_c$), where $s=s_c$ is the
non-universal location of the quantum critical point between these
phases; experimentally, $s$ can be varied by changing the doping
concentration. The action also contains a number of quartic
non-linearities: the RG analysis shows that these are strongly
relevant perturbations about the Gaussian theory, and will play a
crucial role in our analysis below. The coupling $u_2$ selects
between the collinear and spiral SDW states: for $u_2>0$, the
circular spiral state (which has $\Phi_{x\alpha}^2=0$) is
selected, while $u_2 < 0$ prefers a collinear SDW. The couplings
$w_{1,2,3}$ lead to correlations between the orders at ${\bf
K}_{sx}$ and ${\bf K}_{sy}$---if these are attractive, the $s<s_c$
phase will have simultaneous orderings at both wavevectors, and
spatial pattern will have a checkerboard structure.

We have also neglected the couplings to the low energy nodal
quasiparticles, which are additional excitations of the SC phase
carrying spin; their effects are suppressed by the constraints of
momentum conservation, as they can damp the $\Phi$ quanta
effectively only if ${\bf K}_{sx,y}$ equal the separation between
any two nodal points. The case where this nesting condition is
satisfied has been considered earlier \cite{vsvzs}, but we will
not enter into it here for simplicity: essentially all of our
results here on the phase diagram in an applied magnetic field
apply also to the case where the nesting condition is obeyed. For
completeness, in Appendix \ref{Dzyaloshinskii-Moriya} we also
discuss the role of spin symmetry breaking Dzyaloshinskii-Moriya
interaction present in La$_{2-\delta}$Sr$_\delta$CuO$_4$
\cite{Thio}. We show that it helps stabilize collinear SDW order
in a certain direction; however its effect is very small and will
be neglected in the rest of this paper.

For the particular rational value $\vartheta =1/8$, the
U(1)$\times$U(1) sliding symmetry is reduced to a discrete $Z_8
\times Z_8$ symmetry under which $\theta_{x,y}$ in (\ref{sliding})
are only allowed to be multiples of $\pi/4$. This reduced symmetry
allows additional terms in (\ref{SPhi}) whose structure has been
discussed earlier\cite{zachar,pphmf}. Such terms help choose
between site- and bond-centered density waves\cite{pphmf}, and
could also lead to the enhancement of the moment observed by
Wakimoto {\em et al.} \cite{wakimoto2} near $\delta=1/8$. However,
these terms are very high order (eighth) in the $\Phi$ fields, and
consequently they have a negligible effect on the issues we are
interested in here: so we will not consider them further.

It is useful to compare our treatment here of the SC+SDW to SC
transition with others in the literature. It is essential for our
purposes that the spin/charge ordering is taking place in a
background of SC order, as that gaps out the fermionic excitations
except possibly at special points in the Brillouin zone.
Theories\cite{rome,andrey} which consider SDW/CDW order in a Fermi
liquid have additional damping terms in their effective action
which change the universality class of the transition, change the
dynamic exponent to $z=2$, and do not obey strong hyperscaling
properties as the quartic couplings are marginally {\em
irrelevant} in this case. We have also taken a genuinely
two-dimensional view on the SDW/CDW (``stripe'') fluctuations in
our approach. An alternative approach\cite{kfe} assumes there are
intermediate scales on which the physics of the one-dimensional
electron gas applies, although a crossover to similar
two-dimensional  physics occurs on large enough
scales\cite{granath}.

\subsection{Influence of an applied magnetic field}
\label{sec:mag}

An applied magnetic field has a Zeeman coupling to the spin of the
electrons, and this is present for any direction of the applied
field. However, the Zeeman splitting of the magnetic levels has
only a minor effect, and can be safely neglected compared to the
much stronger effects near $s=s_c$ that we consider below. We
discuss the influence of the Zeeman term in Appendix~\ref{zeeman},
and will not consider it further in this paper.

The dominant effect of the field is via its coupling to the
orbital motion of the electrons, which is sensitive only to the
component of the field orthogonal to the layers. The reason for
this strong effect is simple: there is SC order in the orbital
wavefunction of the electrons, and the diamagnetic susceptibility
of the SC state to the applied field is infinite. However, as the
SC order is non-critical across the transition at $s=s_c$, it is
mainly a quiescent spectator and its response can justifiably be
treated in a static, mean-field theory. Consequently, we model the
complex SC order parameter $\psi({\bf r})$ in the familiar
Abrikosov theory with the free energy {\em per layer} (we use
units with $\hbar=k_B=1$ throughout)
\begin{eqnarray}
{\cal F} = \int d^2 r \Bigl[ &-& \alpha |\psi ({\bf r}) |^2 +
\frac{\beta}{2} |\psi ({\bf r}) |^4
\nonumber \\
&+& \frac{1}{2 m^{\ast}} \left| \left( \frac{1}{i} {\bf
\nabla}_{\bf r} - \frac{e^{\ast}}{c} {\bf A} \right) \psi  ({\bf
r}) \right|^2 \Bigr]. \label{of}
\end{eqnarray}
Note that unlike $\Phi_{x,y\alpha}$, $\psi$ is not a fluctuating
variable, and described completely by its mean-value (which will
be ${\bf r}$ dependent). We will work entirely in the limit of
extreme type-II superconductivity (with Ginzburg-Landau parameter
$\kappa_{GL} = \infty$); so there is no screening of the magnetic
field by the Meissner currents, and ${\bf \nabla}_{\bf r} \times
{\bf A} = H \hat{z}$, the applied, space-independent magnetic
field.

To complete the description of the model studied in this paper, we
now need to couple the SC and SDW order parameters together. The
simplest term allowed by symmetry is a connection between the
local modulus of the order parameters:
\begin{equation}
{\cal S}_{\Phi \psi} = \kappa \int d^2 r d \tau |\psi ({\bf r})
|^2 \left(|\Phi_{x \alpha}  ({\bf r},\tau) |^2 + |\Phi_{y \alpha}
({\bf r},\tau) |^2 \right) \label{couple}
\end{equation}
For $\kappa >0$, we can induce a competition between the SC and
SDW orders, in that the SDW order will be enhanced where the SC
order is suppressed and vice-versa. The microscopic origin of the
coupling $\kappa$ is discussed in Appendix~\ref{app:box}.

Although ${\cal S}_{\Phi \psi}$ will be the primary coupling
between the SDW and SC orders, an additional allowed term will be
important for some purposes\cite{pphmf}. To understand this,
notice that all terms in ${\cal S}_{\Phi}$ and ${\cal
S}_{\Phi\psi}$ are invariant under the sliding symmetry
(\ref{sliding}). This means that, with the present terms, the CDW
order is free to slide arbitrarily with respect to any vortex
lattice that may be present in the SC order $\psi$. This clearly
cannot be true, as lattice scale effects should pin the two
modulations with respect to each other. The simplest additional
coupling which will provide this pinning can be deduced by
noticing that there should be a coupling between the charge
modulation in (\ref{e2}) and the local modulus of the
superconducting order; this is induced by the term\cite{pphmf}
\begin{eqnarray}
\widetilde{\cal S}_{\rm lat} &=& -\widetilde{\zeta} \sum_{\bf r}
\int d \tau |\psi ({\bf r})|^2 \mbox{Re}  \left[e^{i {\bf K}_{cx}
\cdot {\bf r}}
\Phi_{x\alpha}^2 ({\bf r}, \tau) \right.\nonumber\\
&~&~~~~~~~~~~~~~~~~~\left.+ e^{i {\bf K}_{cy} \cdot {\bf r}}
\Phi_{y\alpha}^2 ({\bf r}, \tau)\right]. \label{e3}
\end{eqnarray}
Notice that we are now performing a discrete summation over the
lattice sites ${\bf r}$, rather than integrating over a spatial
continuum: this is a direct consequence of the rapidly oscillating
factors $e^{i {\bf K}_{cx} \cdot {\bf r}}$ and $e^{i {\bf K}_{cy}
\cdot {\bf r}}$ which do not have a smooth continuum limit.
Indeed, in regions where $\psi ({\bf r})$ is smoothly varying,
these rapidly oscillating factor will cause the summation over
${\bf r}$ to vanish. So the expression (\ref{e3}) is appreciable
only over regions where $\psi ({\bf r})$ is rapidly varying, and
this happens only in the cores of the vortices. As the centers of
the vortices are identified by the zeros of $\psi ({\bf r})$, and
we are mainly interested in scales larger than vortex core size,
we can replace (\ref{e3}) by the following expression, which is
more amenable to an analysis in the continuum theory \cite{pphmf}:
\begin{eqnarray}
{\cal S}_{\rm lat} &=& -\zeta \sum_{{\bf r}_v, \psi({\bf r}_v) =
0} \int d \tau  \mbox{Re}  \left[e^{i \varpi} \left(
\Phi_{x\alpha}^2 ({\bf r}_v, \tau) \right.\right. \nonumber
\\&~&~~~~~~~~~~~~~~~~~\left.\left. +
 \Phi_{y\alpha}^2 ({\bf r}_v, \tau)\right) \right]. \label{e4}
\end{eqnarray}
Here the summation is over all points ${\bf r}_v$ at which $\psi
({\bf r}_v) = 0$ (these are the centers of the vortices), and
$\varpi$ is a phase which depends upon the microscopic structure
of the vortex core on the lattice scale. The action ${\cal S}_{\rm
lat}$ is not invariant under the sliding symmetry, and so will pin
the CDW order.

We are now in a position to succinctly state the field-theoretic
problem which will be addressed in this paper. We are interested
in the partition function for SDW/CDW fluctuations defined by
\begin{eqnarray}
\mathcal{Z}\left[\psi ({\bf r}) \right] &=& \int {\cal D} \Phi_{x
\alpha} ({\bf r}, \tau) {\cal D} \Phi_{y \alpha} ({\bf r}, \tau)
\nonumber
\\
&~&~\times \exp \left( - \frac{{\cal F}}{T} - {\cal S}_{\Phi} -
{\cal S}_{\Phi \psi} - {\cal S}_{\rm lat} \right), \label{zdef}
\end{eqnarray}
accompanied by the solution of
\begin{equation}
\frac{\delta \ln \mathcal{Z}\left[\psi ({\bf r}) \right]}{\delta
\psi ({\bf r})} = 0 \label{scdef}
\end{equation}
which minimizes $-\ln \mathcal{Z}\left[\psi ({\bf r}) \right]$ to
determine the optimum $\psi ({\bf r})$. Note the highly
asymmetrical treatment of the SDW and SC orders: we include full
quantum-mechanical fluctuations of the former, while the latter is
static and non-fluctuating. This asymmetry is essentially imposed
on us by the perspective of magnetic quantum criticality, and the
fact that we are developing a theory of the SC+SDW to SC
transition. This asymmetry should also be contrasted with the
symmetric treatment of SC and SDW quantum fluctuations in other
approaches\cite{so5}.

\subsection{Physical discussion}
\label{sec:phys}

The primary purpose of this paper is to determine the phase
diagram and low-energy spectrum of SDW and CDW fluctuations of
$\mathcal{Z}$ as a function of the applied field $H$. A summary of
our results has already appeared\cite{prl,sns,pphmf} and detailed
numerical solutions appear in the body of the paper; here we
expand on the central physical idea to provide an intuitive
understanding of our results to readers who do not wish to study
the details in the remainder of the paper. We will initially
ignore the pinning described by $\mathcal{S}_{\rm lat}$, but will
discuss its consequences in Section~\ref{sec:pinsc}.

\begin{figure}
\centerline{\includegraphics[width=3in]{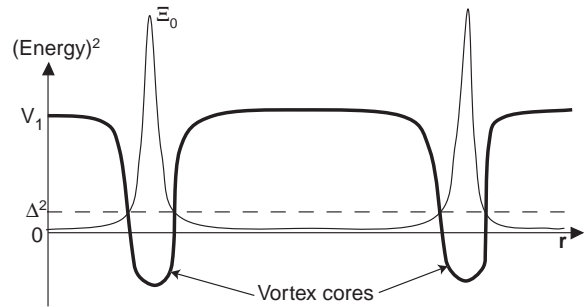}} \caption{A
sketch of the potential ${\cal V}_0 ({\bf r})$ (thick full line)
in the presence of a vortex lattice. Also shown is the exciton
wavefunction $\Xi_0 ({\bf r})$ which solves (\protect\ref{schro})
for ${\cal V} ({\bf r}) = {\cal V}_0 ({\bf r})$ with eigenvalue
$\Delta^2$. Note that there is no drastic change in this picture
as $\Delta^2 \searrow 0$: the peaks in $\Xi_0 ({\bf r})$ remain
exponentially localized within each vortex core, on a length scale
much smaller than the vortex lattice spacing. We argue in the text
that strong interaction corrections to ${\cal V}_0 ({\bf r})$
invalidate this form for $\Xi_0 ({\bf r})$ and the correct
structure is shown in Fig.~\protect\ref{figv2}.} \label{figv1}
\end{figure}

Let us begin in the SC phase with $s>s_c$ and consider the
$\Phi_{x\alpha}$ fluctuations in a simple Gaussian theory (the
considerations of this subsection apply equally to
$\Phi_{y\alpha}$, which we will not mention further). Assume $\psi
({\bf r})$ has been determined by the minimization of ${\cal F}$,
and so takes the standard form in an Abrikosov flux lattice. The
Gaussian fluctuations of $\Phi_{x \alpha}$ are described by the
effective action
\begin{eqnarray}
{\cal S}_G = && \int d^2 r d \tau \Bigl[
   \left| \partial_\tau \Phi_{x \alpha} \right|^2
   + v_1^2 \left| \partial_x \Phi_{x \alpha} \right|^2
   + v_2^2 \left| \partial_y \Phi_{x \alpha} \right|^2
\nonumber \\
&&~~~~~~~~~~~~~~~~~~~~~~~~~+ {\cal V} ({\bf r}) |\Phi_{x
\alpha}|^2 \Bigr] \label{SG}
\end{eqnarray}
To leading order, the effective potential ${\cal V} ({\bf r })$ is
given by ${\cal V} = {\cal V}_0$ where
\begin{equation}
{\cal V}_0 ({\bf r} ) = s + \kappa |\psi ({\bf r})|^2 \label{V0}
\end{equation}
A sketch of the spatial structure of ${\cal V}_0 ({\bf r})$ is
shown in Fig~\ref{figv1}: because $\psi ({\bf r})$ vanishes at the
centers of the vortices, ${\cal V}_0 ({\bf r})$ has well-developed
minima at each such point.
Indeed, there can even be regions in each vortex core where ${\cal
V}_0 ({r}) < 0$, and Arovas {\em et al.}\cite{arovas} and Bruus
{\em et al.}\cite{bruus} argued that superconductivity would
`rotate' or transform into {\em static\/} N\'{e}el order in such a
region. In our treatment of {\em dynamic\/} SDW \cite{csy,prl}, we
see that the structure of the magnetism is determined by the
solution of the Schr\"odinger equation\cite{note}
\begin{equation}
\left( -v_1^2 \partial_x^2 - v_2^2 \partial_y^2 + {\cal V}({\bf
r}) \right) \Xi_0 ({\bf r}) = \Delta^2 \Xi_0 ({\bf r})
\label{schro}
\end{equation}
where $\Xi_0 ({\bf r})$ is the lowest eigenmode of (\ref{schro}),
the eigenvalue $\Delta^2$ is required to be positive for the
stability of the Gaussian theory ${\cal S}_G$. The energy $\Delta$
is the spin-gap, and $\Xi_0 ({\bf r})$ then specifies the envelope
of the lowest energy SDW fluctuations; in other words $\Xi_0 ({\bf
r})$ is the wavefunction of a $S=1$ {\em exciton} associated with
dynamic SDW fluctuations. Note that $\Delta^2$ can be positive
even if there are regions where ${\cal V} ({\bf r})< 0$. A sketch
of the spatial form of $\Xi_0 ({\bf r})$ is shown in
Fig~\ref{figv1} for a particular small value of $\Delta^2$ and
${\cal V}({\bf r})={\cal V}_0 ({\bf r})$. Observe that $\Xi_0
({\bf r})$ is peaked at the vortex centers, but decays rapidly
outside the vortex cores over a SDW localization length $\ell \sim
v_{1,2} /\sqrt{V_1 - \Delta^2}$, where $V_1$ is the value of
${\cal V}_0 ({\bf r})$ outside the vortex cores (see
Fig~\ref{figv1}).

Remaining within the Gaussian theory specified by (\ref{SG}) and
(\ref{V0}), we now consider the consequences of raising the value
of $H$ in the hope of reaching the SC+SDW phase. With increasing
$H$, the vortex cores will approach each other, and we expect that
the value of $\Delta^2$ will decrease. Indeed, the picture of
Fig~\ref{figv1} holds all the way up to the point $\Delta=0$;
beyond this field the Gaussian theory becomes unstable and this
signals the onset of the SC+SDW phase driven by the condensation
of $\Phi_{x \alpha}$. Note that the localization length $\ell \sim
v_{1,2} /\sqrt{V_1 - \Delta^2}$ of the SDW order peaked in the
vortex cores remains {\em finite} all the way up to the critical
point. This localization length $\ell$ must be clearly
distinguished from the spin correlation length, $\xi_s$: the
latter is associated with correlations between different vortices,
and arises because there is an exponentially small coupling
between magnetism in neighboring cores. Thus this simple Gaussian
theory yields a picture of dynamic magnetism appearing first in
the vortex cores, with possible weak correlations between
neighboring cores. Such a viewpoint was also discussed by Lake
{\em et al.} \cite{lake} who proposed ``spins in the vortices''
but noted that the large value of $\xi_s$ implied coupling between
nearby vortices. Following our work\cite{prl}, Hu and
Zhang\cite{hu} also presented a picture of dynamic SDW
fluctuations similar to the one above.

\begin{figure}
\centerline{\includegraphics[width=3in]{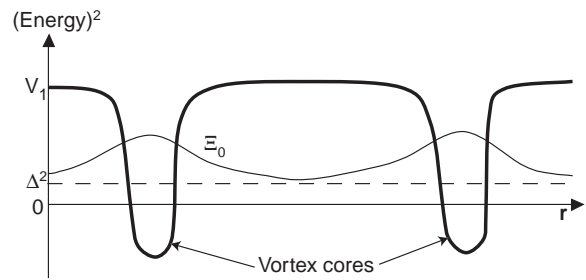}} \caption{A
sketch of the potential ${\cal V}_0 ({\bf r})$ (thick full line)
in the presence of a vortex lattice along with the true form of
the exciton wavefunction $\Xi_0 ({\bf r})$ which solves
(\protect\ref{schro}) with the full potential ${\cal V} ({\bf r})$
in (\protect\ref{V}). The spatial structure of $\Xi_0 ({\bf r})$
as $\Delta^2 \searrow 0$ is characterized by the vortex lattice
spacing. } \label{figv2}
\end{figure}

We now argue that corrections beyond the Gaussian theory
approximation invalidate the above picture when $\Delta$ becomes
small\cite{prl}. Indeed, the picture of nearly-independent,
localized magnetic excitations in each vortex core holds only then
$\Delta$ is of order the spin exchange energy $J$; such high
energy magnetic excitations are expected to strongly damped by the
fermionic quasiparticles. Also, the validity of the present
continuum model is questionable at scales as short as vortex core
size and at energies of order  $J$: a full solution of the BCS
theory of the underlying electrons is surely needed, and
subsidiary order parameters may well develop within the vortex
cores. However, as $\Delta$ is lowered, we will now argue that the
physics is actually dominated by the large region outside the
vortex cores, where the present continuum approach can be used
without fear, and the subtle issues of the short-distance physics
within the core can be sidestepped. The central weakness in the
analysis of the previous paragraph is that it does not account for
the repulsive interactions $u_{1,2}$ between the bosonic
$\Phi_{x\alpha}$ exciton modes that are condensing. As has been
discussed in different contexts long ago \cite{hfa,bm}, such
interactions are crucial in determining the structure of the
lowest energy state in which condensation occurs. In particular,
it is well known that the effect of interactions is to delocalize
the lowest energy states: bosons initially prefer to occupy
strongly localized, low energy states, but then their repulsive
interaction with subsequent bosons drives the energy of such
states up. Bray and Moore \cite{bm} presented an argument
demonstrating that in the vicinity of the condensation, the
localization length must diverge as one approached the bottom of
the band of states of interacting bosons in the presence of an
external potential. To apply their argument in the present
context, we need to replace (\ref{V0}) by
\begin{eqnarray}
&& {\cal V} ({\bf r} ) = {\cal V}_0 ({\bf r}) + \frac{(4u_1 +
2u_2)}{3} \left\langle |\Phi_{x \alpha} ({\bf r},
\tau) |^2 \right\rangle_{{\cal S}_G} \nonumber \\
&& \!\!\!\!\!\!\!\!\!\!= s + \kappa |\psi ({\bf r})|^2 +
\frac{(4u_1 + 2u_2)}{3} \left\langle |\Phi_{x \alpha} ({\bf r},
\tau) |^2 \right\rangle_{{\cal S}_G}; \label{V}
\end{eqnarray}
the additional terms arise from a Hartree-Fock decoupling of the
quartic interaction terms in ${\cal S}_{\Phi}$, and the
expectation values have to be evaluated self-consistently under
the Gaussian action in (\ref{SG}) which itself depends upon ${\cal
V} ({\bf r})$. Note that the perspective of magnetic criticality
requires that we account for the $u_{1,2}$ interactions, as these
are strongly relevant perturbations about the Gaussian theory; so
we are led to (\ref{V}) also by a naive application of the RG
approach. We will present detailed numerical solutions of
equations closely related to (\ref{V}) in the body of the paper.
An adaption of the argument of Bray and Moore \cite{bm} to
(\ref{V}) was given in Ref.~\onlinecite{prl}, and we will not
repeat it here: the main result is that the length scale $\ell$
characterizing the lowest energy state $\Xi_0 ({\bf r})$ cannot
remain finite as $\Delta \searrow 0$. Instead the states around
neighboring vortex cores overlap strongly, and $\Xi_0 ({\bf r})$
is characterized by the vortex spacing itself. A sketch of the
actual structure of $\Xi_0 ({\bf r})$ is shown in Fig~\ref{figv2}.
The spin correlation length, $\xi_s$, does not have a direct
connection with the spatial form of $\Xi_0 ({\bf r})$ itself, but
is instead related to an integral over a band of states which
solve (\ref{schro}) at finite momentum, as we shall discuss in
Section~\ref{sec:pinsc} and later in the paper.

It is worth noting here that the passage from (\ref{V0}) to
(\ref{V}) in zero field is precisely that needed to reproduce the
known properties of magnetic quantum critical points in other
situations. In one dimension, (\ref{V0}) would imply that there is
no barrier to magnetic long-range order, while (\ref{V}) correctly
implies that the presence of the Haldane gap, and reproduces its
magnitude in the semiclassical limit~\cite{book}. At finite
temperature, (\ref{V}) yields the correct crossovers in the
magnetic correlation length in the vicinity of the spin ordering
transition in two dimensions. Although we will not present
detailed solutions on this case here, (\ref{V}) is also expected
to provide a reasonable description of the magnetic crossovers at
finite temperatures in the vicinity of the SC+SDW to SC transition
in the presence of a magnetic field.

With the knowledge of the spatial structure of the exciton
wavefunction $\Xi_0 ({\bf r})$ in Fig~\ref{figv2}, the origin of
our main results\cite{prl} can be easily understood. As the vortex
cores occupy only a small fraction of the system volume, the
magnitude of the energy $\Delta^2$ is influenced mainly by the
structure of $\psi ({\bf r})$ in the remaining space. Here, the
predominant consequence of the magnetic field is the presence of a
superflow with velocity ${\bf v}_s = - \delta {\cal F}/\delta {\bf
A}$ circulating around each vortex core. Focusing on the region
around a single vortex at the origin ${\bf r}=(0,0)$, the
superflow obeys $|{\bf v}_s| \sim 1/r$ in the wide region $\xi_{0}
< r < L_v$ where $\xi_{0} = 1/\sqrt{2 m^{\ast} \alpha}$ is the
vortex core size, $L_v \sim (e^{\ast} H/ c)^{-1/2}$; so the
average superflow kinetic energy is
\begin{equation}
\langle {\bf v}_s^2 \rangle \propto \frac{ \displaystyle
\int_{\xi_{0}}^{L_v} \frac{d^2 r}{r^2}}{ \displaystyle
\int_{\xi_{0}}^{L_v} {d^2 r}} \propto \frac{H}{H_{c2}^{0}}\ln
\left( \frac{H_{c2}^{0}}{H} \right) \label{hlog}
\end{equation}
where $H_{c2}^{0}$ is the upper critical field for the destruction
of the Meissner state at the coupling constant corresponding to
the point M in Fig~\ref{figpd} below. This kinetic energy is a
scalar with the same quantum numbers and symmetry properties as
$|\psi|^2$: hence, via the coupling in ${\cal S}_{\Phi\psi}$ in
(\ref{couple}), the value of (\ref{hlog}) feeds into all the
effective coupling constants in ${\cal S}_{\Phi}$ in (\ref{SPhi}).
The most important modification is that the tuning parameter $s$
gets replaced by
\begin{equation}
s_{\rm eff}(H) = s - {\cal C} \frac{H}{H_{c2}^{0}}\ln \left(
\frac{H_{c2}^{0}}{H} \right) \label{reff}
\end{equation}
where ${\cal C}$ is a constant of order unity. The implication of
(\ref{reff}) is that we may as well replace ${\cal V} ({\bf r})$
in (\ref{SG}) and (\ref{V}) by
\begin{equation}
{\cal V} ({\bf r}) \approx s_{\rm eff} (H) \label{Vs}
\end{equation}
to obtain a first estimate of the consequence of the magnetic
field in the vicinity of the SC+SDW to SC transition. The $H$
dependence in (\ref{reff}) and (\ref{Vs}) is sufficient to
determine our main results\cite{prl}: the small $H$ portion of the
phase diagram in Fig~\ref{figpd}, the intensity of the elastic
scattering Bragg peak in the SC+SDW phase, and the energy of the
lowest energy SDW fluctuation in the SC phase.
\begin{figure}
\centerline{\includegraphics[width=3in]{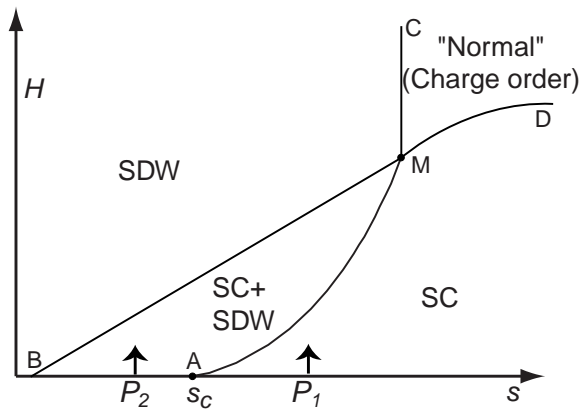}} \caption{Zero
temperature phase diagram as a function of the coupling $s$ and
the magnetic field $H$ in an extreme type II superconductor
described by (\protect\ref{zdef}). The theory is accurate in the
region of small $H$, and only qualitatively correct elsewhere ($H$
is measured in units described in (\protect\ref{trans}). The
phases without SC order are likely to be insulators, and the
``Normal'' phase is expected to have residual CDW order, which is
initially induced by the pinning terms in $\mathcal{S}_{\rm lat}$
as discussed in Section~\protect\ref{sec:pinsc}. The positions of
the phase boundaries are summarized in
Section~\protect\ref{sec:pd}. The path $P_1$ denotes the location
of the original neutron scattering measurements of Lake {\em et
al.}\protect\cite{lake}, and the path $P_2$ the subsequent neutron
scattering measurements of Khaykovich {\em et
al.}\protect\cite{boris} and Lake {\em et
al.}\protect\cite{lake2}. The STM measurements of Hoffman {\em et
al.}\protect\cite{seamus} are also along path $P_1$.}
\label{figpd}
\end{figure}
In particular, it follows directly from (\ref{reff}) that the
small $H$ portion of the AM phase boundary in Fig~\ref{figpd}
between the SC and SC+SDW phases behaves as
\begin{equation}
H \sim \frac{2 (s-s_c)}{\kappa \ln (1/(s-s_c))}. \label{amlog1}
\end{equation}
Note that this phase boundary approaches the $s=s_c$, $H=0$
quantum critical point with {\em vanishing slope}. This implies
that a relatively small $H$ for $s>s_c$ will successfully move the
system close the AM phase boundary, and so produce low energy spin
excitations. This should be contrasted to the corresponding
$H$-dependent phase boundary of the SDW phase in insulators which
is discussed in Appendix~\ref{zeeman}; here, there is no orbital
diamagnetism, only the Zeeman coupling is operative, and the phase
boundary approaches $H=0$ with {\em infinite slope}. Evidently,
Zeeman effects are much weaker and can be justifiably neglected.


We conclude this subsection by a brief discussion of earlier
works\cite{ssvortex,nl,so5,arovas,bruus,ogata,xhu,juneau,andersen,lee,franz,zhu,dden1,dden2}
on vortex magnetism and the change in perspective that has been
offered here by our analysis. It was proposed in by
Sachdev\cite{ssvortex} and Nagaosa and Lee \cite{nl} that vortex
cores in the underdoped cuprates should have spin gap correlations
characteristic of Mott insulators. Zhang\cite{so5} and Arovas {\em
et al.}\cite{arovas} described vortex core correlations in terms
of static N\'{e}el order, and estimated that the field-induced
moment would be proportional to $H$ (in our phase diagram in
Fig~\ref{figpd}, vortices with static moments are only present in
the SC+SDW phase, and as we will review below in
Section~\ref{sdwop}, the average moment increases as\cite{prl}
$H\ln(1/H)$ for small $H$). Our discussion here also uses the SDW
order parameter, but allows it to fluctuate dynamically into a
spin gap state, and so interpolates between these earlier works. A
separate description of vortex cores in terms of `staggered flux'
correlations\cite{dden1,dden2} has also been proposed. One of our
central points here is that while the vortex core correlations may
well be quite complicated (they are dependent on lattice scale
effects, and difficult to distinguish from each other as the
short-distance `order' fluctuates dynamically), these issues can
be sidestepped: a reliable continuum theory can be developed by
considering first the dominant effects arising from the interplay
between superconductivity and magnetism in the superflow region
outside the vortex cores. Spin density wave correlations induced
in these regions may leak into the vortex cores, but our treatment
is not expected to be reliable in the latter region: the nature of
the electronic correlations in the vortex cores remains an open
question.

Our continuum treatment of dynamic and static spin density wave
order differs from earlier works in several key aspects. An
important feature of Refs.~\onlinecite{arovas,bruus} is the static
mean-field treatment of the SDW order in the vortex cores, which
is imposed by their ``SO(5)'' picture of SC order outside the
cores `rotating' into static antiferromagnetism in the
cores\cite{note}. This should be contrasted to our approach, in
which magnetic quantum criticality implies dynamic magnetic
fluctuations while the SC order can be safely considered static.
Further, Refs.~\onlinecite{arovas,bruus} assumed the (near)
equality of the gradient and ``mass'' terms for the SC and
two-sublattice SDW order parameters, as naturally suggested by the
dynamic SO(5) symmetry, which requires a symmetry between the
excited states in the SC and SDW phases. As a result they found
static two-sublattice magnetization induced by the vortex core,
over a scale which was of order the vortex core size, $\xi_0$, and
in a regime where superconductivity was essentially completely
suppressed. This assumption was relaxed in a recent paper
\cite{hu}, where following our work \cite{prl}, the possibility of
a generalized dynamic SDW in regions larger than the
non-superconducting core, and co-existing with well-established
superconductivity, was appreciated.  Hu and Zhang\cite{hu} also
suggested that a small proximity-type coupling between the
magnetic domains centered on the neighboring vortices may be
sufficient to stabilize static long-range magnetic order in a
SC+SDW phase, in which enhancement of the SDW in the vortex cores
was the dominant effect. As we have reviewed at length above, the
strongly relevant exciton self-interactions lead to a different
description of the SC+SDW phase\cite{prl}; in the SC phase, as one
approaches the SC+SDW phase, the SDW order is induced over large
length scales outside the vortex core, and the influence of the
superflow is paramount. Only in the regime where magnetic field is
small and the system is well within the SC phase (and far from the
SC to SC+SDW boundary), can we speak in terms of localized bound
state pulled below the continuum. However, this limit is of little
experimental interest, since it corresponds to high energy
magnetic excitons (which, as discussed above, are probably
strongly overdamped by other excitations associated with the
vortex cores) with a vanishingly small intensity.


\subsubsection{Pinning of charge order in the SC phase}
\label{sec:pinsc}

Our physical discussion has so far neglected the influence of the
pinning potential in ${\cal S}_{\rm lat}$ in (\ref{e4}). We will
continue to neglect this term in most of this paper, apart from
computations in Section~\ref{sec:pin} whose content we briefly
describe here. This analysis is motivated by the STM experiments
of Hoffman {\em et al.}\cite{seamus}.

The SC phase of Fig~\ref{figpd} preserves spin rotation
invariance, and so has $\langle \Phi_{x\alpha} \rangle =0$ and, by
(\ref{e1}), $\langle S_{\alpha} \rangle =0$ (if we were to account
for the small Zeeman term (Appendix~\ref{zeeman}), the analogous
statement holds for the spin density in the plane perpendicular to
the magnetic field). In the absence of ${\cal S}_{\rm lat}$, all
the remaining terms in the partition function $\mathcal{Z}$ in
(\ref{zdef}) are invariant under the sliding symmetry $\Phi_{x
\alpha} \rightarrow e^{i \theta} \Phi_{x \alpha}$, and so we also
have $\langle \Phi_{x\alpha}^2 \rangle =0$ and, by
(\ref{e2},\ref{square}), $\langle \delta \rho \rangle =0$ in the
SC phase. Now if we include the effect of ${\cal S}_{\rm lat}$
perturbatively (which is all we shall do here), the pinning of the
dynamic fluctuations by the vortex cores leads to static CDW order
with $\langle \Phi_{x\alpha}^2 \rangle \neq 0$ and $\langle \delta
\rho \rangle \neq 0$, while the continued preservation of spin
rotation invariance implies that we still have $\langle
\Phi_{x\alpha} \rangle =0$ and $\langle S_{\alpha} \rangle =0$.
(Of course in the other SC+SDW phase, spin rotation symmetry is
broken, and so $\langle \Phi_{x\alpha} \rangle \neq 0$ and
$\langle S_{\alpha} \rangle \neq 0$, along with static CDW order.)

The nucleation of {\em static\/} CDW order, but with {\em
dynamic\/} SDW order, in the SC phase by the vortices was first
predicted in Refs.~\onlinecite{kwon,sns}, where a connection was
also made with lattice scale studies of bond-centered charge order
correlations in superconductors with preserved spin rotation
invariance \cite{vsvzs}. These latter works found a significant
doping range over which the charge order had a period pinned at
four lattice spacings, which is the period observed in the STM
experiments of Hoffman {\em et al.}\cite{seamus} (the same period
also appeared in density matrix renormalization group studies by
White and Scalapino\cite{white}). Here we are interested in the
spatial extent of the {\em envelope} of the period four charge
order. Following Ref.~\onlinecite{pphmf}, here we will compute
this envelope using our present models for dynamic SDW/CDW
fluctuations in the SC phase, and the pinning of a static CDW by
$\mathcal{S}_{\rm lat}$.

After this paper was originally released, we learnt of the
microscopic model of Chen and Ting\cite{chenting} for the STM
experiments, which follows the earlier work of
Ref.~\onlinecite{zhu}. Their model has static order for {\em
both\/} the SDW and CDW, and thus would apply only in the SC+SDW
phase of our phase diagram in Fig~\ref{figpd}. It appears unlikely
to us that the slightly overdoped BSCCO sample used by Hoffman
{\em et al.}\cite{seamus} is in the SC+SDW phase.

The simple model of the field-induced dynamic SDW fluctuations we
have described in this section can be readily extended to compute
the static CDW order induced by $\zeta$ in the SC phase. Indeed,
the upshot of our preceeding discussion of the extended structure
of $\Xi_0 ({\bf r})$ is that we can use the Gaussian theory ${\cal
S}_G$ in (\ref{SG}) with ${\cal V} ({\bf r})$ given by the
constant value in (\ref{Vs}): computing $\langle \Phi_{x \alpha}^2
\rangle$ in the theory ${\cal S}_G + {\cal S}_{\rm lat}$ for this
value of ${\cal V} ({\bf r})$ and to first-order in $\zeta$, we
find\cite{pphmf}
\begin{eqnarray}
\left\langle {\Phi _{x\alpha}^2 \left( {{\bf r},\tau } \right)}
\right\rangle  &=& \sum_{{\bf r}_v} \left( \frac{3}{8 \pi^{3/2}
[s_{\rm eff}(H)]^{1/4} v^{5/2}} \right) \zeta  e^{ - i\varpi }
\nonumber \\
&~&~~~~~~~~~~~~\times \frac{e^{-2 |{\bf r} - {\bf r}_v|
\sqrt{s_{\rm eff} (H)}/v }}{|{\bf r} - {\bf r}_v|^{3/2}},
\label{pcdw}
\end{eqnarray}
where $|{\bf r} - {\bf r}_v| \equiv v \left(
((x-x_v)/v_1)^2+((y-y_v)/v_2)^2 \right)^{1/2}$ and $v=(v_1
v_2)^{1/2}$; the result (\ref{pcdw}) holds for large $|{\bf r} -
{\bf r}_v|$, and the divergence at small $|{\bf r} - {\bf r}_v|$
it cutoff by lattice scale effects. Note that the static CDW order
decays exponentially around each vortex core over a length scale
$\xi_c = v/(2 \sqrt{s_{\rm eff} (H)})$ which has been increased by
the influence of the field-induced superflow (by the decrease of
$s_{\rm eff} (H)$ in (\ref{reff})). Note also that this length
scale is {\em not} related to any localization scale associated
with the SDW state $\Xi_0 ({\bf r})$; indeed, we have argued above
that the latter state is extended. In the present simple Gaussian
calculation, we used the very simple constant potential given in
(\ref{Vs}) in the Schr\"odinger equation for the exciton,
(\ref{schro}); all eigenstates of such an equation are extended
plane-wave states. Instead, the exponential decay in (\ref{pcdw})
arises from the integral over all the oscillating (but extended)
excited states of (\ref{schro}). The body of the paper will show
that the same feature also holds when the full form of ${\cal V}
({\bf r})$ is used, and not just the crude approximation in
(\ref{Vs}) (see Figs~\ref{figcdw02} and~\ref{figcdw05}).

It is useful to make an analogy between the above result and the
phenomenon of Friedel oscillations in a Fermi liquid. A Fermi
liquid state has no static SDW or CDW order, but there are
enhanced fluctuations of these orders at $2 k_F$, the wavevector
which spans extremal points of the Fermi surface. In the presence
of an external impurity, static CDW oscillations at $2 k_F$ are
induced, while full spin-rotation invariance is preserved. The
amplitude of these oscillations decay with a power-law because the
Fermi liquid has gapless spectrum of SDW/CDW excitations.

In the present situation, the physics of the doped Mott insulator
induces a preference for excitonic SDW fluctuations at the
wavevectors ${\bf K}_{sx,y}$ and for CDW fluctuations at the
wavevectors ${\bf K}_{cx,y} = 2 {\bf K}_{sx,y}$. The SC phase has
a spin gap, $\Delta$, at these wavevectors, and so such spin
correlations decay exponentially on the scale $\xi_s = v/\Delta$
(as we have noted, this is not a localization scale of the spin
exciton states, which are all extended). The vortex core pins the
phase the dynamic SDW fluctuations which reside above this
spin-gap, and the resulting ``Friedel oscillations of the
spin-gap'' are manifested by static CDW oscillations at the
wavevectors ${\bf K}_{cx,y}$ whose envelope decays exponentially
over a length scale $\xi_c = \xi_s/2$. These may therefore be
viewed as the Friedel oscillations associated with the excitonic
bound states that are present below the particle-hole continuum of
the $d$-wave superconductor. In a weak-coupling BCS/RPA theory one
can also expect additional Friedel oscillations associated with
the continuum of particle-hole excitations, whose wavevector is
determined by the quasiparticle dispersion. Such a picture may be
appropriate in the strongly overdoped limit in zero magnetic
field, with pinning provided by impurities. However, as one lowers
the doping in the SC phase (to approach the boundary to the SC+SDW
phase), an excitonic bound state appears, and we have focused on
its physics here; the wavevector of this exciton is determined by
strong-coupling effects in the doped Mott insulator. The strength
of this exciton could also be enhanced relative to the
particle-hole continuum in the vicinity of vortices in an applied
magnetic field--this effect requires explicit consideration of the
fermionic quasiparticles, and so is beyond the scope of the
theories considered here.

The outline of the remainder of this paper is as follows. We will
begin in Section~\ref{sec:zero} by a discussion of the phase
diagram of the spin and charge density wave order parameters in
zero magnetic field. More complex phases and phase diagrams are
also possible, associated the composites and `fractions' of these
order parameters, but we will postpone their discussion until
Section~\ref{newphases}. We will turn to the influence of the
magnetic field in Section~\ref{sec:pd}: here we will restrict our
attention to the quantum transition described by
$\mathcal{S}_{\Phi}$, but most of the zero-field transitions
discussed in Section~\ref{newphases} have a related response to an
applied magnetic field. Section~\ref{sec:pd} contains a
description of the phase diagram in the magnetic field, while the
subsequent sections describe the dynamic and static properties of
the two phases on either side of the critical point in some
detail: Section~\ref{SC} describes the SC phase, while
Section~\ref{SC+SDW} describes the SC+SDW phase.
Section~\ref{oldsdw} reviews earlier theoretical and experimental
work on the interplay of magnetism and superconductivity, and
discuss its relationship to our treatment here. We conclude in
Section~\ref{sec:conc} by considering implications of our results
for recent experiments; readers not interested in theoretical
details may skip ahead to Section~\ref{sec:conc} now. A number of
technical and numerical details appear in the appendices.

\section{Phase diagram in zero magnetic field}
\label{sec:zero}

We orient ourselves by discussing the phase diagrams of models
with various types of spin and charge density wave order. We will
restrict our attention in this section to zero external field,
assume that a background SC order is always present in all the
phases. As we have argued above, this implies that we need not
consider the SC order parameter explicitly, and its influence only
serves to renormalize various couplings in the effective actions.
A somewhat different viewpoint, with a more explicit role for the
SC order, has been taken recently by Lee\cite{dunghai}.

Here, we consider phases that are characterized simply by the
condensates of one or more of the order parameters
$\Phi_{x,y\alpha}$ and $\phi_{x,y}$, introduced in
Section~\ref{sec:op}. More complex phases associated with
composites or ``fractions'' of these fields are also possible and
these will be considered later in Section~\ref{newphases}.
However, the remainder of the paper will only deal with the
influence of the magnetic field on phases and phase boundaries
associated with the order parameters $\Phi_{x,y\alpha}$ and
$\phi_{x,y}$; the more complex cases have similar properties which
can be described in an analogous manner.

\begin{figure}
\centerline{\includegraphics[width=3in]{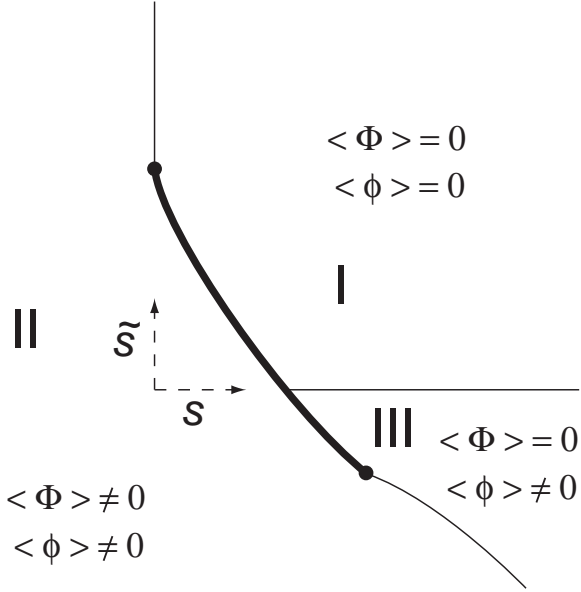}}
\caption{Mean field, zero temperature phase diagram of the zero
magnetic field model ${\cal S}_{\Phi} + {\cal S}_{\phi} + {\cal
S}_{\Phi\phi}$ defined in
(\protect\ref{SPhi},\protect\ref{Sphi},\protect\ref{Spp}), with
$u_2 < 0$.} \label{figzachar}
\end{figure}

To characterize the simple phases we need an effective action
${\cal S}_{\phi}$ for the $\phi_{x,y}$, while that for
$\Phi_{x,y\alpha}$ is ${\cal S}_{\Phi}$ in (\ref{SPhi}); the
former can be written down using a reasoning similar to that for
(\ref{SPhi}), and we obtain
\begin{eqnarray}
&& {\cal S}_{\phi} =\int d^2 r d \tau \Bigl[
   \left| \partial_\tau \phi_{x} \right|^2
   + \widetilde{v}_1^2 \left| \partial_x \phi_{x} \right|^2
   + \widetilde{v}_2^2 \left| \partial_y \phi_{x} \right|^2
\nonumber \\
&& +\left| \partial_\tau \phi_{y} \right|^2
   + \widetilde{v}_1^2 \left| \partial_x \phi_{y} \right|^2
   + \widetilde{v}_2^2 \left| \partial_y \phi_{y } \right|^2
+ \widetilde{s} (\left|\phi_{x}\right|^2
   + \left|\phi_{y}\right|^2)\nonumber \\
&&~~~~~~~~~~
   +\frac{\widetilde{u}_1}{2} ( \left|\phi_{x}\right|^4
   + \left|\phi_{y}\right|^4 ) +
   \widetilde{w}_1 \left|\phi_{x}\right|^2\left|\phi_{y}\right|^2
\Bigr]. \label{Sphi}
\end{eqnarray}
The correspondence (\ref{square}) implies that for ${\bf K}_{cx} =
2 {\bf K}_{sx}$ and ${\bf K}_{cy} = 2 {\bf K}_{sy}$ the SDW and
CDW order parameters are coupled by
\begin{equation}
{\cal S}_{\Phi\phi} = -\lambda \int d^2 r d \tau \left[
\phi_x^{\ast} \Phi_{x \alpha}^2 + \phi_y^{\ast} \Phi_{y \alpha}^2
+ \mbox{c.c.} \right]; \label{Spp}
\end{equation}
without loss of generality, we can assume that the coupling
$\lambda >0$. At the mean-field level, the properties of the
quantum model ${\cal S}_{\Phi}+ {\cal S}_{\phi} + {\cal
S}_{\Phi\phi}$ are essentially identical to the classical models
considered by Zachar {\em et al.}\cite{zachar} for spin and charge
ordering transitions at non-zero temperature; so we can directly
borrow their results, and a characteristic mean-field phase
diagram is shown in Fig~\ref{figzachar}.

Next, we discuss the critical properties of the various
second-order quantum transitions in Fig~\ref{figzachar}.

Near the transition between phase II (SC+SDW) and the symmetric
phase I (SC), the primary order parameters are $\Phi_{x,y
\alpha}$. We can integrate out the non-critical $\phi_{x,y}$
fields and this merely renormalizes the couplings in ${\cal
S}_{\Phi}$. So the theory ${\cal S}_{\Phi}$ is the critical theory
for this transition at $H=0$. This is a model of some complexity,
and the universal critical properties of related simpler models
are the focus of some debate in the
literature\cite{Jones,Bailin,Kawamura,mou,itakura}; these earlier
results are briefly reviewed in Appendix~\ref{rg}. These previous
studies correspond to the case where $\Phi_{x\alpha}$ and
$\Phi_{y\alpha}$ are decoupled ($w_1=w_2=w_3=0$), and weakly
first-order transitions are obtained in some cases. We will
address the generalization of these previous analyses to the case
of non-zero $w_{1,2,3}$ in future work. Here, we will be satisfied
by considering the simplest, and most symmetric, case of a
second-order transition: for the special values $v_1=v_2$,
$u_1=w_1$, $u_2=w_2=w_3=0$ the model ${\cal S}_{\Phi}$ has a O(12)
symmetry, and its properties are identical to that of the
$(N=12)$-component $\varphi^4$ theory ${\cal S}_{\varphi}$ to be
described shortly below. The influence of $H$ on other
second-order or weakly first-order transitions should be very
similar, with the changes only modifying the numerical values of
certain asymptotic critical parameters. Part of our reason for not
expending much effort on this point is that these asymptotic
critical are not particularly relevant for the experimental
situation in $H\neq 0$ anyway: after including the small effects
of $\mathcal{S}_{\rm lat}$ in (\ref{e4}), the ``sliding'' symmetry
of $\mathcal{S}_{\Phi}$ disappears, and the asymptotic critical
properties of the SC+SDW to SC transition in $H \neq 0$ become
identical to the $(N=3)$ component $\varphi^4$ theory ${\cal
S}_{\varphi}$. We will discuss the $H \neq 0$ properties of ${\cal
S}_{\varphi}$ at some length in this paper, and we expect that
closely related results apply to the generalized ${\cal S}_{\Phi}$
and to $\mathcal{S}_{\Phi} + \mathcal{S}_{\rm lat}$.

Near the transition between phases III and I in
Fig~\ref{figzachar}, the roles of $\Phi_{x,y\alpha}$ and
$\phi_{x,y}$ are reversed. Now we can integrate out the
non-critical $\Phi_{x,y\alpha}$, this renormalizes the couplings
in ${\cal S}_{\phi}$, and the renormalized ${\cal S}_{\Phi}$ is
the critical theory for this transition at $H=0$. At non-zero $H$,
a model closely related to the one discussed above applies. We
will not explicitly present the results for this model here, as
most physical properties are essentially identical to those of
${\cal S}_{\Phi} + {\cal S}_{\psi} + {\cal S}_{\Phi\psi}$.

The remaining second order quantum transition in
Fig~\ref{figzachar} is that between phases II and III. Both these
phases have $\langle \phi_{x,y} \rangle \neq 0$, and the charge
order can be viewed as a non-critical spectator to the transition.
For specificity, let us assume that $\langle \phi_x \rangle$ is
real and positive, while $\langle \phi_y \rangle =0$; other cases
lead to similar final results. Now replace $\phi_{x,y}$ by their
expectation values in ${\cal S}_{\Phi} + {\cal S}_{\Phi\phi}$ in
(\ref{SPhi},\ref{Spp}), and examine fluctuations of
$\Phi_{x,y\alpha}$ at the Gaussian level: those of
$\mbox{Re}\left[\Phi_{x\alpha}\right]$ have an energy lower than
all other components. Close to phase boundary between II and III
we can therefore assume that the critical theory involves {\em
only} $\varphi_{\alpha}  ({\bf r}, \tau) \equiv
\mbox{Re}\left[\Phi_{x\alpha}({\bf r}, \tau)\right]$, and all
other components only renormalize the couplings in its effective
action. In this manner, we can conclude that the II to III phase
transition is described by the familiar $(N=3)$-component
$\varphi^4$ field theory, with effective action
\begin{eqnarray}
&& {\cal S}_{\varphi} = \int d^2 r  d \tau \Bigl\{ \frac{1}{2}
\Bigl[ (\partial_{\tau} \varphi_{\alpha})^2 +
v^2 ({\bf \nabla}_{\bf r} \varphi_{\alpha})^2 \nonumber \\
&& ~~~~~~~~~~+ (s + \kappa |\psi ({\bf r}) |^2) \varphi_{\alpha}^2
\Bigr] + \frac{u}{2} (\varphi_{\alpha}^2)^2 \Bigr\}, \label{os}
\end{eqnarray}
where the index $\alpha=1\ldots N$, and the field
$\varphi_{\alpha} ({\bf r}, \tau)$ is real. We have rescaled
spatial co-ordinates to make the velocities $v_{1,2}$ equal to the
common value $v$. For completeness, we have also included the
coupling to the SC order $\psi$ which derives from (\ref{couple}).
An analysis of the properties of the theory ${\cal F}/T + {\cal
S}_{\varphi}$, defined in (\ref{of}) and (\ref{os}), in non-zero
field shall occupy us in most of the remainder of the paper.
Recall also that the $N=12$ case of this theory also describes a
particular case of the I to II transition discussed earlier.

\section{Phase diagram in a magnetic field}
\label{sec:pd}

We now embark on a presentation of the main new results of this
paper: a description of the phase diagram and the dynamic spin
spectra of ${\cal F}/T + {\cal S}_{\varphi}$, defined in
(\ref{of}) and (\ref{os}), as a function of the applied field $H$.
As discussed near (\ref{os}), this theory describes the response
of a number of specific phase boundaries of states with SDW/CDW
order to an applied magnetic field; the number of components of
$\varphi_{\alpha}$ takes the values $N=3,12$ depending upon the
transition of interest, but we expect similar results for all
values of $N\geq 3$. Actually, closely related analyses can be
applied to most of the phases to be discussed in
Section~\ref{newphases}. The basic effect, that all couplings
associated with the non-superconducting order parameter acquire a
$H \ln (1/H)$ depends, is very robust and leads to analogous phase
diagrams in almost all cases.

The theory ${\cal F}/T + {\cal S}_{\varphi}$ has a rather number
of coupling constants, and it is useful to use our freedom to
rescale lengths, times, and field scales to obtain an irreducible
set of parameters whose values control the structure of our
results. First, as is conventional in the standard Ginzburg-Landau
theory of superconductivity, we introduce the superconducting
coherence length, $\xi_0$, and the field scales $H_c$ and
$H_{c2}^0$:
\begin{eqnarray}
\xi_0 &=& \sqrt{\frac{1}{2 m^{\ast} \alpha}} \nonumber \\
H_c &=& \sqrt{\frac{4 \pi \alpha^2}{\beta}} \nonumber \\
H_{c2}^0 &=& \frac{2 m^{\ast} \alpha c}{e^{\ast}}; \label{pars}
\end{eqnarray}
as we noted earlier, $H_{c2}^0$ is the value of the upper critical
field at the point M in Fig~\ref{figpd}, and $H_{c2}^0 = \sqrt{2}
\kappa H_c$ where $\kappa$ is usual the Ginzburg Landau parameter.
We will also see below in Section~\ref{sec:renorm} that the
coupling $\alpha$ acquires a shift renormalization due to its
coupling to $\varphi_{\alpha}$ fluctuations: we assume that
renormalization has already been performed in the definitions
(\ref{pars}). We now use the length $\xi_0$, the velocity $v$, and
the parameters in (\ref{pars}) to set various length, time,
temperature, field, and coupling constant scales; we define the
dimensionless parameters
\begin{eqnarray}
\widetilde{{\bf r}} = \frac{{\bf r}}{\xi_0}~~;~~ \widetilde{\tau}
&=& \frac{v\tau}{\xi_0}~~;~~
\widetilde{T} = \frac{\xi_0 T}{v} \nonumber \\
\widetilde{H} = \frac{H}{H_{c2}^0} ~~;~~ \widetilde{\psi} &=&
\sqrt{\frac{\beta}{\alpha}} \psi ~~;~~
\widetilde{\varphi}_{\alpha} = \sqrt{v \xi_0} \varphi_{\alpha}
\nonumber \\
\widetilde{s} = \frac{\xi_0^2}{v^2} s ~~;~~ \widetilde{u} &=&
\frac{\xi_0}{v^3} u ~~;~~ \widetilde{\kappa} =
\frac{\xi_0^2\alpha}{v^2\beta} \kappa . \label{trans}
\end{eqnarray}
It is evident from the above that we are measuring length scales
in units of $\xi_0$ and energy scales in units of $v/\xi_0$.

Collecting all the transformations, let us restate the problem we
are going to solve; we drop all the tildes, and it is henceforth
assumed that all parameters have been modified as in
(\ref{trans}). The partition function in (\ref{zdef}) is now
simplified to
\begin{equation}
\mathcal{Z}\left[\psi ({\bf r}) \right] = \int {\cal D}
\varphi_{\alpha} ({\bf r}, \tau) \exp \left( - \frac{{\cal F}}{T}
- {\cal S}_{\varphi}\right), \label{zdef1}
\end{equation}
where ${\cal S}_{\varphi}$ is as in (\ref{os}) but with $v=1$,
while ${\cal F}$ is now given by
\begin{equation}
{\cal F} = \Upsilon \int d^2 r \left[- |\psi|^2 + \frac{1}{2}
|\psi|^4 + \left| \left( {\bf \nabla}_{\bf r} - i {\bf A} \right)
\psi \right|^2 \right] \label{f}.
\end{equation}
The dimensionless constant $\Upsilon$ is given by
\begin{equation}
\Upsilon = \frac{H_c^2 \xi_0^3 d}{4 \pi v}, \label{defUp}
\end{equation}
where $d$ is the inter-layer spacing (this factor of $d$ is needed
to make $\Upsilon$ dimensionless, and arises because ${\cal F}$ is
the free energy per layer); in determining $\Upsilon$, a useful
unit of conversion is 1 $(\mbox{Tesla})^2$ = 0.0624 meV
\AA$^{-3}$. The vector potential ${\bf A}$ in (\ref{f}) now
satisfies
\begin{equation}
{\bf \nabla}_{\bf r} \times {\bf A} = H \hat{z}. \label{defah}
\end{equation}

An important property of the continuum theory (\ref{zdef1}) is
that all dependence on the short distance cutoff can be removed by
a single ``mass renormalization'': this amounts to measuring the
tuning parameter $s$ in terms of its deviation from $s=s_c$, the
critical point between the SC+SDW and SC phases at $H=0$.
Consequently all physical properties are functions only of the
dimensionless parameters $u$, $\kappa$, $\Upsilon$, $H/H_{c2}^0$,
and $s-s_c$. We will present numerical results for the frequency
and spatial dependence of various observables below as a function
of $H/H_{c2}^0$ and $s-s_c$ for the simple set of values
$u=\kappa=\Upsilon=1$; we do not expect any qualitative changes
for other values of these last three parameters. Also, it will
occasionally be convenient to exchange the parameter $s-s_c$ for
$\Delta$, the value of the spin gap in the $s>s_c$ SC phase at
$H=0$.

The technical tool we shall use in our analysis of (\ref{zdef1})
is the large $N$ expansion. This approach\cite{csy} is known to
yield an accurate description of the vicinity of spin ordering
quantum critical points in two dimensions, and we expect the same
to hold here in the presence of a non-zero $H$. Details of the
approach will emerge in the following sections: here we summarize
the main $N=\infty$ results for the positions of the phase
boundaries appearing in Fig~\ref{figpd}:
\begin{itemize}
\item The tetra-critical point M where all four phases meet is at $H=1$, $s-s_c= \kappa$.
\item The line BM represents the upper-critical field for the
vanishing of superconductivity in the presence of SDW order; it is
at
\begin{equation}
H = 1 - \frac{\kappa^2}{4u\Upsilon} + \frac{\kappa}{4 u \Upsilon}
(s - s_c). \label{BM}
\end{equation}
\item
The line CM, the boundary for SDW order in the insulator, is at
$s-s_c= \kappa$.
\item The line DM, the upper-critical field for superconductivity
in the absence of SDW order is at
\begin{eqnarray}
H = 1 + \frac{N\kappa}{8 \pi \Upsilon} \Bigl[ \Bigl( \frac{N^2
u^2}{16 \pi^2 } - \kappa + s - s_c \Bigr)^{1/2} - \frac{Nu}{4 \pi}
\Bigr]. \label{DM}
\end{eqnarray}
\item
Experimentally, the most important and accessible phase boundary
is AM, the line representing onset of SDW order in the SC phase.
The position of this line cannot be determined analytically: we
will present detailed numerical results and an expansion in the
vicinity of M; for small $H$ its location behaves as
\begin{equation}
H \sim \frac{2 (s-s_c)}{\kappa \ln (1/(s-s_c))}, \label{amlog}
\end{equation}
as may be readily deduced from (\ref{reff}), and was quoted
already in (\ref{amlog1}).
\end{itemize}

Our numerical as well as analytical studies will be divided into
two parts, one for ``SC'' region of Fig~\ref{figpd} in
Section~\ref{SC}, and the other for ``SC + SDW'' region in
Section~\ref{SC+SDW}.

\section{Physical properties of the SC phase} \label{SC}

This section will describe an analysis of (\ref{zdef1}) in the
regime where spin rotation invariance is preserved with $\langle
\varphi_{\alpha} \rangle =0$. As we discussed earlier at the end
of Section~\ref{sec:phys}, upon including the effect of the
lattice pinning term (\ref{e4}) in a non-zero $H$, this phase does
have static CDW order with $\langle \phi_{x,y} \rangle \neq 0$,
while preserving spin rotation invariance: this will be discussed
in Section~\ref{sec:pin}.

\subsection{Large $N$ saddle point equations}
\label{nsc}

The index $\alpha$ in ${\cal S}_{\varphi}$ in (\ref{os}) extends
over $\alpha=1\ldots N$, and depending upon the transition in
Fig~\ref{figzachar} we are interested in, we have either $N=3$ or
$N=12$. For both cases, it is known that an accurate description
of the physical properties is described by the large $N$
expansion, whose implementation we shall now describe.

First, we introduce an auxiliary field
\begin{equation}
\mathcal{V} ({\bf r}, \tau) = s + \kappa | \psi_H ({\bf r}) |^2 +
2 u \varphi_{\alpha}^2 ({\bf r}, \tau). \label{aux}
\end{equation}
We will often place a subscript $H$ on various quantities (as for
$\psi$ above) to emphasize that they are being evaluated at a
non-zero $H$. Let us also denote
\begin{equation}
s' = s +  \kappa | \psi_H ({\bf r}) |^2. \label{sprime}
\end{equation}
Now we add an innocuous term to ${\cal S}_{\varphi}$, whose only
effect is to multiply the partition function by a constant after a
functional integration over $\mathcal{V} ({\bf r}, \tau)$:
\begin{eqnarray}
{\cal S}_{\varphi} & \to & {\cal S}_{\varphi} - \int d^2 r
\int_0^{1/T} d \tau \frac{1}{8 u}
(\mathcal{V} - 2 u \varphi_{\alpha}^2 - s')^2 \nonumber \\
&=& \int d^2 r \int_0^{1/T} d \tau \Bigl[
\frac{1}{2}(\partial_{\tau} \varphi_{\alpha})^2
+ \frac{1}{2} ({\bf \nabla}_{\bf r} \varphi_{\alpha})^2 \nonumber \\
& & - \frac{1}{8 u} \mathcal{V}^2 + \frac{1}{2} \mathcal{V}
\varphi_\alpha^2 + \frac{1}{4 u} \mathcal{V} s' \Bigr]. \label{rS}
\end{eqnarray}
After integrating out $\varphi_\alpha (\alpha = 1 \cdots N)$, we
have
\begin{eqnarray}
{\mathcal Z}  = \int {\mathcal D} \mathcal{V} ({\bf r}) \exp
\Bigl[&& - \frac{N}{2}
Tr \ln (- \partial_{\tau}^2 - {\bf \nabla}_{\bf r}^2 + \mathcal{V}) \nonumber \\
&& - \frac{1}{4 u} \mathcal{V} s' + \frac{1}{8 u} {\mathcal{V}}^2
\Bigr].
\end{eqnarray}
Now by taking $N \to \infty$ while keeping $N u$ constant, we
obtain the saddle point equation in which $\mathcal{V}$ is a
function of ${\bf r}$ but independent of $\tau$:
\begin{equation}
\mathcal{V}_H ({\bf r}) = s + \kappa | \psi_H ({\bf r}) |^2 + 2 N
u T \sum_{\omega_n}
    G_H({\bf r}, {\bf r}, \omega_n). \label{chi}
\end{equation}
where the $\varphi_{\alpha}$ propagator $G_H({\bf r},{\bf r}',
\omega_n)$ is given by
\begin{equation}
G_H({\bf r},{\bf r}',\omega_n) = \langle {\bf r} |
\left(\omega_n^2 - {\bf \nabla}_{\bf r}^2 + \mathcal{V}_H ({\bf
r}) \right)^{-1} |{\bf r}'\rangle \label{g},
\end{equation}
with $\omega_n$ a Matsubara frequency. In this case, the large-$N$
expansion is equivalent to a self-consistent one-loop calculation.

The saddle point equation for superconducting order parameter
follows from (\ref{scdef}): it is just the conventional
Ginzburg-Landau equation with one additional term from the
$\varphi,\psi$ coupling,
\begin{eqnarray}
\Bigl[-1 &+& \frac{N\kappa}{2\Upsilon} T \sum_{\omega_n} G_H ({\bf
r},{\bf r},\omega_n)
\nonumber \\
&+& |\psi_H ({\bf r})|^2 - ({\bf \nabla}_{\bf r} - i {\bf A})^2
\Bigr] \psi_H ({\bf r}) = 0. \label{gl}
\end{eqnarray}

So the two unknown functions, $\mathcal{V}_H ({\bf r})$, and
$\psi_H({\bf r})$ are to be determined simultaneously by the
solution of (\ref{chi}) and (\ref{gl}). As stated above, the
expressions in these equations depend upon the short distance
cutoff, but we show in (\ref{sec:renorm}) that this can easily be
removed by a simple shift of parameters.

\subsection{Renormalization of parameters}
\label{sec:renorm}

It is first useful to obtain the complete solution of (\ref{chi})
and (\ref{gl}) at $H=0$. Let $s=s_c$ be the point where magnetic
order appears (so that $\langle \varphi_{\alpha} \rangle \neq 0$
for $s<s_c$), where\cite{csy} $\mathcal{V}=0$. Then (\ref{chi})
tells us that
\begin{equation}
0 = s_c + \kappa |\psi_{0c}|^2 + 2 N u \int \frac{d \omega}{2 \pi}
\int \frac{d^2 k }{4 \pi^2} \frac{1}{\omega^2 + k^2}, \label{chi0}
\end{equation}
where $\psi_{0c}$ is the ${\bf r}$ independent value of $\psi
({\bf r})$ at $s=s_c$ and $H=0$, while (\ref{gl}) gives
\begin{equation}
- 1 + |\psi_{0c}|^2 + \frac{N\kappa}{2\Upsilon} \int \frac{d
\omega}{2 \pi} \int \frac{d^2 k }{4 \pi^2} \frac{1}{\omega^2 +
k^2} = 0. \label{gl0}
\end{equation}
It is useful to normalize things so that $\psi_{0c} = 1$ at
$s=s_c$, $H=0$ and $T=0$. This is achieved if we renormalize
$\alpha$ to remove the offending term in (\ref{gl0}). We make the
shift in (\ref{of}) (before the rescalings in (\ref{trans}))
\begin{equation}
\alpha \rightarrow \alpha + \frac{N\kappa\beta}{2\alpha} \int
\frac{d \omega}{2 \pi} \int \frac{d^2 k}{4 \pi^2}
\frac{1}{\omega^2 + v^2 k^2}. \label{alpha}
\end{equation}
Then, after (\ref{trans}), (\ref{gl}) is modified to
\begin{eqnarray}
\Bigl\{-1 &+& \frac{N\kappa}{2\Upsilon} \Bigl[ T \sum_{\omega_n}
G_H ({\bf r},{\bf r},\omega_n) - \int \frac{d \omega d^2 k}{8
\pi^3}
\frac{1}{\omega^2 + k^2}\Bigr] \nonumber \\
&+& |\psi_H ({\bf r})|^2 - ({\bf \nabla}_{\bf r} - i {\bf A})^2
\Bigr\} \psi_H ({\bf r}) = 0 \label{gl1}
\end{eqnarray}
while (\ref{gl0}) becomes simply
\begin{equation}
\psi_{0c} = 1. \label{psi0c}
\end{equation}

Now move to $s>s_c$, where we have a spin gap
\begin{equation}
\Delta_0 \equiv \sqrt{\mathcal{V}_0} > 0. \label{spingap}
\end{equation}
Subtracting (\ref{chi0}) from (\ref{chi}) we get
\begin{equation}
\Delta_0^2 = s-s_c + \kappa(|\psi_0|^2-1) - \frac{N u \Delta_0}{2
\pi} \label{chi1}
\end{equation}
where (\ref{gl1}) yields
\begin{equation}
|\psi_0|^2 = 1 + \frac{N\kappa \Delta_0}{8 \pi \Upsilon}
\label{psi0}.
\end{equation}
Inserting (\ref{psi0}) back into (\ref{chi1}) we obtain
\begin{equation}
\Delta_0^2 + \frac{Nu}{2 \pi } \left( 1 -
\frac{\kappa^2}{4u\Upsilon} \right) \Delta_0 = s-s_c \label{chi2}
\end{equation}

Let us now use the above equations to simplify the equations for
$H \neq 0$ and $T \neq 0$. The new form will be independent of
lattice cutoff.

{}From (\ref{chi},\ref{chi0},\ref{chi2}) we obtain
\begin{eqnarray}
\mathcal{V}_H ({\bf r}) =&& \Delta_0^2 + \kappa \left( |\psi_H
({\bf r})|^2 -|\psi_0|^2 \right)  \nonumber
\\ &&+ 2 N u \Bigl[ T \sum_{\omega_n} G_H({\bf r},{\bf r},\omega_n)
\nonumber \\
&&~~~- \int \frac{d \omega}{2 \pi} \frac{d^2 k}{4 \pi^2}
\frac{1}{\omega^2 + k^2+\Delta_0^2} \Bigl] \label{chif}
\end{eqnarray}
where $|\psi_0|^2$ is given in (\ref{psi0}). Using (\ref{chif}) and
(\ref{gl1}) we obtain
\begin{eqnarray}
&&\Bigl[\left(1- \frac{\kappa^2}{4 u \Upsilon} \right) \left(
|\psi_H ({\bf r})|^2 - |\psi_0|^2 \right) + \frac{\kappa}{4 u
\Upsilon} (\mathcal{V}_H({\bf r}) - \Delta_0^2 )
\nonumber \\
&&~~~~~~~~~~~~~- ({\bf \nabla}_{\bf r} - i {\bf A})^2 \Bigr]
\psi_H ({\bf r}) = 0 \label{glf}
\end{eqnarray}
The expressions (\ref{chif},\ref{glf}) are the main equations we
shall solve for the unknowns $\mathcal{V}_H ({\bf r})$ and $\psi_H
({\bf r})$ in this paper. It can be checked that at $H=0$, $T=0$,
these equations are solved by $\mathcal{V}_H=\Delta_0^2$ and
$\psi_H = \psi_0$. We describe the numerical solution of these
equations for $H\neq 0$ in Appendix~\ref{numsc} and present the
results in the following subsection. A useful step in this
numerical solution is the following parameterization of the
Green's function $G_H ({\bf r}, {\bf r}', \omega_n)$ in (\ref{g})
\begin{equation}
G_H ({\bf r}, {\bf r}', \omega_n) = \sum_{\mu} \int_{1BZ}
\frac{d^2 k}{4 \pi^2} \frac{\Xi_{\mu {\bf k}}^{\ast} ({\bf r})
\Xi_{\mu {\bf k}} ({\bf r}')}{\omega_n^2 + E_{\mu}^2 ({\bf k})},
\label{geigen}
\end{equation}
where $\Xi_{\mu {\bf k}} ({\bf r})$ are the complete set of
eigenfunctions of the analog of the Schr\"odinger equation
(\ref{schro})
\begin{equation}
\left( -\nabla_{{\bf r}}^2 + \mathcal{V}_H ({\bf r}) \right)
\Xi_{\mu {\bf k}} ({\bf r}) = E_{\mu}^2 ({\bf k}) \Xi_{\mu {\bf
k}} ({\bf r}). \label{schro1}
\end{equation}
Here ${\bf k}$ is a `Bloch' momentum which extends over the first
Brillouin zone of the vortex lattice, $\mu$ is a `band' index, and
$E_{\mu}({\bf k})$ are the corresponding energy eigenvalues. All
of our numerical analysis was performed for the values
$u=\kappa=\Upsilon=1$ and $N=3$.

\subsection{Phase boundaries}
\label{sec:pb}

The equations (\ref{chif}) and (\ref{glf}) can be readily solved
to obtain the locations of the CM and DM phase boundaries in
Fig~\ref{figpd}. On DM, the superconducting phase parameter
$\psi_H({\bf r})$ vanishes and all parameters become ${\bf r}$
independent; thus Eqn (\ref{chif}) becomes
\begin{eqnarray}
\mathcal{V}_H &=& \Delta_0^2 - \kappa |\psi_0|^2 + \frac{N u}{2
\pi } (\sqrt{\mathcal{V}_H}
-\Delta_0) \nonumber \\
&=& s - s_c - \kappa - \frac{N u}{2 \pi } \sqrt{\mathcal{V}_H}.
\label{chiMD}
\end{eqnarray}
where we used Eqn (\ref{psi0}) and (\ref{chi1}). Then from Eqn.
(\ref{glf}) we have
\begin{eqnarray}
H &=& 1 - \frac{\kappa^2}{4 u \Upsilon} + \frac{\kappa}{4 u \Upsilon}
(s - s_c - \mathcal{V}_H) \nonumber \\
 &=& 1 - \frac{\kappa^2}{4 u\Upsilon} + \frac{\kappa}{4 u\Upsilon} (\frac{N u}{2 \pi }
 \sqrt{\mathcal{V}_H} + \kappa) \nonumber \\
 &~&\!\!\!\!\!\!\!\!\!\!\!\!\!\!\!\!\!\!\!\!\! = 1 +
 \frac{N \kappa}{8 \pi \Upsilon}\Bigl[\Bigl(\frac{N^2 u^2}{16 \pi^2 }
 -\kappa + s - s_c\Bigr)^{1/2} - \frac{N u}{4 \pi } \Bigr],
\end{eqnarray}
which is the result quoted in (\ref{DM}). Similarly, it is easy to
see that the phase boundary CM is at $s-s_c = \kappa$.

\begin{figure}
\centerline{\includegraphics[width=3.3in]{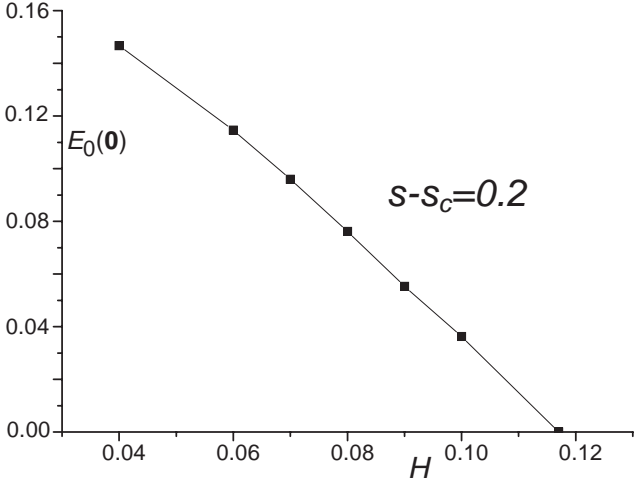}}
\caption{The lowest eigenvalue of (\protect\ref{schro}), $E_{0}
({\bf 0})$ vs. $H$ for $s - s_c = 0.2$. The linear continuation of
the line to solve $E_{0} ({\bf 0}) = 0$ gives us the critical $H$
for this $s$ , which is about $0.117$ with an uncertainty of $\pm
0.002$.} \label{figE00r2}
\end{figure}

\begin{figure}
\centerline{\includegraphics[width=3.3in]{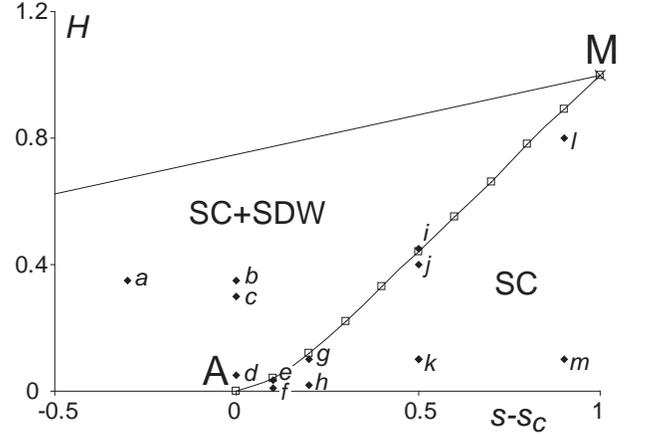}}
\caption{Numerical results for the phase boundary AM in
Fig~\protect\ref{figpd} for $u=\kappa=\Upsilon=1$. Also shown is a
portion of the phase boundary BM whose position is known
analytically from (\protect\ref{BM}). Different aspects of the
physical properties are described in the remainder of the paper at
the points labeled {\em a-l}.} \label{figAM}
\end{figure}

It remains to determine the location of the phase boundary AM,
which is also physically the most interesting one. We determined
this boundary by a full numerical solution of (\ref{chif}) and
(\ref{glf}) for a range of parameters. Stability of the SC phase
requires that all the eigenvalues, $E_{\mu}^2 ({\bf k})$, of
(\ref{schro}) remain positive. The lowest of these eigenvalues is
$E_{0} ({\bf 0})$ and we followed its behavior as a function $H$:
a typical result is shown in Fig~\ref{figE00r2}.
We expect $E_{0} ({\bf 0})$ to vanish linearly in the deviation
from the critical field, as the critical theory is expected to be
in the universality class of the ordinary O(3) $\varphi^4$ field
theory, and the latter has critical exponent $z \nu = 1$ in the
large $N$ limit. So we can determine the critical field by a
linear extrapolation, and this is also shown in
Fig~\ref{figE00r2}. Combining the results of such calculations at
a range of values of $s$, we obtain our numerical result for the
location of the AM boundary shown in Fig~\ref{figAM}.

Some further analytic results on the location of the AM phase
boundary can be obtained in the vicinity of the multi-critical
point M. It can be shown that the deviation of the phase boundary
from M is linear in the large $N$ limit {\em i.e.} it is at $H = 1
- \varrho (\kappa - s + s_c)$, where $\varrho$ is a numerical
constant. We describe these results in Appendix~\ref{amlin},
including the determination of $\varrho$. The results obtained in
this manner are consistent with our complete numerical analysis
described above, and this is a strong check on our numerical
analysis.

Finally, we recall our result (\ref{amlog}) for the behavior of AM
at small $H$ and $s-s_c$. Here there is a crucial logarithm which
follows from (\ref{reff}), and whose physical origin was discussed
in Section~\ref{sec:phys}. The signal of this logarithm are
clearly visible in the phase boundary in Fig~\ref{figAM}.

\subsection{Dynamic spin susceptibility}
\label{dyspin}

In this section we describe the evolution of the dynamic spin
fluctuation spectrum in the SC phase of Fig~\ref{figpd}. This is
clearly specified by the Green's function $G_H({\bf r},{\bf
r}',\omega_n)$ in (\ref{g}) which we computed above in determining
the phase boundary. More specifically, we see from (\ref{e1}) that
the observed dynamic spin susceptibility $\chi ({\bf q}, \omega)$
is given by
\begin{eqnarray}
&& \chi({\bf q}, \omega) \propto  \chi_{\varphi}({\bf q} + {\bf
K}_{sx}, \omega) + \chi_{\varphi} ({\bf q} - {\bf
K}_{sx},\omega) \nonumber \\
&&~~~~~ +  \chi_{\varphi}({\bf q} + {\bf K}_{sy}, \omega) +
\chi_{\varphi} ({\bf q} - {\bf K}_{sy},\omega), \label{totalsus}
\end{eqnarray}
where $\chi_{\varphi}$, the dynamic susceptibility for the field
$\varphi_{\alpha}$, is given by
\begin{eqnarray}
&& \chi_{\varphi} ({\bf k}, \omega)  =  \frac{1}{V} \int d^2 r d^2
r' e^{i {\bf k} \cdot ({\bf r} - {\bf r}')}
G_H({\bf r}, {\bf r}', \omega) \nonumber \\
&&= \sum_{\mu, {\bf G}} \int_{1BZ} d^2 p \  \delta( {\bf p} + {\bf
G} - {\bf k}) \frac{|c_{\mu{\bf G}} ({\bf p})|^2} {E_{\mu}^2 ({\bf
p}) - \omega^2}, \label{sus}
\end{eqnarray}
where $V$ is the volume of the system, the ${\bf p}$ integration
is over the first Brillouin zone of the reciprocal vortex lattice,
${\bf G}$ extends over the reciprocal lattice vectors of the
vortex lattice, $E_{\mu}^2 ({\bf p})$ are the eigenvalues of
(\ref{schro1}) (see also Appendix~\ref{numsc}), and the parameters
$c_{\mu{\bf G}} ({\bf p})$ are defined in (\ref{xirr}). We present
results for $\mbox{Im} [\chi_{\varphi} ({\bf k}, \omega)]$ below.

\begin{figure}
\centerline{\includegraphics[width=2.6in]{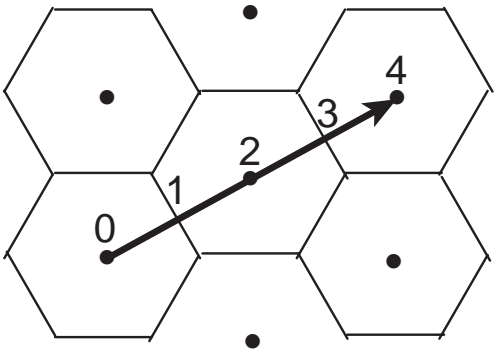}}
\caption{Reciprocal lattice of the vortex lattice. The density
plots in Figs~\protect\ref{figr1b01}, \protect\ref{figr1b035},
\protect\ref{figr9b10}, and \protect\ref{figr9b80} have ${\bf k}$
varying along the arrow shown, with numerical values as shown.}
\label{figk}
\end{figure}

\begin{figure}
\centerline{\includegraphics[width=3.4in]{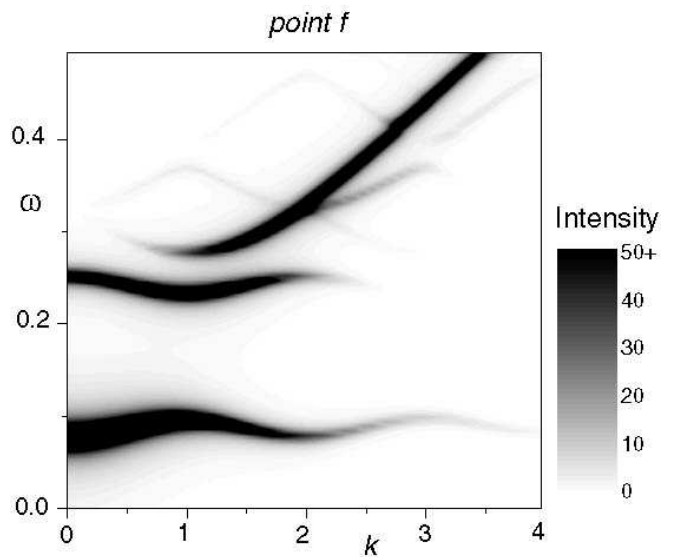}}
\caption{Density plot of $\mbox{Im} \chi_{\varphi}$ in
(\protect\ref{sus}) in the SC phase for momenta along the arrow in
Fig~\ref{figk}. The plot is for $s-s_c=0.1$ and $H=0.01$ (point
$f$ in Fig~\protect\ref{figAM}). In this, and all subsequent plots
of $\mbox{Im} \chi_{\varphi}$, the delta function peaks in
(\protect\ref{sus}) have been broadened into Lorentzians with
energy width 0.01 for display purposes only.} \label{figr1b01}
\end{figure}

\begin{figure}
\centerline{\includegraphics[width=3.4in]{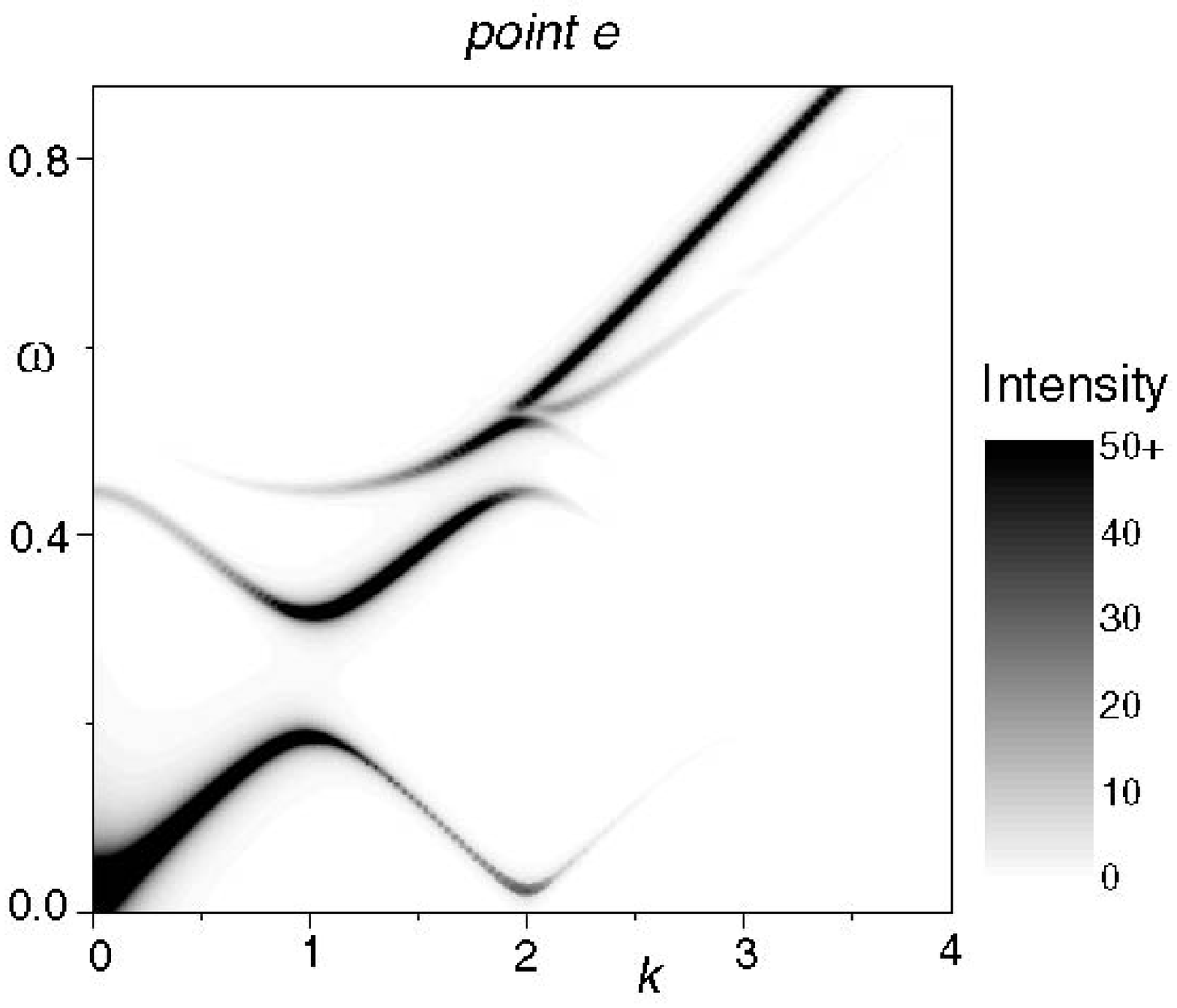}}
\caption{As in Fig~\protect\ref{figr1b01} but for larger
$H=0.035$, which brings the system very close to the AM phase
boundary to the SC+SDW phase (point $e$ in
Fig~\protect\ref{figAM}).} \label{figr1b035}
\end{figure}

\begin{figure}
\centerline{\includegraphics[width=3.4in]{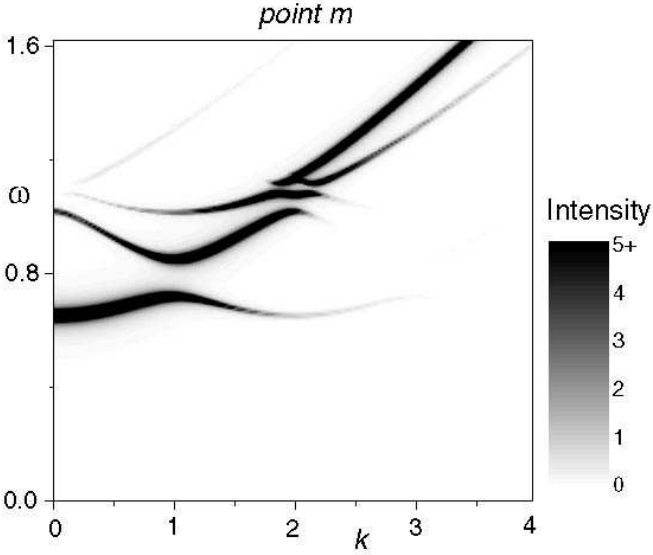}}
\caption{As in Fig~\protect\ref{figr1b01} but with larger $s-s_c$:
$H = 0.1$ and $s - s_c =0.9$ (point $m$ in
Fig~\protect\ref{figAM}).} \label{figr9b10}
\end{figure}

\begin{figure}
\centerline{\includegraphics[width=3.4in]{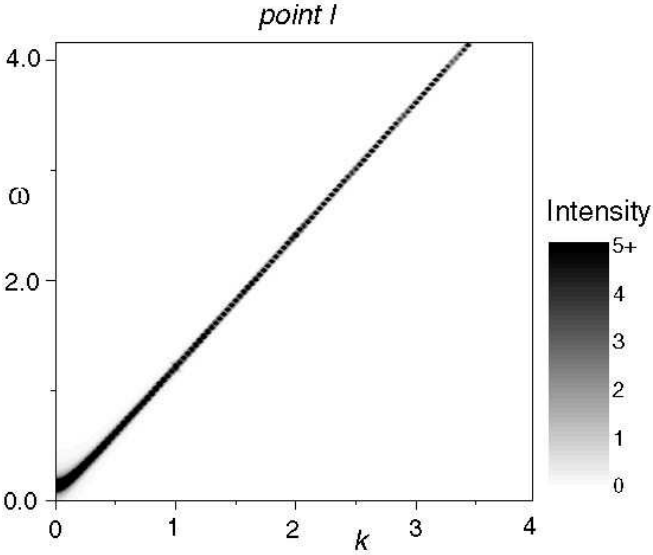}}
\caption{As in Fig~\protect\ref{figr9b10} but with a large
$H=0.8$, which brings the system very close to the AM phase
boundary to the SC+SDW phase (point $l$ in
Fig~\protect\ref{figAM}).} \label{figr9b80}
\end{figure}

It is clear from (\ref{sus}) that in the present large $N$
approximation, the spectrum of $\chi_{\varphi}$ consists entirely
of sharp delta functions. These specify the dispersion of $S=1$
``excitons'' which describe the SDW fluctuations, and are
connected with the zero field ``resonance'' peak discussed early
on in Section~\ref{sec:intro}. The excitons scatter off the vortex
lattice, and our results describe the evolution of the resulting
spectrum as one moves towards the onset of SDW order by increasing
the applied magnetic field. We show the structure of $\mbox{Im}
[\chi_{\varphi} ({\bf k}, \omega)]$ by broadening the delta
functions into sharp Lorentzians, and displaying the results in
density plots. The momentum ${\bf k}$ in these plots varies along
the direction of the reciprocal lattice shown in Fig~\ref{figk}.
The results for a smaller value of $s-s_c$ are shown in
Figs~\ref{figr1b01} and~\ref{figr1b035}, and those for larger
value of $s-s_c$ are in Figs~\ref{figr9b10}, and~\ref{figr9b80}.
Note that for very small $H$, there is less dispersion for the
lowest mode: this is an indication that this excitation is
centered on the vortex core, and there is weaker coupling between
neighboring vortices. As the field is increased, this coupling
increases, and the dispersion looks closer to that of a
nearly-free particle, with weak reflections at the Brillouin zone
boundaries of the vortex lattice. Also, the energy of the minimum
excitation decreases with increasing field, until it vanishes at
the AM phase boundary to the SC+SDW phase.

\begin{figure}
\centerline{\includegraphics[width=2.8in]{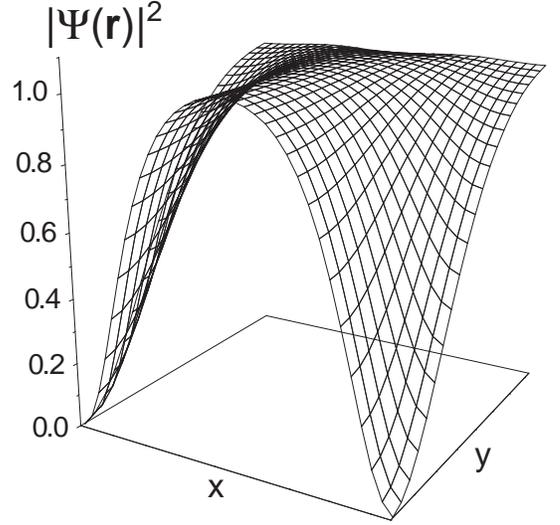}}
\caption{Spatial dependence of the modulus of the superconducting
order parameter $|\psi_H ({\bf r})|^2$ plotted on the rectangular
half unit cell of the vortex lattice indicated by
Fig~\protect\ref{figcell}. This result is for $s-s_c = 0.5$, and
$H=0.1$ (point $k$ in Fig~\protect\ref{figAM}).} \label{figomega}
\end{figure}

We also show in Fig~\ref{figomega} the spatial structure of the
modulus of the superconducting order parameter $|\psi_H ({\bf
r})|^2$.
\begin{figure}
\centerline{\includegraphics[width=2.3in]{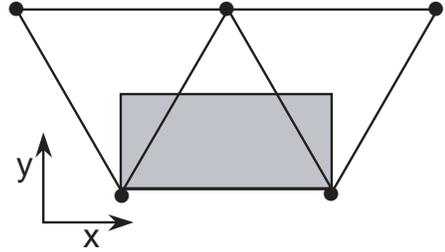}}
\caption{Half unit cell of the triangular vortex lattice in real
space.} \label{figcell}
\end{figure}

The Brillouin zone boundary reflections above arise from the
scattering of the exciton off the potential created by $|\psi_H
({\bf r})|^2$.

\begin{figure}
\centerline{\includegraphics[width=2.8in]{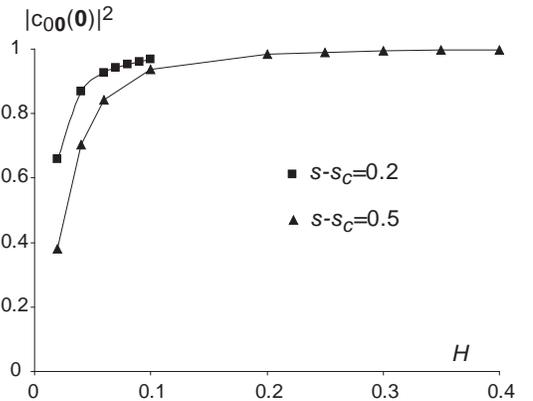}}
\caption{Intensity of the lowest exciton mode in the SC phase,
$|c_{0 {\bf 0}}({\bf 0})|^2$, as a function of $H$ for two values
of $s-s_c$. } \label{c00b}
\end{figure}

Finally, for experimental comparisons, it is useful to plot the
intensity of the lowest exciton mode as a function of the applied
field. From (\ref{sus}) we see that this intensity is $|c_{0 {\bf
0}}({\bf 0})|^2$. We show a plot of this quantity in
Fig~\ref{c00b}.
Observe that except for very small values of $H$, the intensity is
of order unity, which is the behavior expected for an extended
exciton scattering off a periodic potential as in
Fig~\ref{figomega}. As $H \rightarrow 0$, the behavior crosses
over to that expected when the vortex cores are essentially
decoupled, and the lowest mode is associated with a state
localized around each vortex core: in this limit, we
expect\cite{hu} the intensity $\sim H$.

\subsection{Pinning of charge order}
\label{sec:pin}

This section will consider the consequences of the pinning term
${\cal S}_{\rm lat}$ in (\ref{e4}). We argued at the end of
Section~\ref{sec:phys} that this term pins the charge order, and
leads to a static CDW with $\langle \phi_{x,y} \rangle  \neq 0$
(recall (\ref{e2})) in the SC phase, while preserving spin
rotation invariance with $\langle \Phi_{x,y\alpha} \rangle = 0$.
We have recently proposed\cite{pphmf} this as an explanation for
the CDW observed around the vortex in the STM measurements of
Hoffman {\em et al.}\cite{seamus}. Section~\ref{sec:phys}, also
gave an initial estimate (in (\ref{pcdw})) of the spatial
structure of this pinned CDW: here we will obtain a more precise
result, using the full solution of the SDW fluctuations in the
presence of the vortex lattice. Using the relationship
(\ref{square}) between the CDW and SDW orders in the vicinity of
the SC to SC+SDW transition, we conclude that to first order in
$\zeta$
\begin{equation}
\langle \phi_{x,y} ({\bf r} ) \rangle \propto \zeta e^{-i\varpi }
\Omega ({\bf r})
\end{equation}
with
\begin{equation}
\Omega ({\bf r} ) \equiv T \sum_{\omega_n} \sum_{{\bf r}_v} G_H^2
({\bf r}, {\bf r}_v, \omega_n ), \label{defOmega}
\end{equation}
where ${\bf r}_v$ extends over the vortex lattice sites; clearly
$\Omega ({\bf r})$ has the full periodicity of the vortex lattice.

\begin{figure}
\centerline{\includegraphics[width=3in]{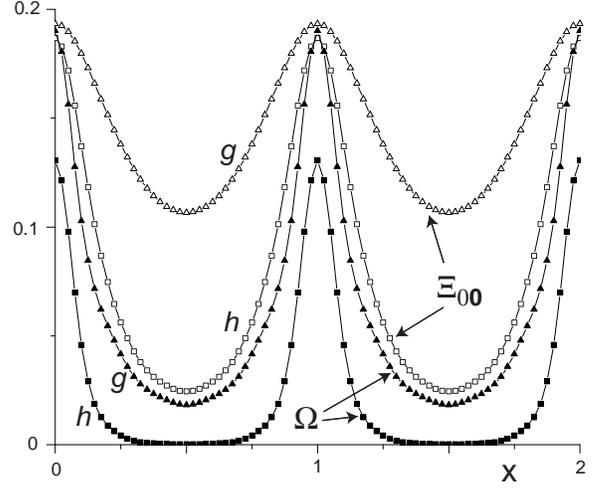}}
\caption{Plots of the function $\Omega ({\bf r})$ (filled symbols)
in (\protect\ref{defOmega}) representing the static CDW order
pinned by the vortices, along with the lowest SDW eigenfunction
$\Xi_{0{\bf 0}}( {\bf r})$ of the dynamic spin fluctuations above
the spin gap (open symbols), at $s-s_c = 0.2$. The spatial
co-ordinate $x$ is along the line connecting two nearest-neighbor
vortices and its scale has been chosen so that the vortex lattice
spacing is unity (see Fig~\protect\ref{figcell}). The field takes
the values $H=0.02$ (squares, point $h$ in
Fig~\protect\ref{figAM}) and $H=0.1$ (triangles, point $g$ in
Fig~\protect\ref{figAM}); the latter field is close to the AM
phase boundary in Fig~\protect\ref{figAM}. Note that the spin
exciton state at point $g$ is well extended through the lattice,
while the charge order remains localized around the vortices. For
point $h$ the localization length of the spin exciton state is
about twice that of the charge order. These results are consistent
with the discussion in Section~\protect\ref{sec:pinsc}. As was
also noted below (\protect\ref{pcdw}), the continuum expression
(\protect\ref{defOmega}) actually has a divergence for ${\bf r}$
equal to any ${\bf r}_v$: our numerical computation uses a finite
momentum cutoff $\Lambda$, and this rounds out the divergence at
distances $|{\bf r} - {\bf r}_v| \lesssim \Lambda^{-1}$; we have
verified this by numerical computations at different $\Lambda$. In
the same units as those for $x$ in the figure, we used $\Lambda
\approx 36$ above.}\label{figcdw02}
\end{figure}

\begin{figure}
\centerline{\includegraphics[width=3in]{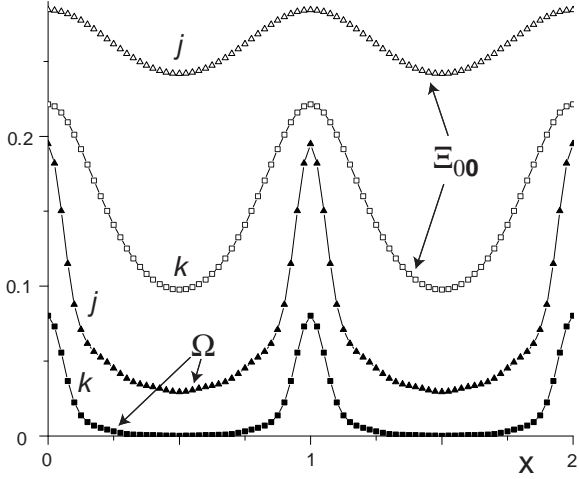}} \caption{As
in Fig~\protect\ref{figcdw02} but for $s-s_c=0.5$. The field takes
the values $H=0.1$ (squares, point $k$ in Fig~\protect\ref{figAM})
and $H=0.4$ (triangles, point $j$ in Fig~\protect\ref{figAM}); the
latter field is close to the AM phase boundary in
Fig~\protect\ref{figAM}. Now both points have extended spin
exciton states (that at point $j$ is essentially a plane wave),
while the charge order is exponentially
localized.}\label{figcdw05}
\end{figure}

We used our numerical solution of (\ref{chif}) and (\ref{glf}) to
compute the function $\Omega ({\bf r})$, which is proportional to
the amplitude of the static CDW induced by the vortex lattice in
the spin gap phase. We show our results for $\Omega ({\bf r})$ in
Figs~\ref{figcdw02} and~\ref{figcdw05} for a representative set of
values in the SC phase.
Also shown in the same figures, for orientation, is the form of
$\Xi_{0{\bf 0}} ({\bf r})$, the lowest energy eigenfunction of the
dynamic SDW equation (\ref{schro}) which appears in the Green's
function (\ref{geigen}). For very small field, both $\Omega ({\bf
r})$ and $\Xi_{0{\bf 0}} ({\bf r})$ are localized around the
vortex centers, with the localization length of the former being
about half that of the latter. However, for larger fields, the
exciton wavefunction $\Xi_{0{\bf 0}} ({\bf r})$ gets delocalized,
while the CDW order {\em remains localized}. This localization
arises from the summation over all the states in (\ref{geigen})
and is in keeping with the discussion at the end of
Section~\ref{sec:pinsc}.

\section{Physical properties of the SC+SDW phase} \label{SC+SDW}

We now turn to the analysis of the partition function
(\ref{zdef1}) in the phase with broken spin rotation invariance
and $\langle \varphi_{\alpha} \rangle = 0$. This phase is reached
when the lowest $S=1$ exciton mode in Section~\ref{SC},
$\Xi_{0{\bf 0}} ({\bf r})$, reaches zero energy ($E_0 ({\bf 0}) =
0$) and then condenses. The presence of the condensate leads to
long-range SDW order. We will adapt our large $N$ computation to
include such a condensate in the following subsection, and then
describe the spatial structure of the condensate and the dynamic
spin excitations.

\subsection{Large-$N$ saddle point equations}
\label{nsdw}

The analysis here is parallel to that in Section~\ref{nsc}. We
introduce the auxiliary field ${\cal V} ({\bf r }, \tau)$ defined
in (\ref{aux}) and write the action in the form similar to
(\ref{rS}). However, to account for the condensate, we have to
select a particular orientation in spin space, and treat the
corresponding spin component in a selective manner. So we write
\begin{equation}
\varphi_\alpha = (\sqrt{N} n, \pi_1, \pi_2, \cdots, \pi_{N-1}),
\label{transverse}
\end{equation}
and integrate out only $\pi_1, \pi_2, \cdots, \pi_{N-1}$ to obtain
\begin{eqnarray}
&&  {\mathcal Z}  = \int {\mathcal D} \mathcal{V} ({\bf r}, \tau )
{\mathcal D} n ({\bf r}, \tau )
\nonumber \\
&&\!\!\!\!\!\!\exp \Bigl[ - \frac{N-1}{2} Tr \ln (-
\partial_{\tau}^2 - {\bf \nabla}_{\bf r}^2 + \mathcal{V}) +
\frac{1}{8 u} {\mathcal{V}}^2
\nonumber \\
&& \!\!\!\!\!\! - \frac{1}{4 u} \mathcal{V} s' - \frac{N}{2}
(\partial_{\tau} n)^2 - \frac{N}{2}  ({\bf \nabla}_{\bf r} n)^2 -
\frac{N}{2} \mathcal{V} n^2 \Bigr],
\end{eqnarray}
where $s'$ was defined in (\ref{sprime}). Now we take $N \to
\infty$ while keeping $N u$ fixed, and ignoring the difference
between $N$ and $N-1$. This leads to saddle point equations for
the time-independent field $\mathcal{V}_H ({\bf r})$ and the SDW
condensate $n_H ({\bf r})$; these equations replace (\ref{chi}),
but contain additional terms due to the spontaneous spin
polarization:
\begin{eqnarray}
&& \mathcal{V}_H ({\bf r}) = s + \kappa | \psi_H ({\bf r}) |^2 \nonumber \\
&&~~~~ + 2 N u T \sum_{\omega_n} G_H({\bf r}, {\bf r}, \omega_n)
    + 2 N u n_H^2 ({\bf r}) \label{mchi}
\end{eqnarray}
and
\begin{equation}
(- {\bf \nabla}_{\bf r}^2 + \mathcal{V}_H ({\bf r})) n_H ({\bf r})
= 0, \label{mzeta}
\end{equation}
where $G_H$ is given by (\ref{g}). Comparing (\ref{mzeta}) and
(\ref{g}) it is easy to see that the spectrum of $G_H$, as defined
in (\ref{geigen}) has one mode with $E_{0} ({\bf k}) \rightarrow
0$ as ${\bf k} \rightarrow 0$; this is, of course, the Goldstone
spin wave mode associated with the spontaneous SDW condensate.

The equation which determined the superconducting order parameter,
$\psi_H ({\bf r})$, was (\ref{gl}), and this is now replaced by
\begin{eqnarray}
\Bigl\{-1 &+& \frac{N \kappa}{2\Upsilon} \Bigl[ T \sum_{\omega_n}
G_H({\bf r}, {\bf r}, \omega_n)
+ 2 N u n_H^2 ({\bf r})   \Bigr] \nonumber \\
&+& |\psi_H ({\bf r})|^2 - ({\bf \nabla}_{\bf r} - i {\bf A})^2
\Bigr\} \psi_H ({\bf r}) = 0. \label{mgl}
\end{eqnarray}

\subsection{Renormalization of parameters}
\label{sec:renorm1}

Now we proceed as in Section~\ref{sec:renorm} to remove all
dependence of (\ref{mchi}), (\ref{mzeta}), and (\ref{mgl}) on the
short-distance cutoff.

First consider the case when $T=0$, $H=0$ and $s=s_c$, where
(\ref{chi0}) and (\ref{gl0}) hold. Now after we shift parameter
$\alpha$ in as in (\ref{alpha}), the Ginzburg-Landau equation
(\ref{mgl}) is modified to
\begin{eqnarray}
\Bigl\{&-& 1 + \frac{N \kappa}{2\Upsilon} \Bigl[ T \sum_{\omega_n}
G_H({\bf r}, {\bf r}, \omega_n) + 2 N u n_H^2 ({\bf r})
\nonumber \\
&-& \int \frac{d \omega}{2 \pi} \int \frac{d^2 k}{4 \pi^2}
\frac{1}{\omega^2 + k^2} \Bigr]
+ |\psi_H ({\bf r})|^2 \nonumber \\
&-& ({\bf \nabla}_{\bf r} - i {\bf A})^2 \Bigr\} \psi_H ({\bf r})
= 0. \label{mgl1}
\end{eqnarray}
Next, subtracting (\ref{chi0}) from (\ref{mchi}) while noticing
that $\psi_{0c}$ is already renormalized to unity, we have
\begin{eqnarray}
&&\mathcal{V}_H ({\bf r}) = s - s_c + \kappa (| \psi_H ({\bf r})
|^2 - 1) + 2 N u
n_H^2 ({\bf r}) \nonumber \\
&&\!\!\!\!\!\!+ 2 N u \Bigl[ T \sum_{\omega_n} G_H({\bf r}, {\bf
r}, \omega_n) - \int \frac{d \omega d^2 k}{8 \pi^3}
\frac{1}{\omega^2 + k^2} \Bigr]. \label{mchir}
\end{eqnarray}
{}From (\ref{mchir}) and (\ref{mgl1}) we have
\begin{eqnarray}
&&\Biggl[\left(1-\frac{\kappa^2}{4u\Upsilon}\right)(|\psi_H({\bf
r})|^2 - 1) + \frac{\kappa}{4u\Upsilon} ( \mathcal{V}_H({\bf r}) -
s + s_c) \nonumber \\
&&~~~~~~~~~~~~- ({\bf \nabla}_{\bf r} - i {\bf A})^2 \Biggr]
\psi_H ({\bf r}) = 0. \label{mglr}
\end{eqnarray}
The final set of equations for the properties of the SC+SDW phase
are (\ref{mchir}), (\ref{mzeta}) and (\ref{mglr}); these are to be
solved for the unknowns $\mathcal{V}_H ({\bf r})$, $\psi_H ({\bf
r})$ and $n_H ({\bf r})$. We describe the numerical solution in
Appendix~\ref{numsdw}.

\subsection{Phase boundaries}
\label{pbsdw}

We have already determined the positions of several phase
boundaries in Fig~\ref{figpd} in Section~\ref{sec:pb}, and it
remains only to determine BM. First notice that at the transition
into a non-superconducting phase, the order parameter $\psi({\bf
r})$ vanishes, and thus $\mathcal{V}_H({\bf r})$ and $\zeta_H({\bf
r})$ are spatially uniform. So from (\ref{mzeta}) we have
$\mathcal{V}_H = 0$. Plugging this into (\ref{mgl1}) we obtain the
position of the phase boundary BM specified in (\ref{BM}).

\subsection{SDW order parameter}
\label{sdwop}

The presence of a static spin condensate implies that the dynamic
spin susceptibility contains sharp Bragg peaks at zero frequency
and at wavevectors separated from the SDW ordering wavevectors by
the reciprocal lattice vectors of the vortex lattice as suggested
by Zhang\cite{so5} and discussed by us in Ref.~\onlinecite{prl};
these are in addition to the dynamic spectra specified in
(\ref{sus}). This means that the dynamic structure factor
$S_{\varphi} ({\bf k}, \omega)$ (which is related to the
susceptibility $\chi_{\varphi} ({\bf k}, \omega)$ in (\ref{sus})
by the usual fluctuation-dissipation theorem) has the
contributions
\begin{equation}
S_{\varphi} ({\bf k}, \omega) = (2 \pi) \delta (\omega) \sum_{\bf
G} |f_{\bf G}|^2 (2 \pi)^2 \delta({\bf k} - {\bf G}) \label{sko}
\end{equation}
where ${\bf G}$ extends over the reciprocal lattice vectors of the
vortex lattice, and
\begin{equation}
f_{\bf G} = \frac{\sqrt{N}}{A_{\cal U}} \int_{{\cal U}} d^2 r
e^{-i {\bf G} \cdot {\bf r}} n_H ({\bf r}) \label{fK}
\end{equation}
where the spatial integral is over ${\cal U}$ the unit cell of the
vortex lattice with area $A_{\cal U}$. Note that, by
(\ref{totalsus}), the physical momentum is related to ${\bf k}$ in
(\ref{sko}) by shifts from the SDW ordering wavevectors ${\bf
K}_{sx}$ and ${\bf K}_{sy}$.

\begin{figure}
\centerline{\includegraphics[width=3in]{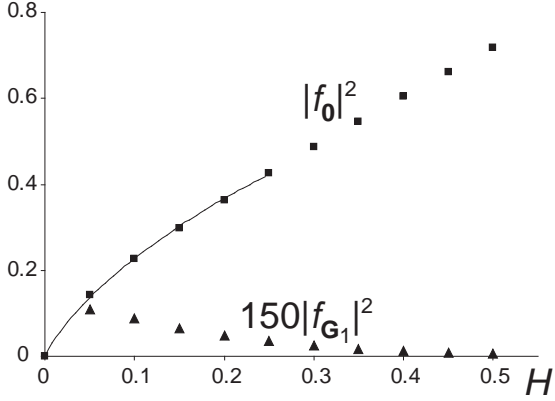}}
\caption{Bragg scattering intensity, $|f_{\bf G}|^2$, as a
function of $H$ at $s - s_c=0$. Shown are the values at ${\bf
G}=0$ (squares) and at ${\bf G}={\bf G}_1$ (triangles), which is
the smallest non-zero reciprocal lattice vector of the vortex
lattice. Note that the intensities at ${\bf G}={\bf G}_1$ have
been magnified by a factor of 150 to make them visible on this
plot. The intensities are zero at $H=0$, because $s=s_c$ is the
quantum critical point in zero field. The line shows $0.63 H
\ln(3.61/H)$, which is the best fit to the functional form in
(\protect\ref{f0ln}).} \label{figbragg}
\end{figure}

\begin{figure}
\centerline{\includegraphics[width=3in]{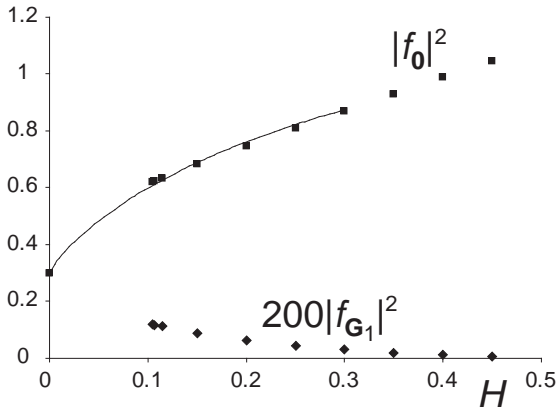}}
\caption{As in Fig~\protect\ref{figbragg1} but for $s - s_c=-0.3$,
showing $|f_{\bf G}|^2$ at ${\bf G}=0$ (squares) and at ${\bf
G}={\bf G}_1$ (diamonds) . Unlike Fig~\protect\ref{figbragg1}, the
intensity $|f_{{\bf 0}}|^2$ is non-zero even at zero field. The
intensities at ${\bf G}={\bf G}_1$ have now been magnified by a
factor of 200. The line is $0.3+0.98 H \ln(2.12/H)$, which is the
best fit to the functional form in (\protect\ref{f0ln}) }
\label{figbragg1}
\end{figure}

Figs~\ref{figbragg} and~\ref{figbragg1} show plots of the Bragg
scattering intensity $|f_{\bf G}|^2$, for the two smallest values
of ${\bf G}$ and two values of $s-s_c$, as a function of $H$.

As argued in Ref.~\onlinecite{prl}, the correspondence
(\ref{reff}) implies that the scattering intensity at zero
wavevector, $|f_{{\bf 0}}|^2$ should increase with as
\begin{equation}
\langle |f_{{\bf 0}}|^2 \rangle \propto H \ln (1/H). \label{f0ln}
\end{equation}
The fits to this functional form in Fig~\ref{figbragg} show that
this works quite well. Notice also that the intensity at the first
non-zero reciprocal lattice vector, ${\bf G}_1$, is quite small,
and that it {\em decreases} with increasing $H$. This suggests
that observation of this satellite peak is best performed at as
small a field as possible--of course, $H$ should be large enough
so that $|{\bf G}_1|$ is large enough to be outside the resolution
window of the peak at ${\bf G} = 0$. It is interesting to observe
here that we can view the Bragg peak at ${\bf G}_1$ as arising
from condensation at the non-zero ${\bf k}$ minimum in
Fig~\ref{figr1b035} of the dispersion of the exciton in the SC
phase.

\begin{figure}
\centerline{\includegraphics[width=3.2in]{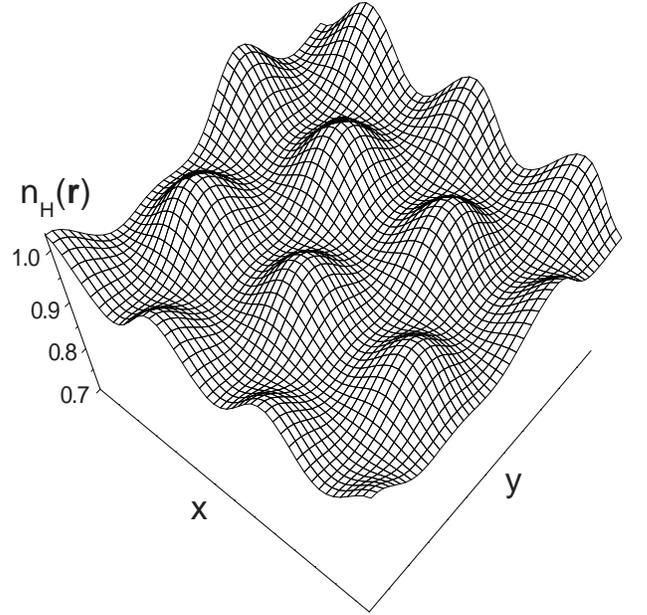}}
\caption{Spatial form of the SDW order parameter $n_H({\bf r})$ in
the SC+SDW phase at $s-s_c=-0.3$, $H=0.35$ (point $b$ in
Fig~\ref{figAM}) over vortex lattice shown in
Fig~\protect\ref{figcell}. Notice that the vertical scale extends
over a rather short range, and the modulation in $n_H ({\bf r})$
is quite small relative to the uniform component.} \label{figzeta}
\end{figure}

\begin{figure}
\centerline{\includegraphics[width=3.2in]{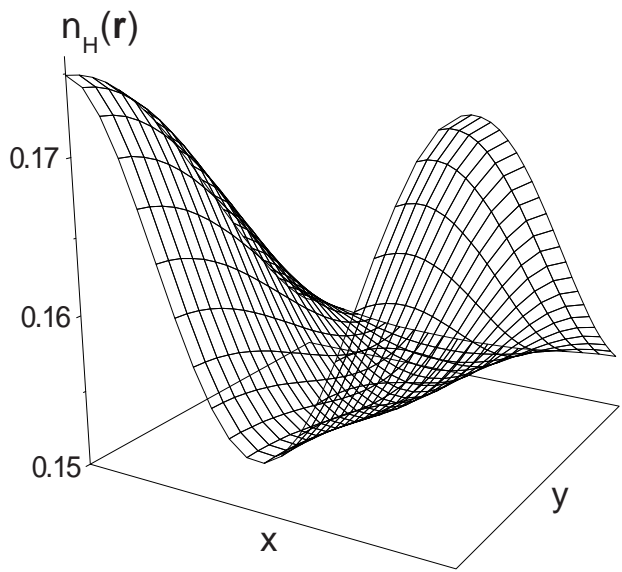}}
\caption{As in Fig~\protect\ref{figzeta}; in the SC+SDW phase at
$s-s_c=0.5$, $H=0.45$ (point $i$ in Fig~\ref{figAM}) over a single
vortex lattice unit cell shown in Fig~\protect\ref{figcell}.}
\label{figzetan}
\end{figure}

\begin{figure}
\centerline{\includegraphics[width=2.6in]{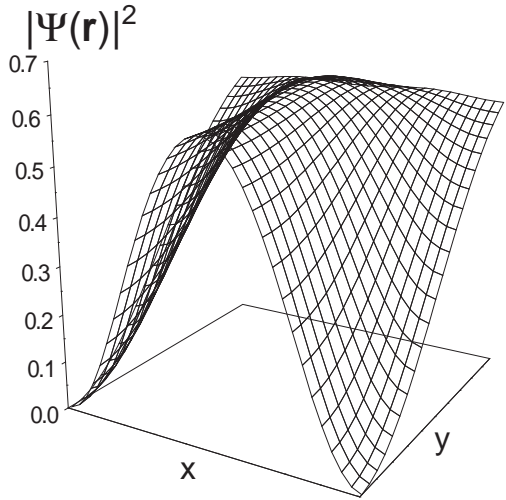}}
\caption{Spatial dependence of the modulus of the superconducting
order parameter $|\psi_H ({\bf r})|^2$ plotted on the rectangular
half unit cell of the vortex lattice indicated by
Fig~\protect\ref{figcell}. As in Fig~\ref{figzeta}, this result is
for $s=s_c$, and $H=0.35$ (point $b$ in Fig~\ref{figAM}).}
\label{figomega2}
\end{figure}

For completeness, we also show the real space form of the
condensate $n_H ({\bf r})$ in Figs~\ref{figzeta}
and~\ref{figzetan} for two points in the SC+SDW phase.
The spatial form of the modulus of the superconducting order
parameter for the first set of parameters is shown in
Fig~\ref{figomega2}.
This last figure is the analog of Fig~\ref{figomega} which was for
the SC phase.

\subsection{Dynamic spin susceptibility}
\label{dysdw}

Finally, we follow the presentation in Section~\ref{dyspin} and
discuss the dynamic spin spectrum in the SC+SDW phase. The
non-zero $\omega$ spectral densities presented here appear along
with the $\omega=0$ contributions in (\ref{sko}). We will restrict
our attention to the susceptibility {\em transverse} to the
ordering direction: this is given by the fluctuations of the last
$N-1$ components in (\ref{transverse}), which are in turn related
to the Green's function $G_H$ in (\ref{mchi}) and (\ref{mgl}). So
the transverse dynamic spin susceptibility is given by a formula
analogous to (\ref{sus}). As before, we present the results by
broadening the delta functions to sharp Lorentzians.

\begin{figure}
\centerline{\includegraphics[width=3.4in]{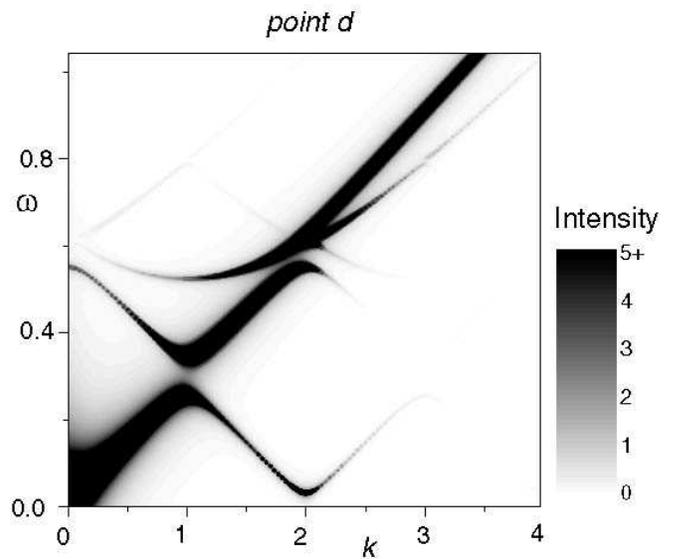}}
\caption{As in Fig~\protect\ref{figr1b01}, but for the transverse
susceptibility in the SC+SDW phase. The parameter values are
$s=s_c$ and $H=0.05$ (point $d$ in Fig~\ref{figAM}).}
\label{figmr0b05}
\end{figure}

\begin{figure}
\centerline{\includegraphics[width=3.4in]{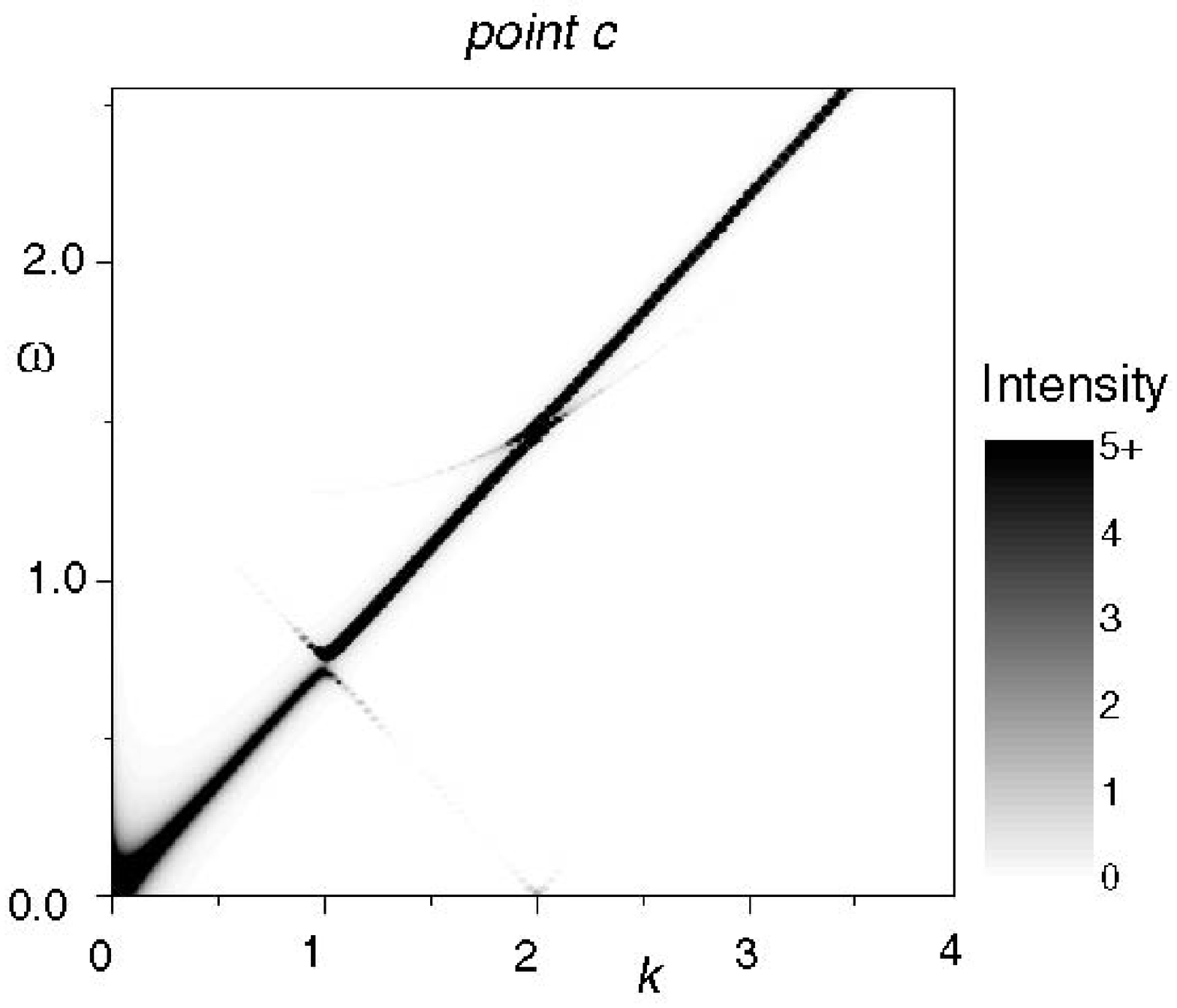}}
\caption{As in Fig~\protect\ref{figmr0b05} but for the values
$s=s_c$ and $H=0.3$ (point $c$ in Fig~\ref{figAM}). }
\label{figmr0b30}
\end{figure}

\begin{figure}
\centerline{\includegraphics[width=3.4in]{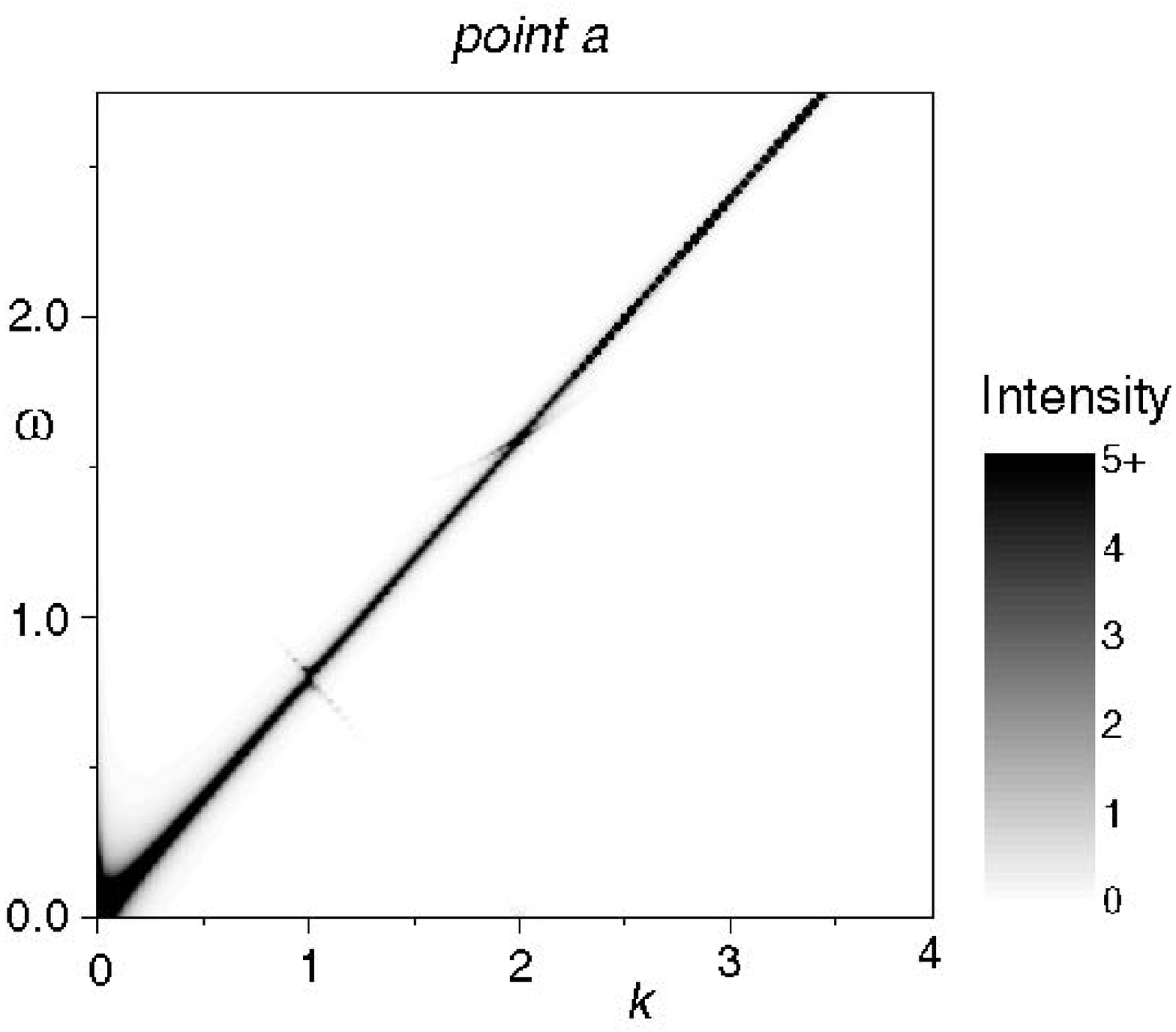}}
\caption{As in Fig~\protect\ref{figmr0b05} but for the values
$s-s_c=-0.3$ and $H=0.35$ (point $a$ in Fig~\ref{figAM}).}
\label{figmrm30b30}
\end{figure}

\begin{figure}
\centerline{\includegraphics[width=3.4in]{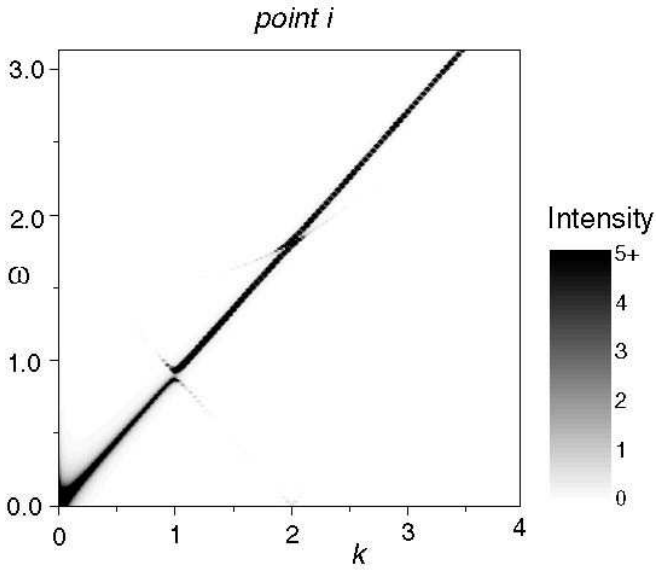}}
\caption{As in Fig~\protect\ref{figmr0b05} but for the values
$s-s_c=0.5$ and $H=0.45$ (point $i$ in Fig~\ref{figAM}).}
\label{figmr5b45}
\end{figure}

Our results for the spectral densities are shown in
Figs~\ref{figmr0b05}, \ref{figmr0b30}, \ref{figmrm30b30}
and~\ref{figmr5b45} for a series of values of $s-s_c$ and $H$ in
the SC+SDW phase.
Note first that there is always a gapless spin-wave mode. In
addition there are features arising from scattering off the vortex
lattice: these are strongest in the vicinity of the quantum
critical point A at zero field.

\section{Other phases in zero magnetic field}
\label{newphases}

The next two sections involve a slight detour from the main flow
of the paper. This section we will examine phases and phase
transitions associated with composites or `fractions' of the
primary order parameters $\Phi_{x,y\alpha}$ and $\phi_{x,y}$. This
is done mainly for completeness. Readers not interested in this
detour may skip ahead to next section without loss of continuity.

\subsection{Phases with nematic order}
\label{sec:nem}

In Section~\ref{sec:op} we argued that a generalized
non-two-sublattice spin density wave order may be associated with
a charge density wave. Another interesting possibility is that of
{\em spin nematic} order, which has been previously discussed in
Refs.~\onlinecite{andreev,chubukov,Gorkov,Demler01}. If the CDW
order parameter may be understood as a spin zero combination of
two $S_\alpha$ operators ($\delta \rho ({\bf r}, \tau) \sim
S_{\alpha}^2 ({\bf r}, \tau)$) , then the spin nematic order
parameter $Q_{\alpha\beta}({\bf r}, \tau)$ corresponds to their
spin two combination
\begin{equation}
Q_{\alpha\beta} ({\bf r}, \tau) \sim S_{\alpha} ({\bf r}, \tau)
S_{\beta} ({\bf r}, \tau) - \frac{\delta_{\alpha\beta}}{3}
S_{\alpha}^2 ({\bf r}, \tau). \label{qn1}
\end{equation}

We pause briefly to also mention here an ``Ising nematic'' order
which has also been considered recently \cite{kfe}. This order
resides in real space associated with the lattice, and is distinct
from the spin-space nematic order we are considering here. Order
parameters with the Ising nematic order are $|\Phi_{x\alpha}|^2 -
|\Phi_{y\alpha}|^2$ and $|\phi_{x}|^2 - |\phi_{y}|^2$, and these
clearly measure a spontaneous choice between the $x$ and $y$
directions of the lattice. Our effective actions for
$\Phi_{x,y\alpha}$ and $\phi_{x,y}$ are rich enough to also allow
such orders.

Returning to our discussion of spin nematic order in (\ref{qn1}),
we see that spin nematic order parameters that are consistent with
the SDW order in (\ref{e1}) may be at wavevectors $(0,0)$ and
${\bf K}_{cx,y}$.
\begin{eqnarray}
&& Q_{\alpha\beta}({\bf r},\tau)= Q_{0\alpha\beta}({\bf r},\tau)
\nonumber \\&&+ \mbox{Re}\,[\, Q_{x\alpha\beta}(r,\tau) e^{i {\bf
K}_{cx}{\bf r}} + Q_{y\alpha\beta}(r,\tau) e^{i {\bf K}_{cy}{\bf
r}}\,] \label{Q}
\end{eqnarray}
It is natural to call $Q_{0\alpha\beta}$ a uniform spin nematic
order parameter, and $Q_{x,y\alpha\beta}$ a spin nematic density
wave (SNDW). Both order parameters are symmetric
($Q_{i\alpha\beta}=Q_{i\beta\alpha}$), but the uniform spin
nematic $Q_{0\alpha\beta}$ must be real, and the spin nematic
density wave $Q_{x,y\alpha\beta}$ may be complex. The uniform spin
nematic couples to the SDW order parameters $\Phi_{x,y\alpha}$ as
\begin{eqnarray}
\mathcal{S}_{Q_0,\Phi}= &-&\lambda_1 \sum_{i=x,y}\, \int d^2 r\,
d\tau \, Q_{0\alpha\beta} \, \Bigl( \Phi^\dagger_{i\alpha}
\Phi_{i\beta} + \Phi_{i\alpha} \Phi^\dagger_{i\beta} \nonumber \\
&~&~~~~~~~~~~~~~-\frac{2}{3} \,\delta_{\alpha\beta}\,
|\Phi_{i\delta}|^2 \Bigr)
\end{eqnarray}
The spin nematic density wave $Q_{x\alpha\beta}({\bf r},\tau)$
couples to  $\Phi_{x\alpha}$ via
\begin{eqnarray}
\mathcal{S}_{Q_x,\Phi_x}= &-&\lambda_2 \int d^2 r d\tau \, [\,\,
Q^\dagger_{x\alpha\beta} \, \big(\, \Phi_{x\alpha} \Phi_{x\beta} -
\frac{1}{3} \,\delta_{\alpha\beta}\, \Phi_{x\delta}^2 \,\big)
\nonumber \\
&+& \mbox{c.c.} \,\,]
\end{eqnarray}
with a similar coupling between $Q_{y\alpha\beta}({\bf r},\tau)$
and $\Phi_{y\alpha}$.

The effective action for the spin nematic order parameters may be
written from the analysis of the symmetries of (\ref{Q}). The
interplay of the spin nematic and spin density wave orders may
produce an extremely rich phase diagram. We will not attempt to
explore its full richness, but restrict ourselves to the
discussion of some simple illustrative examples. It is also worth
pointing out that the appearance of the spin nematic order (either
uniform or SNDW) does not give rise to the additional Bragg peak
at zero energy, but produces a difference in the scattering cross
sections for different neutron polarizations.

\subsubsection{Uniform spin nematic}

To write the effective action for the uniform spin nematic
$Q_{0\alpha\beta}$ we can give essentially the same arguments as
in deriving the Landau free energy for the classical nematics (see
e.g. Ref.\onlinecite{deGennes})
\begin{eqnarray}
\mathcal{S}_{Q_0} &=& \int d^2 r d\tau \, \Bigl[\,\, (\,
\partial_\tau Q_{0\alpha\beta}\,)\,( \partial_\tau Q_{0\beta
\alpha}\,) \nonumber \\ &+& v^2_Q (\, \,{\vec{\nabla}}
Q_{0\alpha\beta}\,)\, (\, {\vec{\nabla}} Q_{0\beta \alpha} \,) +
\frac{1}{2}\, A\, Q_{0\alpha\beta} Q_{0\beta \alpha} \nonumber \\
&+&\frac{1}{3}\, B \, Q_{0\alpha\beta} Q_{0\beta \gamma} Q_{0
\gamma \alpha} +\frac{1}{4}\, C_1\, (\,  Q_{0\alpha\beta}
Q_{0\beta \alpha} \,)^2 \nonumber \\
&+& \frac{1}{4}\, C_2 \, Q_{0\alpha\beta} Q_{0\beta \gamma}
Q_{0\gamma\delta} Q_{0\delta\alpha}  \,\, \Bigr] \label{Q0-action}
\end{eqnarray}

By an appropriate spin rotation, the uniform spin nematic order
parameter may always be brought into the diagonal form (this
follows from the fact that it is a real and symmetric matrix)
\begin{eqnarray}
{Q}_{0\alpha\beta} =  \left(\begin{array}{ccc}
-\frac{1}{2}(q+\eta) & 0 & 0 \\
0 & -\frac{1}{2}(q-\eta) & 0 \\
0 &  0 & q
\end{array}\right)
\label{diagonal_nematic}
\end{eqnarray}
When $\langle q\rangle \neq 0$ but $\langle \eta \rangle =0$ we
have a uniaxial spin nematic, and when both expectation values are
finite we have a biaxial spin nematic.

\begin{figure}
\centerline{\includegraphics[width=3in]{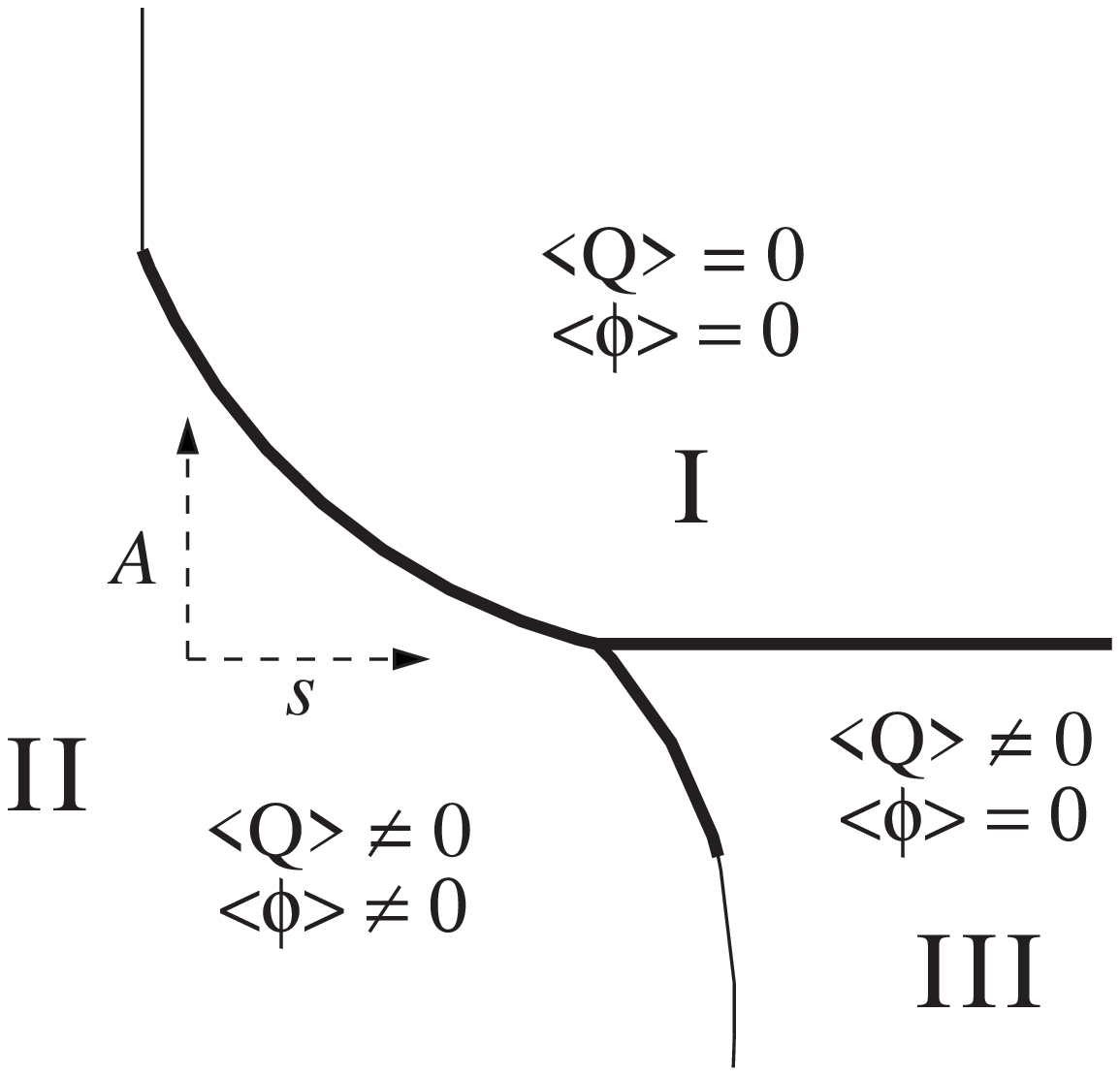}}
\caption{Mean field zero temperature phase diagram of the model
$\mathcal{S}_\Phi+\mathcal{S}_{Q_0} +\mathcal{S}_{Q_0,\Phi}$ in
zero magnetic field for the case $u_2<0$. } \label{uni_nem}
\end{figure}

\begin{figure}
\centerline{\includegraphics[width=3in]{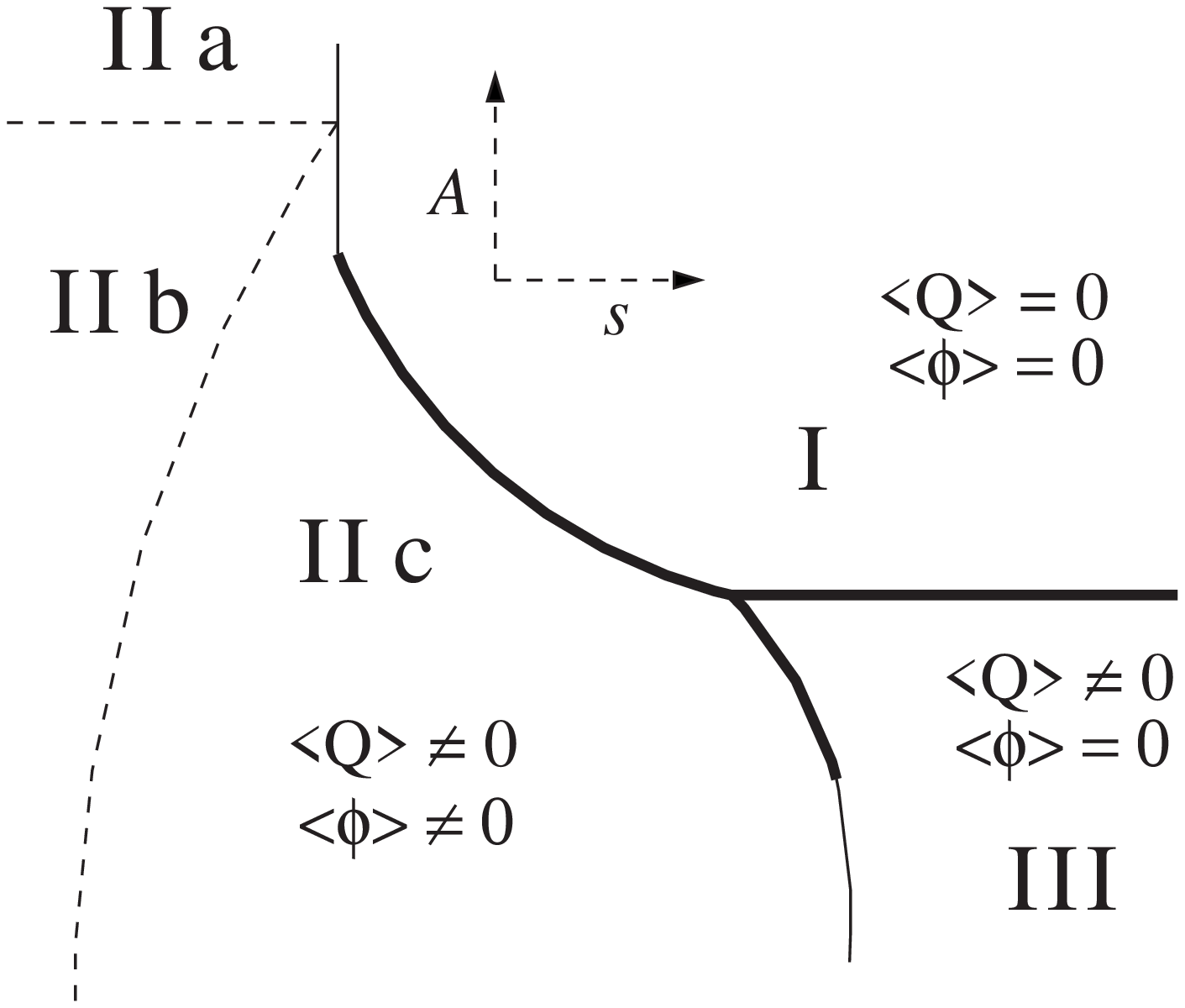}}
\caption{Mean field zero temperature phase diagram of the model
$\mathcal{S}_\Phi+\mathcal{S}_{Q_0} +\mathcal{S}_{Q_0,\Phi}$ in
zero magnetic field for the case $u_2>0$. } \label{uni_ndw}
\end{figure}

Let us start by considering the interplay of the uniform spin
nematic with the collinear SDW (for simplicity we only consider
one of the SDW orders, say $\Phi_{x\alpha}$). A schematic
mean-field phase diagram at $T=0$ for
$\mathcal{S}_\Phi+\mathcal{S}_{Q_0} +\mathcal{S}_{Q_0,\Phi}$ with
$B \lambda_1 <0$ and $u_2<0$ is shown on Fig.~\ref{uni_nem}. Thick
lines correspond to the first-order transitions, and thin lines
correspond to the second order transitions. Phase I (SC) has no
magnetic order of any kind; phase II (SC+SDW) has commensurate SDW
order, which is accompanied by a uniaxial spin nematic order;
phase III (SC+UN) has a uniaxial spin nematic order.
For $B \lambda_1 > 0$ and $u_2<0$ the phase diagram qualitatively
remains the same, however the phase II has a finite expectation
value of both $q$ and $\eta$ in (\ref{diagonal_nematic}), so it
has an SDW order accompanied by the biaxial spin nematic order. A
schematic phase diagram in the case $u_2>0$ is shown on
Fig.~\ref{uni_ndw}. The phase II may now be a circular spiral SDW
(IIa), an elliptic spiral SDW (IIb), and a collinear SDW (IIc).

\subsubsection{Spin nematic density wave}

For the spin nematic density wave the third order terms are
prohibited by symmetry: they carry oscillating factors $e^{\pm i
{\bf K}_{cx,y}{\bf r}}$, and vanish after integrating over space
in the long wavelength limit. Hence,
\begin{eqnarray}
\mathcal{S}_{Q_x} &=& \int d^2 r d\tau \, \Bigl[\,\, (\,
\partial_\tau Q^\dagger_{x\alpha\beta}\,)\,( \partial_\tau
Q_{x\beta \alpha}\,) \nonumber \\ &+& \tilde{v}^2_Q (\,
\,{\vec{\nabla}} Q^\dagger_{x\alpha\beta}\,)\, (\, {\vec{\nabla}}
Q_{x\beta \alpha} \,) + \frac{1}{2}\, \tilde{A}\,
Q^\dagger_{x\alpha\beta} Q_{x\beta \alpha} \nonumber \\
&+&\frac{1}{4}\, \tilde{C}_1\, (\,  Q^\dagger_{x\alpha\beta}
Q_{x\beta \alpha} \,)^2 +\frac{1}{4}\, \tilde{C}_2 \,
Q^\dagger_{x\alpha\beta} Q_{x\beta \gamma}
Q^\dagger_{x\gamma\delta} Q_{x\delta\alpha} \,\, \nonumber \\
&+&\frac{1}{4}\, \tilde{C}_3 \,  Q^\dagger_{x\alpha\beta}
Q^\dagger_{x\beta \gamma} Q_{x\gamma\delta} Q_{x\delta\alpha} \,\,
\Bigr] \label{Sqx}
\end{eqnarray}
and there is a similar action $\mathcal{S}_{Q_y}$.

The order parameter for the spin nematic density wave can be
conveniently written using five complex numbers (see also
Ref.~\onlinecite{Gramsbergen})
\begin{eqnarray}
{Q}_{x\alpha\beta} = \left(
\begin{array}{ccc}
-\displaystyle\frac{\psi_{x1}}{\sqrt{3}}-\psi_{x2} & \psi_{x3} & \psi_{x4} \\
\psi_{x3} & -\displaystyle\frac{\psi_{x1}}{\sqrt{3}} + \psi_{x2} & \psi_{x5} \\
\psi_{x4} & \psi_{x5} & \displaystyle\frac{2\psi_{x1}}{\sqrt{3}}
\end{array} \right)
\label{diagonal_NDW}
\end{eqnarray}
with normalization condition $\sum_{a=1,...,5} |\psi_{xa}|^2 =1$.
This representation makes obvious the connection between the order
parameter for the spin nematic density wave and condensates of
spin 2 particles, for which Ciobanu {\em et.al.}\cite{Ho99} argued
that there exist three distinct phases (not related to each other
by spin rotations), depending on the parameters $\tilde{C}_1$,
$\tilde{C}_2$, and $\tilde{C}_3$.

The phase diagrams of the spin nematic density wave order vs the
SDW order is similar to the case of uniform spin nematic
(Figs.~\ref{uni_nem} and~\ref{uni_ndw}) with the main difference
that the phase boundary between I and III is now second order.

\subsection{Exciton Fractionalization}
\label{sec:frac}

Before concluding the section on the phases in zero field we would
like to point out another interesting possibility for the system
described by the generalizations of ${\cal S}_{\Phi}$. Consider
this model in the regime where the spiral fluctuations are
strongly suppressed, so we need to consider the collinear SDW
order only; this happens in (\ref{SPhi}) for $u_2<0$ and with
$|u_2|$ large. For simplicity we restrict our discussion to a SDW
at wavevector ${\bf K}_{sy}$, $\Phi_{y\alpha}$. As discussed in
the Section~\ref{sec:op}, the collinear SDW can  be written in the
form in (\ref{q1}), which we reproduce here for completeness:
\begin{equation}
\Phi_{y\alpha}({\bf r},\tau)=e^{i \theta({\bf r},\tau)}
n_{\alpha}({\bf r},\tau). \label{ptn}
\end{equation}
We also noted below (\ref{q1}) that such a separation of the
physical order parameter $\Phi_{y\alpha}$ into the phase $\theta$
and the real vector $n_{\alpha}$ has an implicit ambiguity as we
can simultaneously change the sign of both without altering
$\Phi_{y\alpha}$. Formally this means that, for incommensurate
${\bf K}_{sy}$, the order parameter $\Phi_{y \alpha}$ belongs to
the space $(S_2 \times S_1)/Z_2$. For commensurate ${\bf K}_{sy} =
2\pi p'/(p a)$, where $p'$, $p$ are relatively prime integers,
higher order terms not contained in (\ref{SPhi}) (but mentioned
below it) imply $\theta$ prefers a discrete set of
values\cite{sns,pphmf} and the space is restricted to $(S_2 \times
Z_p)/Z_2$. Also, if full SU(2) spin rotation symmetry is absent,
and the spins have an easy-plane restriction, then the first $S_2$
factor changes to $S_1$.

The $Z_2$ quotient in the order parameter space can be explicitly
implemented as an Ising gauge symmetry, and it puts important
constraints on the effective low energy theory. The lattice model
consistent with such symmetry has the form
\begin{eqnarray}
\mathcal{S}_{I} = \sum_{\langle ij\rangle} J^s \sigma_{ij}
n_{i\alpha} n_{j\alpha} +\sum_{\langle ij \rangle} J^c \sigma_{ij}
\cos (\theta_i - \theta_j) \label{lattice-action}
\end{eqnarray}
where $i$ and $j$ are sites on the space-imaginary time lattice,
the sum over $\langle ij \rangle$ extends over nearest neighbor
links of this lattice, $J^s$ and $J^c$ are couplings imposing the
propagation of SDW and CDW order respectively,
$n_{i\alpha}=n_{\alpha}({\bf r}_i,\tau_i)$, $\theta_i =
\theta({\bf r}_i,\tau_i)$, and $\sigma_{ij}= \pm 1$ is an Ising
gauge field that lives on the links of the lattice. One can easily
see that the lattice action (\ref{lattice-action}) is invariant
under the $Z_2$ gauge transformation
\begin{eqnarray}
n_{i\alpha} &\rightarrow& \sigma_i  n_{i\alpha} \nonumber\\
\theta_i   &\rightarrow& \theta_i +\frac{\pi}{2}(1-\sigma_i)
\nonumber\\
\sigma_{ij} &\rightarrow& \sigma_i \sigma_{ij}  \sigma_j
\end{eqnarray}
for $\sigma_i = \pm 1$.

Models of the kind (\ref{lattice-action}) have been discussed
earlier in various contexts
\cite{Jalabert91,Lammert93,Senthil99,Zhou,Demler01,Sachdev01}. It
was pointed out, for example, that another term allowed by
symmetry is a Maxwell term for the lattice gauge field
\begin{eqnarray}
\mathcal{S}_\sigma = -K \sum_{\square} \left[\prod_\square
\sigma_{ij}\right],
\end{eqnarray}
where the sum on $\square$ extends over the plaquettes of a 2+1
dimensional lattice. Such a term may be generated by integrating
out the high energy degrees of freedom or may be present due to
certain frustrating terms in the original microscopic Hamiltonian
\cite{Senthil99,Zhou,Sachdev01}. This term has a striking effect
on the properties of the model (\ref{lattice-action}): it gives
rise a phase in which the exciton $\Phi_{y\alpha}$ fractionalizes,
and fluctuations of $n_\alpha$ are separated from the fluctuations
of $\theta$. Loosely speaking, the SDW and the CDW fluctuations
get decoupled.

It is useful to discuss the consequence of the
confinement-deconfinement in the symmetric phase in which global
symmetries are preserved: the models of this paper are invariant
under SU(2) spin rotations, and the sliding U(1) symmetry (for
commensurate values of ${\bf K}_{sy}$, the U(1) symmetry is
reduced to a discrete $Z_p$ `clock' symmetry, but essentially
unchanged considerations apply nevertheless\cite{sns,pphmf}). The
immediate manifestation of the confinement-deconfinement
transition in such a symmetric phase is the change in the
degeneracy of the lowest energy excitations. In the confining
phase their degeneracy is 6: this 6-fold degenerate excitation
corresponds to the quanta of the exciton field $\Phi_{y \alpha}$,
which have 6 real components. In contrast, in the deconfining
phase we have separate excitations with degeneracies of 3 and 2,
corresponding to quanta of $n_{i \alpha}$ and $\theta_i$
respectively. This may be understood by noting that the unbroken
symmetry ground state of the  model
$\mathcal{S}_{I}+\mathcal{S}_{\sigma}$ is a singlet ground state
of the $SO(3)\times SO(2)$ rotors, where in the confining phase
the angular momenta of the two rotors
$(l_1,l_2)=(L_{SO(3)},L_{SO(2)})$ are bound by the constraint
$l_1+l_2=even$, but this constraint is not present in the
deconfining phase. Hence, in the confining phase the lowest
excitation has $(l_1=\pm 1,l_2=1)$, which gives the degeneracy of
6. In the deconfining phase we can have excitations $(l_1= \pm
1,l_2=0)$ and $(l_1=0,l_2=1)$, and these have degeneracies 3 and 2
respectively. We point out that the exact degeneracy of $l_1 = \pm
1$ states requires the absence of the Berry's phase for the the
$SO(2)$ rotor, and comes from the inversion symmetry of the
system, as was noted below (\ref{SPhi}). It is not related to the
possible particle-hole symmetry of the underlying microscopic
model.

It is worth emphasizing that the {\em exciton\/} fractionalization
discussed above has a very different physical interpretation from
that of {\em electron\/} fractionalization discussed in `RVB'
theories of doped Mott insulators\cite{Senthil99}: in the latter
there are elementary $S=1/2$ spinons which do not appear in our
fractionalized states above. Instead our exciton fractionalization
is within the sector of spin and charge density waves, and the
collective spin excitations only have integer spin.

Zaanen {\em et al.} \cite{zaanen2} have recently discussed
fractionalization in a microscopic picture of spin and charge
order in ``fluctuating stripe'' states: the physical content of
their analysis is quite similar to that of our discussion above.
However their proposed effect action does not include the CDW
phase field $\theta_i$, and we believe this is essential for a
complete description of stripe physics.

We have implicitly assumed above that the exciton
fractionalization transition occurs in a background of SC order.
However, a similar transition is also possible within a Fermi
liquid. We believe that such a quantum critical point is a
promising candidate for describing the finite temperature
crossovers in the normal state of the cuprates. Ordinary SDW/CDW
transitions in a Fermi liquid\cite{rome} have the unsatisfactory
(in our view) feature of flowing to a free field fixed point
because they are in their upper-critical dimensions. In contrast,
the exciton fractionalization transition may well remain strongly
coupled even in the presence of Fermi surface. Corresponding
speculations of fractionalization influencing finite temperature
quantum criticality were also made by Zaanen {\em et al.}
\cite{zaanen2}. Again, their and our proposals should be
distinguished from those associated with electron
fractionalization made in {\em e.g.} Ref.~\onlinecite{sf2}.

\subsection{Topological defects}
\label{sec:topo}

An alternative picture of fractionalization, and of the various
order parameters above, may be given in the language of the
topological defects of the SDW phase; the condensation of distinct
defects in the SDW state distinguishes the new phases that appear.
To simplify the presentation of this subsection we will describe
the case of an easy plane antiferromagnet, in which the vectors
$\Phi_{i\alpha} = e^{ i \theta_i} n_{i\alpha}$ may only be in the
$x$-$y$ plane, but will also state the results for systems with
full SU(2) spin rotation symmetry. A related discussion of defects
in SDW states also appears in Ref.~\onlinecite{brazov}.

\begin{figure}
\centerline{\includegraphics[width=2in]{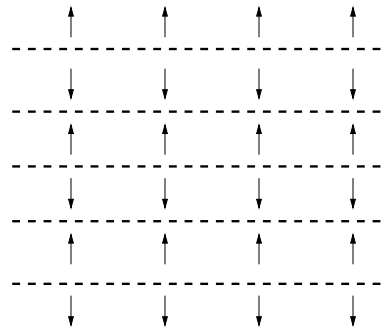}} \caption{ A
schematic picture of the non-two-sublattice collinear SDW order
and associated CDW as a periodic array of anti-phase domain walls
in N\'{e}el order at $(\pi/a,\pi/a)$. Arrows show the change of
sign of the N\'{e}el order across a hole rich domain wall. The
fields $n_{i \alpha}$ and $\theta_i$ are {\em space-independent}
in the above configuration.} \label{collinear}
\end{figure}

We start by giving a simple cartoon\cite{zaanen,zaanen2,tranquada}
of the non-two-sublattice SDW order $\Phi_{y\alpha}=\mbox{const}$
and the associated CDW in Fig.~\ref{collinear}. Hole rich stripes
(indicated by the dashed lines) act as antiphase domain walls for
the hole poor antiferromagnetic domains. The N\'{e}el order shown
by arrows changes sign when crossing such domain walls (the
N\'{e}el order should not be confused with the vector
$n_{i\alpha}$ which appears in the definition $\Phi_{i\alpha} =
e^{i \theta_i} n_{i\alpha}$; the former oscillates as shown in
Fig~\ref{collinear}, while $n_{i\alpha}$ is constant in this
configuration.).

\begin{figure}
\centerline{\includegraphics[width=2in]{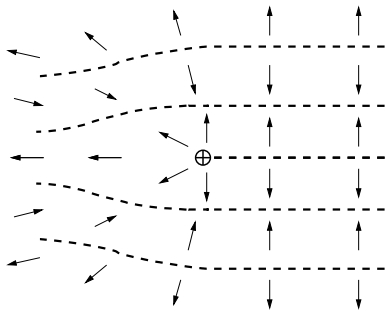}} \caption{
Elementary topological excitation of the collinear SDW phase: a
composite of $1/2$ vortex in $\theta_i$ and $\pi$-disclination in
$n_{i\alpha}$. Both $e^{i \theta_i}$ and $n_\alpha$ change sign
when going around this topological defect, but the physical order
parameter $\Phi_{y \alpha i} = e^{i \theta_i} n_{i \alpha}$ is
single valued.} \label{topo_a}
\end{figure}

\begin{figure}
\centerline{\includegraphics[width=2in]{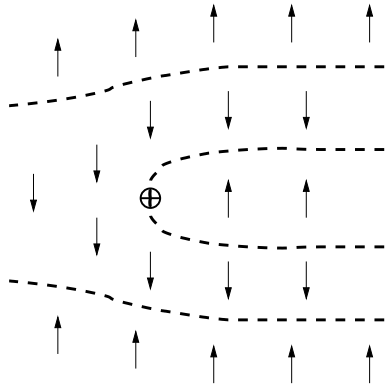}} \caption{
Elementary topological excitation of the collinear SDW phase: a
vortex in $\theta_i$. The circulation of $\theta$ is equal to $2
\pi$. } \label{topo_b}
\end{figure}

\begin{figure}
\centerline{\includegraphics[width=2in]{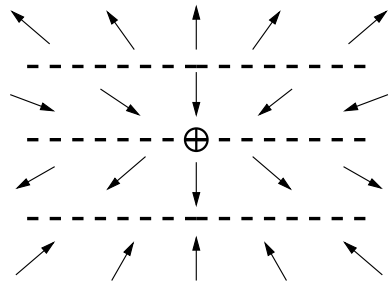}} \caption{
Elementary topological excitation of the collinear SDW phase: a
meron of  $n_{i \alpha}$. Such an object is stable only in systems
with an easy-plane symmetry. Far away from the vortex core $n_{i
\alpha}$ winds in the plane like a usual vortex. Closer to the
vortex center it may acquire an out of plane component. Systems
with full SU(2) spin rotation symmetry only have point-like,
instanton defects in spacetime: hedgehogs.} \label{topo_c}
\end{figure}

Schematic pictures of the topological defects of the collinear SDW
state are shown on Figures (\ref{topo_a}) - (\ref{topo_c}) with
crosses indicating the locations of the centers of defects (see
also Ref.~\onlinecite{zaanen2}). These defects can also be
formally classified by computing the homotopy groups; for systems
with an easy-plane spin symmetry the relevant homotopy
group\cite{homotopy} is $\pi_1 ((S_1\times S_1)/Z_2)=Z \times Z$,
while for full SU(2) spin symmetry it is $\pi_1 ((S_2\times
S_1)/Z_2)=Z$. These mathematical statements actually obscure some
of the physical content, as will become clear from our discussion
below.

We first discuss the physical content of the defect classification
for the easy-plane case. Consider the most elementary topological
defect: this is a composite of a $1/2$ vortex for the phase
$\theta_i$ and a $\pi$-disclination (i.e. $1/2$ a meron) for the
vector $n_{i\alpha}$ (see Fig~\ref{topo_a}); this defect is also a
central actor in the discussion of Zaanen {\em et
al.}\cite{zaanen2}. When circling around such a defect both $e^{ i
\theta}$ and $n_{i\alpha}$ change sign, however the physical order
parameter $\Phi_{i\alpha} = e^{ i \theta_i} n_{i\alpha}$ is
uniquely defined. Given the circulations in $\theta_i$ and
$n_{i,\alpha}$, we label this defect $(1/2,1/2)$. Actually, we can
make four such elementary defects by changing the signs of the
circulation of $\theta_i$ and $\pi$-disclination and taking all of
such combinations: we label these as $(\pm 1/2, \pm 1/2)$ in an
obvious manner. Pairs of such elementary defects may be combined
to give a full vortex for $\theta$, which is trivial in the
$n_{i\alpha}$ sector (see Fig~\ref{topo_b}; this is the defect
$(1,0)$) and a meron for the $n_{i\alpha}$, that it is trivial in
the $\theta_i$ sector (see Fig~\ref{topo_c}; this is the defect
$(0,1)$).
Continuing in this manner, we see that all defects are labelled
$(m_1/2, m_2/2)$ with $m_1$, $m_2$ integers such that $m_1 + m_2$
is even. These labels lie on the analog of a FCC lattice in two
dimensions; this is equivalent to a square lattice after a
rotation by 45 degrees, and hence the homotopy group is $Z \times
Z$. This mathematical statement hides the fact that there is a
fundamental physical difference between the $(\pm 1/2,\pm 1/2)$
and the $(1,0),(0,1)$ defects, which we have discussed above.

Next we turn to the case with full SU(2) symmetry. Now the
$1/2$-meron in $n_{i \alpha}$ is actually equivalent to the $-1/2$
meron (they are both better called $\pi$ disclinations), and so
there is no distinction between $(1/2,1/2)$ and $(1/2,-1/2)$;
moreover, the $(0,1)$ defect is topologically trivial.
Consequently the spacetime line defects can simply be labeled
$m_1/2$, where $m_1$ is an integer representing the phase winding
of $\theta_i$, and hence the homotopy group is $Z$. However, there
continues to be a fundamental physical distinction between the
cases with $m_1$ odd and even. For $m_1$ odd, there must be a
corresponding $\pi$ disclination in $n_{i \alpha}$, while for
$m_1$ even the $n_{i \alpha}$ configuration can be constant. The
SU(2) case also has point defects in spacetime, the `hedgehogs',
which proliferate at spin disordering transitions.

The various phases discussed above can be easily understood using
the picture of topological defect condensation in a phase with
conventional SDW order (the SC+SDW phase):
\begin{itemize}
\item When the elementary $1/2$ vortex - $\pi$ disclination composites condense
we have a conventional (unfractionalized) disordered phase (the SC
phase).
\item When vortices and merons (or hedgehogs) condense, but
$1/2$ vortex - $\pi$ disclination composites remain gapful
excitations, we find exciton fractionalization as discussed above.
The uncondensed $1/2$ vortex - $\pi$ disclination composites
correspond to the finite energy `visons' \cite{Senthil99} of the
fractionalized phase of the $Z_2$ gauge theory.
\item When only the merons (or hedgehogs) condense we find the CDW phase
with no spin order.
\item When only the $\theta$-vortices condense we get the spin nematic phase
with no CDW order.
\end{itemize}

\section{Earlier work on SC and SDW orders}
\label{oldsdw}

As we noted earlier, this section is a detour from the main flow
of arguments in this paper. For completeness, we review earlier
theoretical and experimental work on the interplay of magnetism
and superconductivity, and discuss connections to our treatment
here. Less specialized readers may skip ahead to the conclusions
if they wish.

Early neutron scattering measurements~\cite{hayden,keimer1} of the
evolution of the magnetic order in
La$_{2-\delta}$(Sr,Ba)$_{\delta}$CuO$_4$ with $\delta$ observed
spectra which were interpreted~\cite{sy} as evidence for the
proximity of a quantum critical point at which the SDW order
vanished, and which obeyed strong hyperscaling properties. It was
proposed\cite{sy,sp,csy} that such a quantum critical point (with
dynamic exponent $z=1$) controlled physical properties over a
range of doping concentrations. Further support for such a
proposal appeared in the NMR experiments of Imai and
collaborators\cite{imai} which displayed crossovers characteristic
of the vicinity of a magnetic quantum critical point, with the
critical point at a doping concentration $\delta=\delta_c \approx
0.12$; similar evidence was presented recently by Fujiyama {\em et
al.} \cite{fujiyama} (for a review of the NMR data in this
context, see Ref.~\onlinecite{sciencereview}). The concentration
$\delta_c = 0.12$ is well within the superconducting phase, and so
the magnetic transition takes place within a background of
superconducting order {\em i.e.} there is a second order
transition between a phase with co-existing SC and SDW order (the
SC+SDW phase) and an ordinary superconductor (the SC phase). As we
noted in Section~\ref{sec:intro}, the neutron scattering
measurements of Aeppli {\em et al.} \cite{aeppli} at $\delta=0.14$
provided rather direct evidence for such a magnetic quantum
critical point. Additional evidence for microscopic co-existence
of SC and SDW orders has appeared in a number of recent
experiments\cite{katano,ylee,lake,boris,vaknin,miller,pana,sonier}.

(For completeness, we also note here the additional phases present
at very low $\delta$ which were {\em not\/} the subject of study
in this paper: in La$_{2-\delta}$Sr$_{\delta}$CuO$_4$, the
three-dimensional, two-sublattice, insulating N\'{e}el state is
present for $\delta < 0.02$, and is followed by an insulating SDW
state with its wavevector polarized along the diagonal $(1, \pm
1)$ directions\cite{wakimoto,wakimoto2}. As noted in
Section~\ref{sec:intro}, at $\delta=0.055$ there is a first-order
insulator-to-superconductor transition to the SC+SDW
phase\cite{wakimoto,wakimoto2}, which has the SDW oriented along
the $(1,0)$, $(0,1)$ directions; we discussed the properties of
this SC+SDW phase in this paper.)

A significant implication of the existence of a magnetic critical
point at $\delta=\delta_c$ is that remnants of the magnetic
excitations should be visible in the SC phase at
$\delta>\delta_c$. As originally discussed in
Ref.~\onlinecite{csy}, for such critical points there is a sharp,
gapped $S=1$ collective mode (a {\em spin exciton}) which would
appear as a `resonance' in the neutron scattering cross-section.
This resonance should appear at the SDW ordering wavevector in
(\ref{Kval}), and recent evidence for gapped, low energy spin
excitations at such a wavevector is in
Refs.~\onlinecite{bourges,mook8}. Strong resonant scattering is
also seen at the N\'{e}el order wavevector, $(\pi, \pi)$ in the SC
phase\cite{bourges,rossat1,mook,tony3}: this remains at relatively
high energies and may be viewed as a remnant of commensurate
correlations at short length scales \cite{sns}. Batista {\em et
al.} \cite{batista} have argued that the strong gapped response at
$(\pi,\pi)$ is due to the superposition of the response at the two
neighboring SDW ordering wavevectors at $\pm \vartheta$ in
(\ref{Kval}).


Another perspective on this quantum critical point, which was
useful in our analysis, was provided by Zhang's SO(5)
theory\cite{so5}. This theory goes beyond the picture of competing
SC and SDW orders in the ground state and adopts  a  stronger
assumption of a microscopic dynamic symmetry between them; this
has been supported by analytic \cite{prldz,prbdz,exactdz,annalsdz}
and numeric \cite{numerics1,numerics2} studies of a number of
models. The generator of the enlarged SO(5) symmetry is the
$\pi$-excitation, a $S=1$ collective mode with charge 2 and
momentum $(\pi,\pi)$ \cite{prldz,prbdz}. A sharp distinction
between the models with and without the $\pi$ excitation is
possible in the weak interaction limit of a generalized BCS-RPA
theory, where by going to the normal state one can check for the
existence of a sharp collective mode with the quantum numbers of
the $\pi$-particle \cite{bazaliy}. However, a clear distinction is
absent in the physically relevant strong coupling regime. For
example, in the SC phase charge is only conserved modulo 2, and
this charge 2 particle is in fact indistinguishable from the
neutral $S=1$ exciton in earlier theories \cite{csy} of the SDW
ordering transition (see also Ref~\onlinecite{tnc}). This exciton
is smoothly connected to the $S=1$ excitation in a paramagnetic
Mott insulator\cite{lt}, and an interpretation of its 'resonance
peak' as a generator of SO(5) rotations does not hold. In zero
applied magnetic field, it is possible to formulate a theory of
the exciton\cite{csy,sbv}, and the associated SDW fluctuations,
without any explicit reference to the SC order; the SC
correlations only serve to modify various couplings in the
effective action for the SDW order. What we abstract from the
analysis of Zhang \cite{so5} is the idea that the strength of the
SC order itself should be viewed as a parameter which tunes the
system across the magnetic quantum critical point: this emphasizes
a local competition between the SC and SDW orders.

We also mention that these SO(5) models naturally describe a
competition between the SC and the two sublattice SDW (N\'{e}el)
phases. Non-two sublattice SDW can then appear as a result of the
competition between phase separation and long range Coulomb
interaction \cite{so5stripe,pryadko}, across a first-order
transition from the SC to the SDW phase. In this paper we will
describe effective models for the non-two sublattice SDW directly,
across a second-order transition from the SC to SC+SDW state.

The precise nature of the interplay of SC and SDW orders in the
cuprates at {\em non-zero} temperatures in {\em three}-dimensional
models been a controversial subject (this paper dealt with
two-dimensional quantum models at $T=0$, and so the issues in this
paragraph are only peripherally related to our main discussion).
Following earlier general analyses\cite{fisherliu},
Zhang\cite{so5} pointed out four generic possibilities for the
phase diagram,  proposed the appearance of exact SO(5) symmetry in
the classical theory of a finite temperature bi-critical point
(this symmetry is actually only present in the equal-time
correlators\cite{ftso5}), and suggested that this is the situation
most likely realized for the cuprates. In the presence of such a
bi-critical point, there is a first-order transition between the
SC and SC+SDW phases at low temperatures, and the energy of the
exciton (or $\pi$ particle) remains relatively large in the SC
phase. The possibility of a critical point that is best described
as corresponding to the regime exactly on the border between the
bicritical and tetracritical behavior was suggested in
Ref.~\onlinecite{projectedso5} (the projected SO(5) models
discussed in that paper lead to such fine tuning for the effective
theories). Other critical points, including a tricritical one,
have been suggested recently by Kivelson {\em et al.} \cite{pnas}.
We have argued here, instead, that many features of the
experiments require the energy of the exciton to vanish at a
quantum critical point describing a second order transition
between the SC and SC+SDW phases; this appears when the finite
temperature multi-critical point is {\em tetra-critical} ({\em
i.e.} the four phases, SC, SDW, SC+SDW, and ``normal'' all meet at
one finite temperature point) and has strongly broken equal-time
SO(5) symmetry. We also note that Aharony\cite{amnon} has recently
shown, by an exact renormalization group analysis of fluctuations,
that the finite temperature multicritical point has a `decoupled'
structure, which does indeed exhibit tetra-critical behavior. A
finite coexistence region between the superconducting and
antiferromagnetic phases in the cuprates has been also recently
discussed by Martin {\it et.al}\cite{martin}.

We have also mentioned the recent study of Kivelson {\em et al.}
\cite{pnas} of a variety of finite temperature multicritical phase
diagrams in three dimensions involving the SC and SDW order
parameters. They pay particular attention to the possibility of a
two-phase co-existence of SC and SDW order parameters, which
should be distinguished from the homogenous SC+SDW phase we have
discussed in this paper. In the presence of a finite field in the
two-phase co-existence case, we would expect that the SC component
has a $H \ln (1/H)$ term in its free energy, while the SDW
component only has an analytic $H^2$ correction. Consequently,
with increasing $H$, the fraction of the SDW component will grow
at the expense of the SC component with an $H \ln (1/H)$
dependence.

We mention that several other proposals for the experimental
consequences of the competition between the SC and SDW orders in
the cuprates may be found in
Refs.~\onlinecite{sas1,sas2,sas3,goldbart1,goldbart2,drew}.

\section{Conclusions}
\label{sec:conc}

The primary purpose of this paper has been a description of the
phase diagram in Fig~\ref{figpd} and of the static and dynamic
properties of its low field phases. The point of departure of our
work was the existence of a second-order quantum transition
between the SC and SC+SDW phases in zero applied magnetic field
(our methods can also be extended to weakly first-order
transitions, but we did not discuss this here): we reviewed in
Section~\ref{sec:intro} the early theoretical proposals and the
experimental evidence in support of such a transition. In a
non-zero field we found that this transition extended into a line
of second-order transitions indicated by AM in Fig~\ref{figpd}.
This transition line approaches the $H=0$ axis with a vanishing
derivative, which implied that relatively small fields could have
a significant effect on the low energy spin fluctuation spectrum:
this is our qualitative explanation for the field-induced
enhancement of low energy SDW correlations observed by Lake {\em
et al.}\cite{lake}. Our analysis also showed that the critical
properties of the transition in finite field were in all cases
described by the familiar O(3) symmetric $\varphi^4$ field theory:
these have already been described in some detail in earlier
work\cite{csy,book}. This mapping to the simple O(3) continuum
field theory occurs when the spin correlation length becomes
larger than the vortex lattice spacing (as is always the case
close enough to AM), and accounts for the fact that
$\mathcal{S}_{\rm lat}$ pins the charge order fluctuations and so
reduces the order parameter to a real, 3-component vector. In
principle, the Zeeman coupling to the O(3) field theory modes
should also be included in the asymptotic critical region, but
existing work\cite{dsz,book} has shown how to do this. We believe
that experimental discovery of the critical field along the phase
boundary AM is an exciting possibility for future investigations.
Such a study should begin with a sample with its $s$ value
slightly larger than $s_c$; application of a field should then
allow tuning of the system across the quantum critical behavior
associated with the AM phase boundary. The precise experimental
control available over the value of $H$ should allow unprecedented
access to an interesting, interacting quantum critical point in
two dimensions. In the following subsection we discuss a number of
very recent experimental studies, and compare them to our results
to the extent possible: we also mention proposals for future
experiments.

\subsection{Implications for experiments}
\label{newexp}

So far, the most direct connection of our results with experiments
is provided by neutron scattering measurements of the field
dependence of the ordered moment in the SC+SDW phase. Two such
experiments have been performed\cite{boris,lake2} in different but
related compounds, and both show a reasonable fit to the
predicted\cite{prl} $H\ln(1/H)$ dependence. The experiment of
Khaykovich {\em et al.}\cite{boris} appears to be in a parameter
regime similar to that of Fig~\ref{figbragg1}: there is an
appreciable ordered moment at zero field, and the elastic
scattering intensity roughly doubles in a field about a quarter of
$H_{c2}$. This is an important consistency check on our entire
approach, as all numerical parameters in our computation had
physically reasonable values. As is clear from
Fig~\ref{figbragg1}, the intensity of the satellite peaks
associated with the reciprocal lattice vectors of the vortex
lattice is quite small for these parameters: this explains why
such a satellite peak was not seen in the experiments even though
they had the requisite wavevector resolution. The experiments of
Lake {\em et al.} \cite{lake2} are in a regime similar to that of
Fig~\ref{figbragg}: they had quite a small moment at zero field,
but this grew rapidly with field with a clear $H\ln(1/H)$
dependence. Again, as Fig~\ref{figbragg} shows, the satellite
vortex lattice peaks have a very small intensity, and this is
presumably why they were not observed. This experimental sample
appears to be rather close to $s=s_c$, and we hope that a future
experiment will move just past $s_c$ and study the transition
across the AM phase boundary in Fig~\ref{figpd}.

Our theoretical computations also suggest an approach by which the
vortex (reciprocal) lattice may be detected in the spin
fluctuation spectrum. While its influence on the elastic Bragg
peaks\cite{so5,arovas,prl,hu} was found to be very small in
Figs~\ref{figbragg} and~\ref{figbragg1}, the spectra in
Figs~\ref{figr1b01}-\ref{figr9b10} and~\ref{figmr0b05} show a more
significant influence in the inelastic neutron scattering
cross-section. These plots may be viewed as the 'band structure'
of the exciton moving in the vortex lattice, and the exciton
dispersion shows clear features at the Bragg reflection planes in
the reciprocal lattice of the vortex lattice. So we predict that a
careful study of the {\em inelastic\/} neutron scattering spectrum
may more easily yield evidence for the presence of the vortex
lattice.

Next, we turn to the recent STM measurements of Hoffman {\em et
al.}\cite{seamus}. These authors have observed signals of charge
order in the vortex lattice of BSCCO in the electron density of
states at sub-gap energies. The charge order is at wavevectors
${\bf K}_{cx} = (\pi/(2a),0)$ and ${\bf K}_{cy} = (0,\pi/(2a))$
(period of four lattice spacings), is peaked at the vortex cores,
and extends about to a distance which is about a quarter of the
inter-vortex spacing. These measurements are most likely in the SC
phase, where the SDW order is dynamically fluctuating. The
nucleation of charge order by vortices in such a phase (but with
the spins remaining dynamic) was predicted in
Refs.~\onlinecite{kwon,sns}. Lattice scale theories\cite{vsvzs} of
charge order in superconductors with preserved spin rotation
invariance also found a substantial doping range of bond-centered
charge order with a period of four lattice spacings, as did
density matrix renormalization group studies\cite{white}. The
spatial extent of the envelope of this charge order in the SC
phase has been computed in the present paper: the length scale in
the observations is quite similar to that in our computations in
Figs~\ref{figcdw02} and~\ref{figcdw05}. These computations were
carried out for the {\em same} set of parameters (only the value
of $s-s_c$ was changed to tune the doping level) used to obtain
general quantitative consistency with the neutron scattering
experiments above. The data of Hoffman {\em et al.} seems rather
similar to the result for $\Omega ({\bf r})$ at point $k$ in
Fig~\ref{figcdw05}, and the location of this point in the phase
diagram of Fig~\ref{figAM} is very reasonable, given the optimal
doping of their sample and of their $H$ value. This agreement
suggests to us that the system studied by Hoffman {\em et al.} has
dynamic spin excitons, above a spin gap, which extend throughout
the vortex lattice, as in Figs~\ref{figcdw02} and~\ref{figcdw05};
the charge order is then a signal of the pinning of these excitons
by terms like those in $\mathcal{S}_{\rm lat}$. An alternative
model, in which the spin order was confined only to the region
where charge order has been observed in STM, would have difficulty
explaining the neutron scattering experiments: spin order so
confined should yield easily observable satellite elastic Bragg
peaks at the wavevectors of the reciprocal of the vortex lattice.

Our computations also offer explanations for other features of the
STM data which would be difficult to understand in terms of charge
order nucleated independently in each vortex core: there is a
noticeable correlation between the phase and orientation of the
charge order between different vortices, which extends across the
entire experimental sample. We believe this correlation is induced
by the extended spin exciton states above the spin gap. Our model
for the STM experiments can therefore be summarized as follows:
the superflow in the vortex lattice reduces the energy of extended
spin exciton states, and the sliding degree of freedom associated
with spin density is then pinned by the vortex cores; this results
in static CDW around each vortex, but the SDW order remains
dynamic and gapped. A particular strength of our model is that it
consistently explains the STM and neutron scattering experiments
using the same set of parameters.

For the future, our theory suggests that neutron scattering and
STM studies of SDW/CDW order should be carried out in systems
where a uniform superflow has been induced directly by a current
source, with no magnetic field penetrating the sample. This will
eliminate the vortex cores, but the superflow should still enhance
the tendency for SDW/CDW order. Charge order can be pinned near
impurities/defects of various kinds ({\em e.g.\/} dislocations,
grain boundaries, surfaces), and so become visible to STM.

We briefly comment on the high field phases (SDW and ``Normal'')
in Fig~\ref{figpd}, in which superconductivity is destroyed by the
magnetic field. This regime may be of relevance to the experiments
of Boebinger {\em et al.}\cite{gregb}. Dynamic fluctuations of the
superconducting order surely become important as we approach these
phases, and so the theory of the present paper is not complete.
Nevertheless, given the nucleation of charge order near the vortex
cores in the SC phase (and its observation in the STM
experiments\cite{seamus}), it is natural to presume that this
charge order survives into the ``Normal'' phase. The transport
properties of the non-superconducting phases remain a very
interesting topic for future research, but our naive expectation
is that they are insulators.


Another  interesting type of experiments on superconductors in the
vortex state has been performed recently  by Curro {\it
et.al.}\cite{curro} and Mitrovi\'{c} {\it
et.al.}\cite{halperin,halperin2}. They measured the local field
dependence of the $^{17}$O spin-lattice relaxation rate ($1/T_1$)
and spin-echo decay rate ($1/T_2$), this allowed them to deduce
the rates as a function of position in the vortex lattice. Below
we suggest how these experiments can be interpreted in our picture
of the mixed state of the cuprates.
The spin-lattice relaxation rate $1/T_1$ measures the rate at
which nuclear spins are overturned as a result of interaction with
electron spins. In the BCS picture of vortices in a $d$-wave
superconductor\cite{curro,halperin,wortis,morr}, this quantity is
proportional to $N(0)^2$ and therefore increases dramatically
close to the vortex cores due to suppression in the
superconducting gap. On the other hand, as discussed in detail
earlier in this paper, for the not too overdoped cuprates, charge
density waves are nucleated around the vortex cores, which should
lead to a suppression in the local quasiparticle density of
states, and hence $1/T_1$. This effect appears to have been
observed in the experiments of Ref. ~\onlinecite{halperin}.
Another mechanism for the nuclear spin relaxation is via the
collective excitations of the electron system. In particular, the
excitonic SDW excitations provide a large number of low energy
$S=1$ excitations for flipping the nuclear spins. We suggest that
a strong increase in the the high field part of $1/T_1$
(corresponding to the vortex cores) with increasing magnetic field
in the experiments of Mitrovi\'{c} {\it et.al.}\cite{halperin}
reflects the growth of the SDW correlations and the corresponding
increase in the excitonic susceptibility. It would be interesting
to study this enhancement quantitatively and compare it with the
$H~\ln (1/H)$ behavior observed in neutron scattering experiments
and derived theoretically in this work. We mention that the
non-two-sublattice SDW makes this mechanism more effective for
relaxing the $^{17}$O nuclear spins, in contrast to the
$(\pi,\pi)$ electron magnetism which leads to a magnetic field on
the oxygen sites only through the Dzyaloshinskii-Moriya
interaction and weak ferromagnetism.
The echo decay rate $1/T_2$ is related to the inhomogeneity of the
local magnetic fields. The appearance of the local SDW order (or
sufficiently slow fluctuations) should therefore contribute  to
the increase in $1/T_2$. The SDW  enhancement is relatively
stronger around the vortex cores, which should give rise to the
enhancement in $1/T_2$ in this region; this agrees with the
experimental observations in Ref.~\onlinecite{curro}. The analysis
of our paper suggests that the difference in $1/T_2$ will not
become very large upon approaching the SC to SC+SDW boundary, as
the SDW excitations become extended close to this phase boundary.
As the magnetic field is increased, the SDW fluctuations should
become more pronounced, so we expect that $1/T_2$ will increase
for all values of the local field. By contrast, in the BCS theory,
one would expect that $1/T_2$ decreases with increasing magnetic
field, since the field becomes more uniform.

\begin{acknowledgments}
We thank Gabriel Aeppli, Robert Birgeneau, Antonio Castro Neto,
Rava da Silveira, Cristiane De Morais Smith, Seamus Davis, Mark
Kastner, Boris Khaykovich, Steven Kivelson, Bella Lake, Kristine
Lang, Young Lee, Andrew Millis, Anatoli Polkovnikov, Nick Read,
Matthias Vojta, Jan Zaanen, and Shou-Cheng Zhang for numerous
fruitful discussions. This research was supported by US NSF Grant
DMR 0098226.
\end{acknowledgments}

\appendix

\section{Dzyaloshinskii-Moriya Interaction}
\label{Dzyaloshinskii-Moriya}

An orthorombic distortion of  $La_{2-\delta} Sr_\delta Cu O_4$
results in the Dzyaloshinskii-Moriya (DM) interaction for the Cu
spins,
\begin{eqnarray}
{\cal H}_{DM} = \lambda \sum_{i,\delta} (-)^{i} \vec{d}\, \cdot
\vec{S}_i \times \vec{S}_{i+\delta}, \label{HamDM}
\end{eqnarray}
where the sum over $\delta$ extends over all the nearest neighbors
of site $i$, and $\vec{d}$ is a unit vector in the direction of
the orthorombic ${\bf a}$ axis (i.e. a diagonal $(1,1)$ direction)
\cite{Thio}. In this appendix, we study the effect of the DM
interaction on the non-two sublattice SDW ,and for simplicity we
consider a SDW at one wavevector only. The Hamiltonian
(\ref{HamDM}) mixes wavevectors ${\bf q}$ and ${\bf Q}+{\bf q}$,
where ${\bf Q}= (\pi/a,\pi/a)$. In this case we need to modify
(\ref{e1}) to
\begin{eqnarray}
\vec{S}({\bf r},\tau) = \mbox{Re} \left[e^{i {\bf K}_{sx} \cdot
{\bf r}} \vec{\Phi}_x ({\bf r}, \tau)+  e^{i ({ \bf K}_{sx} + {\bf
Q} ) \cdot {\bf r}} \vec{M}_x ({\bf r}, \tau)\right].
\end{eqnarray}
Straightforward algebra shows that the contribution of the DM
interaction to the action is
\begin{eqnarray}
\widetilde{\mathcal{S}}_{DM}+\mathcal{S}_M&=& \int d^2 d\tau \,
\big\{\,
  \lambda({\bf K}_s) \,\,\vec{d}\,\cdot
[\,\, \vec{\Phi}_{x} \times \vec{M}^*_x + \mbox{c.c.}]
\nonumber\\
&+& \frac{| \vec{M} |^2}{2 \chi} \,\big\}
\end{eqnarray}
where $\lambda({\bf K}_s) = 2 \lambda ( \cos ({\bf K}_{sx} {\bf
a}_x) +  \cos ({\bf K}_{sy} {\bf a}_y) )$ and the last term comes
from the fact that spin fluctuations $\vec{M}_x$ are massive. We
can now integrate $\vec{M}$ out, and find the anisotropy term for
the SDW order parameter
\begin{eqnarray}
\mathcal{S}_{DM}= - \frac{\lambda^2({\bf K}_s) \chi}{2} \int d^2
d\tau \, |\, \vec{\Phi}_{x} \times \vec{d}\,\, |^2 \label{DM2}
\end{eqnarray}

{}From (\ref{DM2}) we see that the DM interaction favors the
collinear SDW, with direction of $\vec{\Phi}$ perpendicular to
$\vec{d}$, {\em i.e.\/} along the orthorombic ${\bf b}$ axis
(direction of the SDW ordering is always in the CuO plane). We
also expect that the anisotropy becomes weaker with increasing
doping due to a decrease of $\lambda({\bf K}_s)$. However, the
typical scale for the anisotropy is small \cite{Thio}, and so we
expect that the quartic $u_2 | \vec{\Phi}_x^2|^2$ term in
(\ref{SPhi}) plays a dominant in selecting the collinear SDW at
low temperatures. We note that $\mathcal{S}_{DM}$ is quadratic, so
it will favor the collinear SDW fluctuations even above the
transition temperature.

\section{Zeeman coupling to the magnetic field}
\label{zeeman}

This appendix briefly discusses the effect of the Zeeman coupling
to the magnetic field on the action ${\cal S}_{\Phi}$ in
(\ref{SPhi}) for the SDW fluctuations. We will see that the
effects are weaker than those considered in the body of the paper,
especially near the critical point A at $s=s_c$ in zero field (see
Fig~\ref{figpd}).

\begin{figure}
\centerline{\includegraphics[width=2.9in]{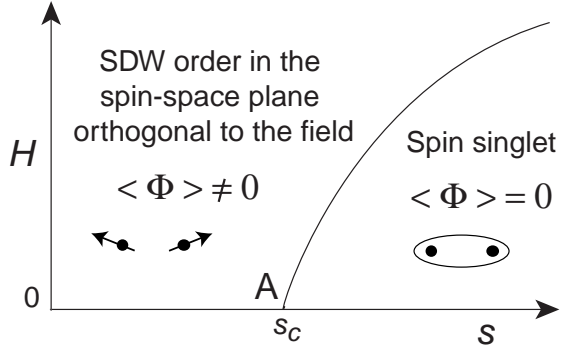}}
\caption{Phase diagram of ${\cal S}_{\Phi}$ in
(\protect\ref{SPhi}) including the Zeeman coupling in
(\protect\ref{rotate}). The point A is the same as the
corresponding point in Fig~\protect\ref{figpd}. The central
argument of Appendix~\ref{zeeman} is that it requires a much
larger field for $s>s_c $ near A to induce SDW order above, than
in Fig~\protect\ref{figpd}.} \label{figzeeman}
\end{figure}

As reviewed in Ref.~\onlinecite{conserve}, in systems without an
overdamped particle-hole continuum of spin excitations (as is the
case here hear the ordering momenta ${\bf K}_{sx,y}$), we can
deduce the coupling to the external field using simple gauge
invariance arguments. In particular, the primary consequence of
the external field is to rotate the spins uniformly about the
field axis, and this can be accounted for by the following
replacement to all temporal gradient terms:
\begin{equation}
\partial_{\tau} \Phi_{x \alpha} \rightarrow \partial_{\tau}
\Phi_{\alpha} - i \epsilon_{\alpha\beta\gamma} H_{\beta} \Phi_{x
\gamma}, \label{rotate}
\end{equation}
and similarly for $\Phi_{y \alpha}$. Here $H_{\alpha}$ is the
three vector in spin space representing the external field. The
resulting ${\cal S}_{\Phi}$ is closely related to models that have
been studied in some detail \cite{dsz,book} in the context of
double layer quantum hall systems. From this work, we can deduce
the phase diagram sketched in Fig~\ref{figzeeman}.
The most important property of this phase diagram is that zero
field phase transition at $s=s_c$ moves to finite field as $H \sim
(s-s_c)^{z \nu}$ where the exponent $z \nu=1/2$ in mean-field
theory. Fluctuation corrections will slightly increase this value,
but the present critical field will nevertheless remain {\em
larger} than the field in (\ref{amlog}) associated with the
corrections arising from the superflow. In particular, the phase
boundary in Fig~\ref{figzeeman} approaches the $H=0$ line with an
infinite slope. Consequently, the Zeeman shift is subdominant to
the stronger effects discussed in the body of the paper.

\section{Microscopic theory for coupling between SC and SDW order parameters}
\label{app:box}

In this Appendix we discuss the microscopic origin of the
effective interaction $\kappa$ between the SC and SDW order
parameters in (\ref{couple}). We will argue that repulsive $\kappa
> 0$ is a remarkable property of doped Mott insulators, but weakly
interacting electron systems quite possibly have $\kappa < 0$.

We start by considering a {\em weakly coupled} Fermi liquid of
electrons $c_{i \sigma}$ moving on the sites, $i$, of a square
lattice which is close to superconducting and commensurate
antiferromagnetic instabilities:
\begin{eqnarray}
{\cal Z} &=& \int {\cal D} c^\dagger {\cal D} c e^{S[c]}
\nonumber\\
S[c] &=& \int_0^\beta d\tau \left( \sum_i c^\dagger_i
\partial_\tau c_i - {\cal H}[c] \right)
\nonumber\\
{\cal H}[c] &=& \sum_{k\sigma} \epsilon_k c^\dagger_{k\sigma}
c_{k\sigma} + {\cal H}_{\rm imp}
\nonumber\\
&-&\sum_k ( \Delta_k c^\dagger_{k\uparrow} c^\dagger_{k\downarrow}
 + h.c. ) + \frac{\Delta_0^2}{ 2 \lambda_{SC}}
\nonumber\\
&-&\vec{\Phi} \sum_k  c^\dagger_{k+Q\alpha}
\vec{\sigma}_{\alpha\beta} c^\dagger_{k\beta} +
\frac{\vec{\Phi}^2}{ 2 \lambda_{AF}} \label{demlerweak}
\end{eqnarray}
Here $\Delta_k = \Delta_0 (\cos k_x - \cos k_y )/2 \equiv \Delta_0
d_k$ is the superconducting $d$-wave order parameter, and we
assume a nearest-neighbor tight binding dispersion of the
electrons $\epsilon_k = -2 t ( \cos k_x + \cos k_y ) - \mu$,
$Q=(\pi,\pi)$, and everywhere in this section momentum integrals
go over the first Brillouin zone. ${\cal H}_{\rm imp}$ describes
the static potential of the impurities which gives rise to a
finite quasiparticle lifetime
\begin{eqnarray}
\frac{1}{\tau} = 2 \pi n_{\rm imp} N(0) V^2,
\end{eqnarray}
where $N(0)$ is the density of states on the fermi level and $V$
is the impurity potential.

\begin{figure}
\centerline{\includegraphics[width=3in]{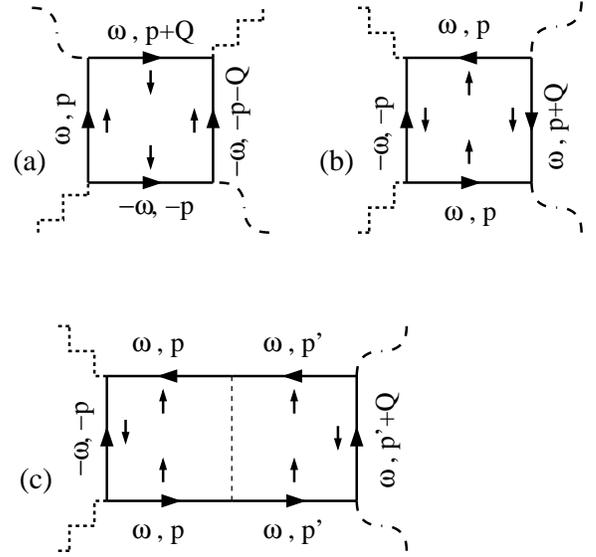}}
\caption{Diagrams that contribute to the effective interaction
between the superconducting and antiferromagnetic orders for Fermi
liquids. Solid lines correspond to the quasiparticle propagators,
zigzag lines correspond to the $d$-wave superconducting order
parameter and contribute a factor $d_p$; wavy line describe the
SDW, and a dashed line describes the static disorder potential. }
\label{figBox}
\end{figure}

Assuming that $\Delta_0$ and $\vec{\Phi}$ are small we can
integrate out the fermions and obtain
\begin{eqnarray}
{\cal Z} &=& e^{-\beta F}
\nonumber\\
F &=& F_{GL}[\Delta_0] + F_{AF}[\vec{\Phi}] + \kappa | \Delta_0
|^2 \vec{\Phi}^2
\end{eqnarray}
The diagrammatic representations of the terms that contribute to
$\kappa$ are shown on Fig.~\ref{figBox}.
Solid lines correspond to the quasiparticle propagators
\begin{eqnarray}
G(p,\omega_n)=(i \omega_n -\epsilon_p -i/(2\tau) \mbox{sgn}
(\omega)),
\end{eqnarray}
zigzag lines correspond to the $d$-wave superconducting order
parameter and contribute a factor $d_p$; wavy line describe the
SDW, and dashed line describes the static disorder potential. We
have
\begin{eqnarray}
{\rm (a)} &=&  \frac{1}{\beta} \int \frac{d^2 p}{(2 \pi)^2} ~d_p
d_{p+Q} ~\sum_{\omega_n} G(\omega_n,p) G(-\omega_n,-p) \nonumber\\
&\times & G(\omega_n,p+Q) G(-\omega_n,-p-Q)
\nonumber\\
{\rm (b)} &=&  \frac{1}{\beta} \int \frac{d^2 p}{(2 \pi)^2}
~d_p^2~ \sum_{\omega_n} G(\omega_n,p) G(-\omega_n,-p) \nonumber\\
&\times & G(\omega_n,p+Q) G(-\omega_n,-p-Q)
\nonumber\\
{\rm (c)} &=&  n_{\rm imp} V^2 \frac{1}{\beta}
 \sum_{\omega_n} L(\omega_n) M(\omega_n)
\label{kappa}
\end{eqnarray}
with
\begin{eqnarray}
L(\omega_n) &=& \int \frac{d^2 p}{(2 \pi)^2} ~d_p^2~ G(\omega_n,p)
G(-\omega_n,-p) G(\omega_n,p)
\nonumber\\
M(\omega_n) &=& \int \frac{d^2 p}{(2 \pi)^2} G(\omega_n,p)
G(-\omega_n,-p) G(\omega_n,p+Q) \nonumber \\ ~\label{LM}
\end{eqnarray}

It is useful to note that if we define the static spin
susceptibility at momentum $Q$ in the superconducting state
\begin{eqnarray}
\chi(Q) &=& -\frac{1}{\beta} \sum_{\omega_n} \int \frac{d^2 p}{(2
\pi)^2} \large\{ G_{sc}(\omega_n,p) G_{sc}(\omega_n,p+Q)
\nonumber\\
&~&~~~~~~~~~~~~~+ F(\omega_n,p) F(\omega_n,p+Q) \large\},
\label{chiAF}
\end{eqnarray}
with the Green's functions in the superconducting state  defined
in the usual manner \cite{mahan}, then
\begin{eqnarray}
\kappa = -\frac{\partial^2 \chi(Q)}{\partial \Delta_0^*
\partial \Delta_0 } \large|_{\Delta_0=0}
\end{eqnarray}
which agrees with (\ref{kappa}) and (\ref{LM}).

In the limit $\mu \tau \gg 1$ the main contribution to $\kappa$
comes from the diagram (a) and we find for $T \rightarrow 0$
\begin{eqnarray}
\kappa &=& -\pi \langle d_p^2 \rangle N(0) \frac{1}{\beta}
\sum_{n>0} \frac{1}{ (\omega_n+ 1/2\tau)(\mu^2 + (\omega_n+
1/2\tau)^2)}
\nonumber\\
&\approx& - \frac{1}{2\mu^2} \ln (\mu \tau) \label{kappa2}
\end{eqnarray}
It is important to note that in deriving the expression
(\ref{kappa2}) we relied on the fact that we have a $d$-wave
superconductor with $d_{p+Q}=-d_p$ and took the average value of
$\langle d_p^2 \rangle$ on the Fermi surface to be 1. Hence, such
Fermi liquids on the square lattice have an effective
``attraction'' between the antiferromagnetic and superconducting
orders, which can be traced back to the enhancement of the
antiferromagnetic susceptibility (\ref{chiAF}) in the $d$-wave
superconducting state.

We have so far examined the interplay between SC and SDW orders
near the boundary of their instability to a weakly interacting
Fermi liquid. Now let us turn to the same interplay, but in the
vicinity of a Mott insulator. Strong interactions are required to
produce the Mott insulator, and so the perturbative approach of
(\ref{demlerweak}) cannot be directly applied. Instead, we have to
turn to alternative strong coupling approaches, in which the
existence of the Mott insulator is built in at the outset. Such
approaches have been discussed recently, and these are expressed
in terms of collective degrees of freedom which are natural in the
vicinity of of Mott insulator. Electron spin singlet states, spin
one magnons, Cooper pairs of holes, and fermionic quasiparticles
are introduced as individual excitations, and interactions between
them are obtained from the microscopic $t$-$J$ Hamiltonian
\cite{kwon,auerbach} (phenomenological models of just the bosonic
degrees of freedom have also been considered\cite{projectedso5}).
All these papers find strong repulsion between magnon and hole
pair states, arising from the constraint on the allowed Hilbert
space. The origin of this repulsion therefore lies in the {\em
short distance, lattice-scale physics of allowed low energy states
near a Mott insulator}, rather than effects near the Fermi surface
in the weak-coupling analysis discussed earlier. The Cooper
pair-magnon repulsion immediately implies repulsion between the
superconducting and antiferromagnetic orders, since the
superconducting and antiferromagnetic phases correspond to the
condensates of the corresponding particles. As an example, see
Fig~2 in Ref.~\onlinecite{kwon}: the pairing amplitude is weak in
the region with magnetic order, but rises rapidly once the
magnetic order is suppressed.

\section{Renormalization group analysis of complex vector fields}
 \label{rg}

This appendix will briefly review existing theoretical results for
the critical properties of field theories which are similar to
${\cal S}_{\Phi}$, but simpler. The analysis of the full ${\cal
S}_{\Phi}$ theory will be addressed in future work.

The simplification made here is to consider a field theory with
only one complex vector field, $\Phi_{\alpha}$, with $\alpha = 1
\ldots m$; the original model has two such fields $\Phi_{x
\alpha}$ and $\Phi_{y, \alpha}$. For the case of only one such
field, we can always rescale $x$ and $y$ co-ordinates to make all
velocities unity; then in $d$ space dimensions we are interested
in the field theory with action
\begin{eqnarray}
{\cal S}_{c} &=& \int d^d r\, d\tau \,  \Bigl[\,\,
|\,\partial_\tau \Phi_{\alpha} |^2 + |\nabla_r  \Phi_{\alpha} |^2
+ s  | \,
\Phi_{\alpha} |^2  \nonumber \\
&~&~~~+ \frac{u_1}{2} | \, \Phi_{\alpha} |^4 + \frac{u_2}{2} | \,
\Phi_{\alpha}^2 |^2 \Bigr] \label{S-phi-x}
\end{eqnarray}

This theory has upper critical dimension $d=3$, and can be studied
in an expansion in $\epsilon=3-d$. Renormalization group equations
for the quartic terms were obtained to $O(\epsilon^2)$ by D. Jones
{\it et. al.} \cite{Jones,Bailin,Kawamura}
\begin{eqnarray}
\frac{du_1}{dl}&=& \epsilon\, u_1 -  K_d\, [\, (m+4) u_1^2 + 4 u_1
u_2 + 4 u_2^2\,] \nonumber \\
&~&~~+2 K_d^2 \,\Bigl[\,\frac{3}{2} \,(3m+7) u_1^3 +22 u_1^2 u_2
\nonumber \\&~&~~~~~~+ (5m +24) u_1 u_2^2 +4 (m+2) u_2^3\,\Bigr]
\nonumber\\
\frac{du_2}{dl}&=& \epsilon\, u_2 -  K_d\, [\, m u_2^2 + 6 u_1 u_2
\,] \nonumber \\
&~&~~- 2 K_d^2 \,\Bigl[\, (m-4) u_2^3 -2 (5+3m) u_2^2 u_1
\nonumber \\
&~&~~~~~~-\frac{1}{2} (5m+41) u_2 u_1^2\,\Bigr] \label{flow}
\end{eqnarray}
where $K_d=2^{-d+1}/[\pi^{d/2}\Gamma(d/2)]$. These flow equations
always have two unstable fixed points: the Gaussian point
$u^*_1=u^*_2=0$  and the isotropic O$(2m)$ Heisenberg fixed point
\begin{eqnarray}
u_1^* &=& \frac{\epsilon}{ K_d}\,\, \frac{1}{m+4} +O(\epsilon^2)
\nonumber\\
u_2^* &=& 0
\end{eqnarray}
For sufficiently large or small $m$ there may also be two other
fixed points
\begin{eqnarray}
u^* &=& \frac{\epsilon}{6 K_d} B_m [\, 3 m^2 - 12 m + 144 \mp 3m
R^{1/2}_m]\, \epsilon + O(\epsilon^2)
\nonumber\\
v^* &=& \frac{\epsilon}{ K_d} B_m [\, m^2+m-12 \pm 3 R^{1/2}_m]\,
\epsilon + O(\epsilon^2)
\end{eqnarray}
where $B_m^{-1} = m^3 + 4m^2 - 24 m + 144$ and $R_m = m^2 - 24 m +
48$. The last two fixed points are absent in the case of $m=3$. We
note, however, that for $m=2$ and in the large $m$ limit a stable
fixed point (the so-called chiral fixed point, see
Ref.~\onlinecite{Kawamura}) is possible for $u_2>0$, so it may
control the transition to the spiral order. When $u_2 <0$, the
system always flows towards strong coupling $u_2 \rightarrow -
\infty$, so we expect the transition to collinear order to be
weakly first-order.

\section{Numerical solution in the SC phase}
\label{numsc}

We will use the methods and notation described in
Brandt\cite{brandt}.

First, assume we know $\mathcal{V}_H ({\bf r})$. Write its Fourier
expansion in the form
\begin{equation}
\mathcal{V}_H ({\bf r}) = \sum_{{\bf G}} d_{{\bf G}} e^{i{\bf G}
\cdot {\bf r}} \label{n1}
\end{equation}
where $d_{{\bf G}}=d_{-{\bf G}}$ are both real, and ${\bf G}$ are
the reciprocal lattice vectors of the triangular vortex lattice.
Unlike the convention followed by Brandt, the sum over ${\bf G}$
always includes ${\bf G}=0$, unless stated otherwise explicitly.
In Brandt's notation, (\ref{n1}) can be inverted by $d_{{\bf G}} =
\langle \mathcal{V}_H ({\bf r}) \cos ({\bf G} \cdot {\bf r})
\rangle$, where the angular bracket denotes a spatial average.
Because of the symmetry we can work on only half a unit cell of
the vortex lattice, and for simplicity we choose the half unit
cell to be the one plotted in Fig. \ref{figcell}.

To obtain $G_H$, we want all the eigenvalues and eigenfunctions of
the Schr\"odinger equation (\ref{schro1}). As in the usual Bloch
theory, these are labeled by a wavevector ${\bf k}$ in the first
Brillouin zone, and a band index $\mu$. The explicit form of these
are
\begin{equation}
\Xi_{\mu {\bf k}} ({\bf r}) = \frac{e^{i {\bf k} \cdot {\bf r}}
}{\sqrt{A_{\cal U}}} \sum_{{\bf G}} c_{\mu {\bf G}} ({\bf k}) e^{i
{\bf G} \cdot {\bf r}} \label{xirr}
\end{equation}
where $A_{\cal U}$ is the area of the unit cell, and the $c_{\mu
{\bf G}} ({\bf k})$ are normalized so that
\begin{equation}
\sum_{{\bf G}} |c_{\mu {\bf G}} ({\bf k}) |^2 = 1
\end{equation}
If we choose $M$ values of ${\bf G}$ (also as in Brandt), then
$\mu=1 \ldots M$, and the $c_{\mu {\bf G}} ({\bf k})$ are the
orthonormal eigenvalues of the $M \times M$ matrix ${\cal M}_{{\bf
G}, {\bf G}'} ({\bf k})$ where
\begin{eqnarray}
&& \sum_{{\bf G}'} {\cal M}_{{\bf G}, {\bf G}'} ({\bf k}) c_{\mu
{\bf G}'} ({\bf k})
= E_\mu^2 ({\bf k}) c_{\mu {\bf G}} ({\bf k})\nonumber \\
&& {\cal M}_{{\bf G}, {\bf G}'} ({\bf k}) =  ({\bf k} + {\bf G})^2
\delta_{{\bf G}, {\bf G}'} + d_{{\bf G}-{\bf G}'} \label{M}
\end{eqnarray}
After this diagonalization we obtain the Fourier components of
(\ref{n1}) as
\begin{eqnarray}
&& d_{\bf 0} =  s - s_c + \kappa (\sum_{\bf G} a_{\bf G} - 1) +
\frac{N u}{N_k A_{\cal U}} \times
\nonumber \\
&&\!\!\!\!\sum_{{\bf k}, {\bf G}} \Bigl[ \frac{\coth(E_{\bf
G}({\bf k}) / (2 T))}{E_{\bf G}({\bf k})} - \frac{1}{\sqrt{ ( {\bf
k} + {\bf G} )^2 + \Delta_0^2 }}\Bigr], \label{d0}
\end{eqnarray}
and for ${\bf G} \ne 0$
\begin{eqnarray}
d_{\bf G} =&& -\kappa a_{\bf G} + \frac{N u}{2 N_k A_{\cal U}}
\sum_{{\bf k}, {\bf G'}, \mu} c_{\mu {\bf G'}}
\Bigl[c_{\mu ({\bf G'} + {\bf G})}({\bf k}) \nonumber \\
&& + c_{\mu ({\bf G'} - {\bf G})}({\bf k})\Bigr] \cdot
\frac{\coth(E_{\mu}({\bf k}) / (2 T))}{E_{\mu}({\bf k})},
\label{dK}
\end{eqnarray}
where the sum over ${\bf k}$ is over $N_k$ points which average
over the first Brillouin zone.  Also note that $c_{\mu {\bf G}}
({\bf k}) = c_{\mu,{\mathcal R}({\bf G})} ({\mathcal R}({\bf k}))$
where ${\mathcal R}$ denotes a rotation by $\pi/3$.  This can be
used to cut the number of ${\bf k}$ points in $1/6$.

The iteration of (\ref{d0}) and (\ref{dK}) will produce the
solution to (\ref{chif}) for a given $\psi_H({\bf r})$.

The next step is to solve (\ref{glf}), given $\mathcal{V}_H ({\bf
r})$ in (\ref{n1}). This is done just as in Brandt. His Eqn (9) is
replaced by
\begin{eqnarray}
&& (-{\bf \nabla}_{\bf r}^2  + 2 ) \omega = 2 \Bigl\{ \Bigl[ 1 +
\left(1 - \frac{\kappa^2}{4u\Upsilon}
\right) |\psi_0|^2 -\nonumber \\
&& \!\!\!\!\!\!\!\!\!\! \frac{\kappa}{4 u\Upsilon} (\mathcal{V}_H
- \Delta_0^2) \Bigr] \omega - \left( 1 -
\frac{\kappa^2}{4u\Upsilon} \right) \omega^2 - \omega Q^2 - g
\Bigr\}
\end{eqnarray}
and a corresponding change to Brandt's Eqn (11). The new form of
Brandt's Eqn (12) is
\begin{eqnarray}
&& a_{\bf G} := a_{{\bf G}} \times \nonumber \\
&& \!\!\!\!\!\!\!\!\!\!\!\!\!\!\!\!\frac{\left\langle \Bigl[
\left(1 - \frac{\kappa^2}{4 u\Upsilon}\right)|\psi_0|^2 -
\frac{\kappa}{4u\Upsilon} (\mathcal{V}_H - \Delta_0^2) \Bigr]
\omega - \omega Q^2 - g \right\rangle}{ \langle \omega^2 \rangle
(1 - \kappa^2 /(4 u\Upsilon))}.
\end{eqnarray}

After determining $\omega$ from above, we use this result to
obtain new $\mathcal{V}_H({\bf r})$ by solving (\ref{chif}), and
so on. By iteration of Eqn. (\ref{chif}) and (\ref{glf}), we will
be able to have the final solution to both of them.

Note that in order to get our numerical results, we did use a
finite momentum cutoff. However, the equations have been designed
to be cutoff independent and we did find that the Fourier
components of $\psi_H({\bf r})$ and $\mathcal{V}_H({\bf r})$
decreases rapidly upon going to higher momenta.

\section{Spin ordering phase boundary near M}
\label{amlin}

Here we discuss the analytical solution of (\ref{chif}) and
(\ref{glf}) in the vicinity of the multi-critical point M in
Fig~\ref{figpd}, with the aim of determining the location of the
AM phase boundary in its vicinity. Analytical progress is possible
because the amplitude of the superconducting order, $|\psi_H ({\bf
r})|^2$, is small in this region. Our analysis will show that in
this region AM behaves as $H = 1 - \varrho (\kappa - s + s_c)$,
where $\varrho$ is a numerical constant. The earlier full
numerical solution in Section~\ref{sec:pb} led to the estimate
$\varrho \approx 1.2$, and we shall find a consistent result here.

In addition to (\ref{n1}), we use the Fourier expansions
\begin{eqnarray}
T \sum_{\omega_n} G_H ( {\bf r}, {\bf r}, \omega_n ) - \int
\frac{d \omega d^2 k}{8 \pi^3} \frac{1}{\omega^2 +
 k^2+ \Delta_0^2} \nonumber \\
=  \sum_{{\bf G}} b_{{\bf G}} e^{i{\bf G} \cdot {\bf r}},
\end{eqnarray}
\begin{equation}
|\psi_H ({\bf r})|^2 = \sum_{{\bf G}} a_{{\bf G}} e^{i{\bf G}
\cdot {\bf r}}. \label{s1}
\end{equation}
Note that this notation for $a_{{\bf G}}$ is slightly different
from that above and in Brandt.

Then (\ref{chif}) becomes
\begin{eqnarray}
d_0 &=& \Delta_0^2 + \kappa ( a_0 - |\psi_0|^2 ) + 2 N u b_0 \nonumber \\
d_{{\bf G}} &=& \kappa a_{{\bf G}} + 2 N u b_{{\bf G}}
~~~~;~~~~{\bf G} \neq 0 \label{s2}
\end{eqnarray}

Second, we can solve (\ref{g}) by a Feynman graph expansion in
$d_{{\bf G} \neq 0}$. This yields
\begin{eqnarray}
b_0 &=& \int_0^{\infty} \frac{k dk}{2 \pi} \Bigl[ \frac{\coth
(\sqrt{ k^2 + d_0}/2T)}{ 2 \sqrt{ k^2 + d_0}}
\nonumber \\
&&- \frac{1}{2 \sqrt{ k^2 + \Delta_0^2}} \Bigr] + {\cal O}
(d_{{\bf G} \neq
0}^2) \nonumber \\
&=& \frac{\sqrt{d_0}-\Delta_0}{4 \pi } + {\cal O} (d_{{\bf G} \neq
0}^2)~~~~~~~\mbox{at $T=0$} \label{s3}
\end{eqnarray}
and
\begin{eqnarray}
b_{{\bf G}} &=& - \frac{d_{{\bf G}}}{4 \pi^2  } \int_0^{\infty}
\frac{d^2 k}{ ({\bf k}+{\bf G})^2 - k^2} \Bigl[ \frac{\coth
(\sqrt{ k^2
+ d_0}/2T)}{ 2 \sqrt{ k^2 + d_0}} \nonumber \\
&&-  \frac{\coth (\sqrt{ ({\bf k}+{\bf G})^2 + d_0}/2T)}{ 2 \sqrt{
({\bf k}+{\bf G})^2 +
d_0}}\Bigr] + {\cal O} (d_{{\bf G} \neq 0}^2) \nonumber \\
&=& -\frac{d_{{\bf G}}}{8 |{\bf G}|} + {\cal O} (d_{{\bf G} \neq
0}^2)~~~~\mbox{at $T=0$}~~;~~{\bf G} \neq 0\label{s4}
\end{eqnarray}
Now we can solve (\ref{s2},\ref{s3},\ref{s4}) for the  $d_{{\bf
G}}$ in terms of the $a_{{\bf G}}$.

Finally, we need to determine the $a_{{\bf G}}$ by solving
(\ref{glf}). This can be done with the realization that for small
$\psi_H$, the functional form of the superconducting order
parameter can be assumed to be equal to the Abrikosov solution. So
we assume
\begin{equation}
a_{{\bf G}} = -\frac{a_0 a^{A}_{{\bf G}}}{2}~~~~;~~~~{\bf G} \neq
0 \label{s5}
\end{equation}
where $a^A_{{\bf G}}$ is given in (8) of Brandt. Now, it remains
to obtain a single additional equation to determine $a_0$. This we
determine by multiplying (\ref{glf}) by $\psi_H^{\ast} ({\bf r})$
and averaging over all space. Using the property of the Abrikosov
solution for $\psi_{H} ({\bf r})$, we obtain
\begin{eqnarray}
\left(1- \frac{\kappa^2}{4 u \Upsilon} \right) \left( \sum_{{\bf
G}} a_{{\bf G}}
a_{-{\bf G}} - |\psi_0|^2 a_0 \right) \nonumber \\
+ \frac{\kappa}{4 u \Upsilon} \left( \sum_{{\bf G}} d_{{\bf G}}
a_{-{\bf G}} - \Delta_0^2 a_0 \right) - H a_0 = 0 \label{s6}
\end{eqnarray}

Eqns (\ref{s2},\ref{s3},\ref{s4},\ref{s5},\ref{s6}) are now simple
equations that can be easily solved to obtain all the Fourier
coefficients. The line AM corresponds to $d_0 = 0$. Our analytical
result of the slope of AM near M point is $\varrho \approx 1.1$,
which is in acceptable agreement with that obtained from the full
numerical solution.

\section{Numerical solution in the SC+SDW phase}
\label{numsdw}

Here we will describe the solution of the equations (\ref{mchir}),
(\ref{mzeta}) and (\ref{mglr}) for the unknowns $\mathcal{V}_H
({\bf r})$, $\psi_H ({\bf r})$ and $n_H ({\bf r})$. First, as
(\ref{mzeta}) is linear in $n_H({\bf r})$, it is convenient to
rescale
\begin{equation}
n_H ({\bf r}) \rightarrow  n_H ({\bf r})/\sqrt{2Nu},
\label{scalen}
\end{equation}
and these equations become:
\begin{eqnarray}
&& \mathcal{V}_H ({\bf r}) = s - s_c + \kappa (| \psi_H ({\bf r}) |^2
- 1) + n_H^2 ({\bf r}) \nonumber \\
&& \!\!\!\!\!\!\! + 2 N u \Bigl[T \sum_{\omega_n} G_H({\bf r},
{\bf r}, \omega_n) - \int \frac{d \omega d^2 k}{8 \pi^3} \frac{1}
{\omega^2 + k^2}\Bigr], \label{mchif}
\end{eqnarray}
\begin{equation}
(- {\bf \nabla}_{\bf r}^2 + \mathcal{V}_H ({\bf r})) n_H ({\bf r})
= 0, \label{mzetaf}
\end{equation}
\begin{eqnarray}
[\left(1-\frac{\kappa^2}{4 u \Upsilon}\right)(|\psi_H({\bf r})|^2
- 1) + \frac{\kappa}{4u \Upsilon} ( \mathcal{V}_H({\bf r}) -
s + s_c ) \nonumber \\
- ({\bf \nabla}_{\bf r} - i {\bf A})^2 ] \psi_H ({\bf r}) = 0.
\label{mglf}
\end{eqnarray}

We use two-step iteration to self-consistently solve the equations
(\ref{mchif}, \ref{mzetaf}, \ref{mglf}). The first step consists
of solving Eqn. (\ref{mchif}) and (\ref{mzetaf}), and the second
step is solving (\ref{mglf}).

For the first step, we use a four-substep iteration. First, define
and calculate
\begin{eqnarray}
&&\eta_H ({\bf r}) = s - s_c + \kappa (| \psi_H ({\bf r}) |^2 - 1) \nonumber \\
&& \!\!\!\!\!\!\!+ 2 N u \Bigl[ T \sum_{\omega_n} G_H({\bf r},
{\bf r}, \omega_n) - \int \frac{d \omega d^2 k}{8 \pi^3}
\frac{1}{\omega^2 +  k^2} \Bigr]. \label{mi1}
\end{eqnarray}
Second, define and calculate the inverse of operator
\begin{equation}
{\mathcal A} = -  {\bf \nabla}_{\bf r}^2 + \eta_H({\bf r}).
\label{mi2}
\end{equation}
Third, calculate $n_H({\bf r})$ which satisfies
\begin{equation}
n_H({\bf r}) = - {\mathcal A}^{-1} n_H^3 ({\bf r}). \label{mi3}
\end{equation}
Last, calculate $\mathcal{V}_H({\bf r})$ using
\begin{equation}
\mathcal{V}_H({\bf r}) = \eta_H({\bf r}) + n_H^2({\bf r}).
\label{mi4}
\end{equation}

Choosing proper initial value for $\mathcal{V}_H({\bf r})$ and
$n_H({\bf r})$ and iterate (\ref{mi1}), (\ref{mi2}), (\ref{mi3}),
(\ref{mi4}) will produce the solution to both (\ref{mchif}) and
(\ref{mzetaf}).

In practice, the above steps are performed in momentum space. If
we let
\begin{eqnarray}
\mathcal{V}_H({\bf r}) &=& \sum_{\bf G}d_{\bf G}e^{i {\bf G} \cdot {\bf r}}, \nonumber \\
n_H({\bf r}) &=& \sum_{\bf G}\widetilde{f}_{\bf G}e^{i {\bf G} \cdot {\bf r}}, \nonumber \\
\eta_H({\bf r}) &=& \sum_{\bf G}g_{\bf G}e^{i {\bf G} \cdot {\bf
r}}, \label{fkgk}
\end{eqnarray}
where ${\bf G}$ are the reciprocal lattice vectors of the vortex
lattice (note that $\widetilde{f}_{\bf G}$ differs slightly from
$f_{\bf G}$ in (\ref{fK}) because of the rescaling
(\ref{scalen})), then (\ref{mi1} - \ref{mi4}) become
\begin{eqnarray}
&&g_{\bf 0} = s - s_c + \kappa (\sum_{\bf G} a_{\bf G} - 1) \nonumber \\
&&\!\!\!\!\!\!\!\!\!\!\!\!\! + \frac{N u}{N_k A_{\cal U}}
\sum_{{\bf k} + {\bf G} \ne 0}\Bigl[ \frac{\coth(E_{\bf G}({\bf
k}) / (2 T))}{E_{\bf G}({\bf k})} - \frac{1}{ | {\bf k} + {\bf G}
| }\Bigr] , \label{mik11}
\end{eqnarray}
\begin{eqnarray}
g_{\bf G} &=& -\kappa a_{\bf G} + \frac{N u}{2 N_k A_{\cal U}}
\sum_{{\bf k}, {\bf G'}}\sum_{\mu, E_{\mu}({\bf k}) \ne 0} c_{\mu
{\bf G'}}
\Bigl[ c_{\mu ({\bf G'} + {\bf G})}({\bf k}) \nonumber \\
&+& c_{\mu ({\bf G'} - {\bf G})}({\bf k}) \Bigr] \cdot
\frac{\coth(E_{\mu}({\bf k}) / (2 T))}{E_{\mu}({\bf k})},
\label{mik12}
\end{eqnarray}
\begin{eqnarray}
{\mathcal A}_{\bf G G'} &=&  {\bf G}^2 \delta_{{\bf G}, {\bf G'}}
+ g_{{\bf G} - {\bf G'}} ,\\
\nonumber\\
\widetilde{f}_{\bf G} &=& - \sum_{\bf G'} {\mathcal A}^{-1}_{\bf G
G'}
\langle n_H^3 ({\bf r}) \cos({\bf G'} \cdot {\bf r}) \rangle, \label{mik3} \\
\nonumber\\
d_{\bf G} &=& g_{\bf G} + \langle n_H^2 ({\bf r}) \cos({\bf G'}
\cdot {\bf r}) \rangle.
\end{eqnarray}

Note that in the substep (\ref{mik3}) the equation is solved by
another smaller iteration.

The second step is very similar to the case with no magnetic order
as in Appendix~\ref{numsc}.  The equation (\ref{mglf}) can be
solved by a two-substep iteration of the following equations
\begin{eqnarray}
(-{\bf \nabla}_{\bf r}^2 + 2) && \omega = 2 \Bigl[ (1 + G({\bf r})) \omega \nonumber \\
&& - \left(1 - \frac{\kappa^2}{4u \Upsilon}\right) \omega^2 -
\omega Q^2 - g \Bigr]
\end{eqnarray}
where
\begin{equation}
G({\bf r}) = \left(1 - \frac{\kappa^2}{4u \Upsilon}\right) -
\frac{\kappa}{4 u \Upsilon}(\mathcal{V}_H({\bf r}) - s + s_c),
\end{equation}
and
\begin{equation}
a_{\bf G} = a_{\bf G} \cdot \frac{\langle \omega G - \omega Q^2 -g
\rangle}{\langle \omega^2 \rangle \left(1 - \kappa^2 / (4 u
\Upsilon\right))}.
\end{equation}
{}From the iteration results we are able to determine
$\mathcal{V}_H({\bf r})$, $n_H({\bf r})$ and $\psi_H({\bf r})$.


\begin{thebibliography}{}
\bibitem{homotopy} For completeness, we note that the
corresponding homotopy group for the case with circular spiral SDW
($u_2 > 0$ in $\mathcal{S}_{\Phi}$) is $\pi_1 (S_3/Z_2) = Z_2$.
\end{thebibliography}
\end{document}